\documentclass[10pt,letterpaper]{article}
\usepackage{amsmath,amsfonts}
\begin{document}

\newcommand{\HH}{\Hset\mathrm{H}^1}
\newcommand{\Bergman}{\Cset \mathrm{H}^2}
\newcommand{\Bergmanf}{\Csetf \mathrm{H}^2}
\newcommand{\AdS}{\mathrm{AdS}}
\newcommand{\Squashedsphere}{\tilde{\mathrm{S}}^3}
 
\newcommand{\Acal}{\mathcal{A}}
\newcommand{\Bcal}{\mathcal{B}}
\newcommand{\Gcal}{\mathcal{G}}
\newcommand{\Kcal}{\mathcal{K}}
\newcommand{\Ecal}{\mathcal{E}}

\newcommand{\Dzero}{\Delta_0}
\newcommand{\Done}{\Delta_1}
\newcommand{\Dim}{{\grl}}
\newcommand{\vx}{{\vec{x}}}

\newcommand{\zp}{\zeta_{+}}
\newcommand{\zm}{\zeta_{-}}
\newcommand{\ra}{\rightarrow}
\newcommand{\xb}{x}
\newcommand{\tr}{\theta_r}
\newcommand{\ts}{\theta_s}
\newcommand{\vone}{\mathrm{v}_1} 
\newcommand{\vtwo}{\mathrm{v}_2}
\newcommand{\grl}{\lambda}
\newcommand{\grL}{\Lambda}
\newcommand{\ff}{\varphi}
\newcommand{\mn}{{\mu\nu}}
\newcommand{\fr}{\varphi_r}
\newcommand{\fs}{\varphi_s}
\newcommand{\comment}[1]{}
\newcommand{\nceq}{\doteq}
\newcommand{\lra}{\leftrightarrow}

\newcommand{\twobytwo}[4]{\left(\begin{array}{cc} #1 &#2 \\ #3 & #4 \\ \end{array}\right)}
\newcommand{\threebythree}[9]{\left(\begin{array}{ccc} #1&#2&#3\\#4&#5&#6\\#7&#8&#9\\ 
\end{array}\right)}

\newcommand{\Gmk}{G^{(\mu,k)}}
\newcommand{\Kmk}{K^{(\mu,k)}}
\newcommand{\Ylmn}{\mathrm{Y}^{l}_{mn}}
\newcommand{\Plmn}{P^l_{mn}}
\newcommand{\Plnn}{P^l_{nn}}
\newcommand{\Pl}{P^l}
\newcommand{\half}{\frac{1}{2}}
\newcommand{\diff}{\mathrm{d}}
\newcommand{\sx}{\sigma_x}
\newcommand{\sy}{\sigma_y}
\newcommand{\sz}{\sigma_z}
\newcommand{\sone}{\sigma_1}
\newcommand{\stwo}{\sigma_2}

\newcommand{\sthree}{\sigma_3}
\newcommand{\Qrm}{\mathrm{Q}}
\newcommand{\Ocal}{\mathcal{O}}
\newcommand{\Rcal}{{\mathcal R}}
\newcommand{\Dcal}{{\mathcal D}}
\newcommand{\Mcal}{{\mathcal M}}
\newcommand{\Ncal}{{\mathcal N}}
\newcommand{\Lcal}{{\mathcal L}}
\newcommand{\Scal}{{\mathcal S}}

\newcommand{\Zset}{{\mathbb Z}}
\newcommand{\Rset}{\mathrm{I}\kern-.18em\mathrm{R}}
\newcommand{\Cset}{{\,\,{{{^{_{\pmb{\mid}}}}\kern-.47em{\mathrm C}}}}}
\newcommand{\Csetf}{\pmb{\mathbb{C}}}
\newcommand{\Hset}{\mathrm{I}\kern-.18em\mathrm{H}}

\newcommand{\Prm}{{\mathrm P}}
\newcommand{\Erm}{{\mathrm E}}
\newcommand{\Frm}{{\mathrm F}}
\newcommand{\Grm}{{\mathrm G}}
\newcommand{\Orm}{{\mathrm O}}
\newcommand{\Urm}{{\mathrm U}}
\newcommand{\Srm}{{\mathrm S}}
\newcommand{\SO}{\mathrm{SO}}
\newcommand{\Sp}{\mathrm{Sp}}
\newcommand{\SU}{\mathrm{SU}}

\newcommand{\p}{\partial}
\newcommand{\vt}{\vartheta}
\newcommand{\vf}{\varphi}

\numberwithin{equation}{section}

\begin{titlepage}

\begin{flushright}
 {\tt YITP-SB-04-13}
\end{flushright}

\mbox{}
\bigskip
\bigskip
\bigskip

\begin{center}
{\Large \bf A Conformally Invariant Holographic Two--Point Function \\ on the Berger Sphere}\\
\end{center}

\bigskip  
\bigskip  
\bigskip  
\bigskip 
  
\centerline{\bf Konstantinos Zoubos}

\bigskip  
\bigskip  
\bigskip

\centerline{\it C. N. Yang Institute for Theoretical Physics}
\centerline{\it State University of New York at Stony Brook}
\centerline{\it Stony Brook, New York 11794-3840}
\centerline{\it U. S. A.}
\bigskip

\centerline{\small \tt  
  kzoubos@insti.physics.sunysb.edu}  
  
\bigskip  
\bigskip  
\bigskip

\begin{center} {\bf Abstract}\end{center} 

\vskip 4pt We apply our previous work on Green's functions for the four--dimensional quaternionic Taub--NUT manifold
to obtain a scalar two--point function on the homogeneously squashed three--sphere (otherwise known as the Berger sphere), 
which lies at its conformal infinity. Using basic notions from conformal geometry and the theory of boundary value problems, 
in particular the Dirichlet--to--Robin operator, we establish that our two--point correlation function is conformally 
invariant and corresponds to a boundary operator of conformal dimension one. It is plausible that the methods we use
could have more general applications in an AdS/CFT context.

\end{titlepage}

\setcounter{footnote}{0}  
\noindent
\tableofcontents

\section{Introduction}

 The AdS/CFT correspondence \cite{Maldacena98,Gubseretal98,Witten98} 
provides a very precise example of the holographic phenomenon, whereby a $d+1$--dimensional gravitational
theory has an equivalent description as a $d$--dimensional theory without gravity. In AdS/CFT this ``dual''
theory is actually a gauge theory. In the limit where
on one side of the duality we have weakly--coupled gravity the gauge theory on the other side is 
expected to be strongly--coupled and to have a large number $N$ of colours. 
  Essentially, as first observed in \cite{Witten98}, in this limit one is simply dealing with a classical 
boundary problem for fields (perhaps with nonzero mass or spin) in AdS, whose (appropriately rescaled) 
boundary values are interpreted as sources for the dual fields in the gauge theory. The gauge theory can 
thus be thought of as living on the boundary of AdS. The bulk fields obey their equations of motion, 
so they can be constructed using Green's functions of the appropriate bulk laplacian. This
allows the calculation of correlation functions in the boundary theory. We refer to \cite{Aharonyetal00,DHokerFreedman02}
for reviews and a survey of the literature.

The Einstein metric on Euclidean Anti--de--Sitter space, or hyperbolic space, is the prototypical example of a 
\emph{conformally compact} Einstein metric on a compact riemannian manifold with boundary, defined as a complete metric 
 in the interior of the manifold which blows up with a double pole at the boundary, and can thus be extended to a
boundary metric via a suitable choice of defining function\footnote{The condition of conformal compactness, together with
the Einstein condition, implies that the metric is \emph{asymptotically hyperbolic}, that is, all sectional curvatures 
approach $-1$ near the boundary \cite{Anderson01}. (For hyperbolic space, the sectional curvature is $-1$ everywhere). So 
metrics of this type are often also called asymptotically hyperbolic, or asymptotically AdS.}. A defining function is
any function $\rho$ which is positive in the bulk, but has a simple zero at the boundary, so that $\rho^2 g$ extends smoothly
to the boundary. Since any such function can play the role of a defining function, 
the induced metric at the boundary is not unique. Thus we obtain    
an induced boundary conformal structure rather than a boundary metric. This is of course crucial for AdS/CFT to work, 
since any theory defined on the boundary will automatically be conformally invariant (modulo possible conformal
anomalies in even boundary dimensions). However, this fact is true 
for any theory conjectured to be dual to gravity on a
conformally compact manifold, so given the relative scarcity of conformal field theories in more than
two dimensions, it is of interest to see how AdS/CFT can be extended to this setting and possibly provide
nontrivial examples of CFT's. A first step in this direction would clearly involve solving the bulk laplacian to
obtain Green's functions, which would then lead to CFT correlation functions, in a similar way to the AdS case.

One way to approach this problem is the method of holographic renormalisation, reviewed in 
\cite{Skenderis02}, which uses extensively the framework of Fefferman and Graham \cite{FeffermanGraham85}, 
in which the bulk metric is expanded in a series of the radial distance from the boundary.  One can do the same 
for any bulk field, by prescribing a source field on the boundary and then performing an iterative near--boundary analysis, 
in which the bulk field equations are solved algebraically and substituted back, thus providing higher terms in the
radial series. Holographic renormalisation is a very general method, applicable to any asymptotically hyperbolic manifold,
and is ideally suited to calculating the local counterterms needed to regularise and renormalise the divergent bulk action.
However, its usefulness in calculating \emph{correlation} functions in the boundary is limited, because the necessary coefficient 
in the radial expansion, being nonlocally related to the prescribed source, cannot be fully determined by the 
near--boundary analysis.   
At that stage the analysis needs to be supplemented by an \emph{exact} solution of the bulk field equations for the field
in question. So although one could go rather far in analysing the dual field theory using holographic renormalisation, it
is still important to obtain exact bulk Green's functions for asymptotically hyperbolic manifolds. 

 In \cite{Zoubos02}, we discussed a particular example of such an asymptotically hyperbolic manifold, 
the Quaternionic Taub--NUT space (QTN). It is a four--dimensional manifold with $\SU(2)\times\Urm(1)$ 
isometry group, whose boundary conformal structure corresponds to a \emph{Berger sphere}, which is a homogeneous squashing
of the three-sphere preserving an $\SU(2)\times\Urm(1)$ isometry group out of the round three--sphere's $\SO(4)$. 
QTN can be thought of as a deformation of hyperbolic space $\HH$ (Euclidean AdS$_4$), but a very
special deformation in that it is \emph{conformally self--dual}, meaning that the Weyl tensor (zero for
$\HH$) is self--dual. Although a non--zero Weyl tensor means that, unlike the case of $\HH$ and its boundary the round sphere, 
the boundary manifold 
is not guaranteed to be conformally flat (and indeed it isn't), and thus deprives us of the standard trick of supposing 
the boundary is really just flat space, keeping the Weyl tensor self--dual is extremely useful in that there is a \emph{twistor}
description of the manifold. This means that we can translate differential--geometric questions about
the four--manifold (in this case, finding Green's functions) into complex--analytic questions about its 6--dimensional 
twistor space, hoping that the more powerful techniques thus available will allow us to answer them. 

 The outcome of \cite{Zoubos02} was the explicit calculation of bulk--to--bulk and 
bulk--to--boundary propagators for a \emph{conformally coupled} scalar on QTN. Assuming there is a holographically dual field theory,
we can apply the holographic prescription \cite{Witten98} to immediately find two-point correlation functions of the 
dual gauge--invariant operator, which (as will be discussed) is expected to have (mass) conformal dimension $\Dim=1$ in the 
strongly--coupled boundary theory. We will write down the two--point function in section \ref{Twopoint}. 
Checking that this two--point function, which we call $M(x,x')$,  is indeed the ``correct'' one 
(i.e. verifying that it is conformally invariant, and corresponds to a $\Dim=1$ operator)  will occupy the rest of this article. 

 Why is it not immediately obvious what conformal dimension our correlator corresponds to? After all, 
in flat space one only has to look at the power of the $1/|x-x'|$ term in a scalar two--point function 
and read off the conformal dimension of the corresponding operator.  However, recall that
all this information arises through the analysis of representations of 
the conformal group. As is well known, on the round sphere (or, equivalently, flat space) this group is large enough to allow a 
complete characterisation of two--point and three--point correlation functions (see e.g. \cite{OsbornPetkos94}). As an illustration, 
suppose $M(x,x')$ is a two--point function (in $d$ dimensions) of the scalar operator $\Ocal$, which has conformal 
dimension $\Dim$. Then, if $\upsilon$ is a conformal Killing vector, we have the following condition on the correlation 
function\footnote{We will very briefly indicate how this equation comes about in section \ref{Conformal}. A detailed discussion 
of similar formulas can be found in \cite{Erdmenger97}.}:
\begin{equation} \label{conformaltransformation}
\left[\Lcal_{\upsilon(x)}+\Lcal_{\upsilon(x')}+\Dim (\sigma_\upsilon(x)+\sigma_\upsilon(x'))\right]M(x,x')=0
\end{equation}
where $\Lcal_{\upsilon}$ is the Lie derivative, and  $\sigma_\upsilon(x)$ is defined through the conformal Killing equation
\begin{equation}
\Lcal_{\upsilon(x)} g_\mn-2\sigma_\upsilon(x) g_\mn=0
\end{equation}
to be $\sigma_\upsilon=\nabla\cdot\upsilon/d$. 
As we will discuss in section \ref{Conformal}, if we denote by $s$ the order of the differential operator $\Dcal_x$ 
associated to $M(x,x')$, in the sense that
\begin{equation} \label{DM}
\Dcal_x M(x,x')=\frac{1}{\sqrt{g}}\delta^{(d)}(x,x')
\end{equation}
(in other words, $\Dcal$ is the inverse of $M(x,x')$) then the quantities $d$, $\Dim$ and $s$ are related by
\begin{equation} \label{conformaldimension}
2\Dim=d-s
\end{equation}
As an illustration of (\ref{conformaltransformation}), pick a flat--space conformal Killing vector, for instance 
the dilatation vector 
$\upsilon=x^\mu \p_\mu$, and apply it to the well--known flat--space conformal correlation function $M(x,x')\sim 1/[(x-x')^2]^\Dim$. 
We get (notice here $\sigma_\upsilon(x)=\sigma_\upsilon(x')=1$) 
\begin{equation} 
[\Lcal_{\upsilon(x)}+\Lcal_{\upsilon(x')}]M(x,x')=-2\Dim M(x,x')
\end{equation}
corresponding to the well--known fact that $M(x,x')$ has conformal dimension $\Dim$. If we put $s=2$, i.e. $M(x,x')$ is the
Green's function of the laplacian, we get the dimension $\Dim=d/2-1$ of a free scalar field.  
  
 Now let us turn to the Berger sphere. By Obata's theorem \cite{Obata71} the only \emph{compact} manifolds 
where the conformal group is larger than the isometry group are the spheres, equipped with the standard 
bi--invariant metric. So in the case of the Berger sphere, which is only left--invariant, the conformal group is just the 
isometry group, in other words there are no conformal Killing vectors apart from the Killing vectors themselves\footnote{Note
however that in the limit of extreme prolate squashing the left--invariant sphere degenerates to a 
CR structure, which is not a smooth manifold, Obata's theorem does not apply, and indeed (as discussed
in \cite{Britto--Pacumioetal99}, where it was realised as the boundary of the Bergman space $\Bergmanf$) 
 this space allows conformal Killing vectors. Lee \cite{Lee96} and Schoen \cite{Schoen95}
have analysed this case (the unit sphere in $\Csetf^n$), which turns out to be the unique CR manifold to admit a 
nontrivial conformal group. The analogue of conformal flatness in this case is the CR equivalence to the Heisenberg group.}. 
So the method just outlined would not work for the Berger sphere, and we need
to look for more sophisticated ways of showing that our correlation function has the form that is required by conformal
invariance on the Berger sphere.

 It turns out that a suitable framework to discuss this problem is that of \emph{conformal geometry}, a
field pursued in recent years by several mathematicians, and which has deep connections to 
the ideas of Fefferman and Graham on conformal invariants. Thus section \ref{Conformal} will be devoted to a lightning
review of the main ideas. Then, motivated by (\ref{DM}), we will be led to ask the question: ``Since our correlator is  
a Green's function, which is the operator it is the Green's function of?''. In other words, given our
two--point function $M(x,x')$, we will look for an operator $\Dcal$ on the Berger sphere such that (\ref{DM}) holds. The 
idea is that, although (\ref{conformaltransformation}) is not useful anymore in constraining $M(x,x')$, if we can
find an operator $\Dcal$, with \emph{known conformal transformation properties},  such that equation (\ref{DM})  
holds, then it can be used to constrain $M(x,x')$. Of course this hinges on the existence of an independent way of 
finding $\Dcal$. Note that if  $M(x,x')$ indeed corresponds to $\Dim=1$, then from (\ref{conformaldimension}) we should expect 
$\Dcal$ to be a first--order operator. 

In section \ref{DirichlettoNeumann} we will see that this $\Dcal$ turns out to be the Dirichlet--to--conformal--Neumann 
operator (we will mostly call it Dirichlet--to--Robin for short), which for hyperbolic space takes the simple form:
\begin{equation} \label{DtN}
\Bcal=\sqrt{\Delta_B+1/4}
\end{equation}
where $\Delta_B$ is the (positive) laplacian on the boundary. So $\Bcal$ is a pseudodifferential operator of order one 
on round $\Srm^3$, as anticipated\footnote{It should be clear 
that $\Bcal$ has nothing to do with the Dirac operator. Apart from the fact that 
$\Bcal^2=\Delta_B+1/4=\Delta_B+\Rcal_B/6$ (the scalar curvature of $\Srm^3$ is $\Rcal_B=3/2$ in our 
conventions---see appendix \ref{BergerSphere}), while (according to Lichnerowicz) 
${\Dcal\!\!\!\!\slash\,}^2=\Delta_B+\Rcal_B/4$, it is crucial that $\Bcal$ has to be a \emph{nonlocal} 
operator, as will be discussed in section \ref{DirichlettoNeumann}.}. Before explaining why, on general principles, 
 $\Bcal$ is expected to be the inverse of $M(x,x')$, we will discuss a way to construct 
this operator for a given manifold with boundary, give some background on the kind of problems it appears in, and go 
over its properties as a \emph{conformally covariant} operator.  

The outcome of section \ref{DirichlettoNeumann} is  that we can verify that $M(x,x')$ is conformally invariant 
by simply checking (\ref{DM}) with
$\Dcal=\Bcal$. We find it instructive to demonstrate this  first in the simple case of hyperbolic space where 
everything is clear and can be easily understood. This is the purpose of section \ref{Roundsphere}. To act 
with $\Bcal$ on $M(x,x')$ we will find it  convenient to expand  $M(x,x')$ in eigenfunctions, which is trivial 
to do on the round sphere. 

The confirmation we have been looking for will finally appear in section \ref{Bergersphere}, where we apply our
methods to Quaternionic Taub--NUT, and its conformal boundary the Berger sphere. Here too the fact
that the eigenfunctions of the laplacian are known will be crucial in avoiding the use of heavy
functional--analytic machinery. The drawback is that expanding $M(x,x')$ is a rather more involved technical exercise.

Section \ref{Conclusions}  contains some conclusions and possibilities for future work. We have also
included an appendix collecting various facts on the squashed three--sphere which we will use at various
places in this paper, as well as an appendix with some facts and references on pseudodifferential operators.

\section{The conformally invariant correlation function} \label{Twopoint}

In this section, after  first discussing some properties of the QTN metric, we state the results of the analysis
in \cite{Zoubos02}, namely the bulk--to--bulk and bulk--to--boundary propagators for a conformally coupled scalar. 
Then we can easily find a two--point function on the boundary Berger sphere, which we argue should correspond to a
dual operator of conformal dimension $\Dim=1$. We take the AdS limit and check that it is so in that special case.

\subsection{The Pedersen metric for QTN}

Our starting point is the Pedersen metric \cite{Pedersen86} for quaternionic Taub--NUT 
\footnote{There are several other forms of the metric for this manifold, which is often called AdS--Taub--NUT in the 
physics literature. See \cite{Zoubos02} for references and more details.}:
\begin{equation} \label{Pedersen}
 \diff s^2=\frac{4}{(1-k^2r^2)^2}\left[\frac{1-\mu^2 r^2}{1-\mu^2 k^2r^4}
\diff r^2+r^2 (1-\mu^2 r^2)(\sone^2+\stwo^2)+\frac{r^2(1-\mu^2 k^2 r^4)}
{1-\mu^2 r^2}\sthree^2\right]\,.
\end{equation}
The parameter $\mu^2$ can take values in the range $[k^2,-\infty)$, while the parameter $k^2$ is related
to the (negative) cosmological constant through $\grL=-3k^2$. Keeping $\mu^2\in[0,k^2]$ corresponds to
a prolate squashing of the three-sphere at the boundary, while $\mu^2\in[0,-\infty)$ to an oblate squashing. We will
restrict our attention to the prolate case in this article. In this range QTN has two limits which are special, in the
sense that they are symmetric spaces: $\mu^2=0$ is just 4--dimensional hyperbolic space (in other words, 
Euclidean $\AdS_4$) which we  denote by $\HH$, and $\mu^2=k^2$ is an Einstein--K\"ahler manifold known
as the Bergman space, often denoted by $\Bergman$, a notation we will adopt\footnote{In \cite{Zoubos02} the
Bergman space was denoted by $\widetilde{\Csetf \Prm^2}$.}. As cosets, these manifolds are $\HH=\SO(1,4)/\SO(4)$ and 
$\Bergman=\SU(2,1)/\Urm(2)$. 
 
As for the radial coordinate $r$, we take it to lie in the range $[0,1/k)$. This choice ensures that the
metric is complete. The boundary is at $r\ra 1/k$ where the metric blows up and smoothly induces a boundary
conformal structure. Finally, the $\sigma_i$ are the left--invariant $\SU(2)$ one--forms, which are given 
explicitly in appendix \ref{BergerSphere}. Taking $(1-k^2r^2)$ as our defining function, and calling $i,j$ the 
non--radial directions, the induced metric at the boundary is 
\begin{equation}
g^{(3)}_{ij}=\lim_{r\ra1/k}(1-k^2r^2)^2g_{ij}\quad\Rightarrow\quad \diff s^2=\frac{1}{k^2}\left[(1-\mu^2/k^2)(\sone^2+\stwo^2)+
\sthree^2\right]
\end{equation}
This is the left--invariant metric of the Berger sphere, which is examined in some detail in appendix \ref{BergerSphere}. 

The twistor space of QTN was constructed in \cite{Pedersen86}. 
Knowledge of the twistor space was essential in \cite{Zoubos02}, since it allowed us to make an 
educated ansatz (following a similar construction of Page \cite{Page79} for Euclidean Taub--NUT) 
for the Green's function in the case of a conformally coupled scalar field. Using this ansatz we were led to 
an explicit form for the Green's function. In this article we will start with this result, so no
twistor methods will be used. 

We now wish to consider free scalar fields propagating on the QTN geometry\footnote{That is, we ignore 
self--interactions as well as any backreaction on the geometry.}, 
having a \emph{conformal} coupling to the curvature. We recall from \cite{Zoubos02} the scalar laplacian 
corresponding to the metric (\ref{Pedersen}):
\begin{equation} \label{Pedlaplacian}
\begin{split}
\nabla^2&=\frac{1}{\sqrt{g}}\partial_\mu\sqrt{g}g^\mn\partial_\nu\\
\\
&=\frac{(1-k^2r^2)^2}{(1-\mu^2 r^2)}
\left[\frac{1}{4}\left(1-\mu^2k^2r^4\right)\p_{rr}
 +\frac{1}{4r(1-k^2r^2)}\left(3+k^2r^2-7\mu^2k^2r^4+3\mu^2 k^4r^6\right)\p_r\right.\\
\\
&\left.+\frac{1}{r^2}\left\{\p_{\theta\theta}+\cot\theta\p_\theta
+\csc^2\theta\left[\p_{\ff\ff}-2\cos\theta\p_{\ff\psi}
+\left(\sin^2\theta\frac{(1-\mu^2r^2)^2}{1-\mu^2k^2r^4}+\cos^2\theta\right)
\p_{\psi\psi}\right]\right\}\right]\;\;.\\
\end{split}
\end{equation}
The conformal laplacian $Y=\nabla^2-\Rcal/6$ corresponds, since the scalar curvature is 
$\Rcal=4\grL=-12k^2$, to the operator $Y=\nabla^2+2k^2$. We should now supplement this operator with appropriate
boundary conditions on the boundary of QTN.
 
Note that for massless scalars, with an arbitrary curvature coupling
$\xi$, satisfying Dirichlet boundary conditions, the presence of the bulk $\Rcal$ term leads us to consider the 
following action (see \cite{MincesRivelles01}): 
\begin{equation} \label{Scalaraction}
S=-\half \int_\mathcal{M}\diff^{d+1}x\sqrt{g}\left[\p^\mu\phi\p_\mu\phi+\xi \Rcal \phi^2\right]
+\xi\int_{\p\mathcal{M}}\diff^d x \sqrt{h} \mathcal{K} \phi^2\;\;.
\end{equation}
where $\Kcal=-\nabla_r N^r$ is the trace of the extrinsic curvature ($N^r$ is the outward unit normal vector) 
and $h_{ij}$ is the (unrescaled, thus divergent) restriction of the metric 
to the boundary. The term with the extrinsic curvature plays a similar role to the Gibbons--Hawking
term \cite{GibbonsHawking77} in the Einstein--Hilbert action, i.e. it ensures that the variational problem is well 
defined, because with this addition:
\begin{enumerate}
\item $S$ is stationary under $ g_\mn\ra g_\mn+\delta g_\mn, $ with the metric and its normal derivatives fixed on the boundary, and 
\item $S$ is also stationary under $ \phi\ra \phi +\delta\phi, $ where $\phi$ satisfies Dirichlet boundary 
conditions $ \delta\phi|_{\p\mathcal{M}}=0$.
\end{enumerate}
Now recall that, as first discussed in \cite{KlebanovWitten99b} (see also \cite{MincesRivelles01,MuckViswanathan99}),  
for low values of the conformal dimension 
we have a choice of boundary conditions in AdS/CFT, in the sense that the same solution for the bulk field can correspond to 
two different boundary operators, depending on the asymptotic behaviour we wish to impose.  
The ``regular'' boundary condition (which can be imposed for any bulk mass) leads to a boundary 
operator with the usual conformal dimension $\Dim_+=d/2+\sqrt{d^2/4+m^2}$, while the ``irregular'' boundary condition leads 
to $\Dim_-=d/2-\sqrt{d^2/4+m^2}$. Massless, conformally coupled fields (where effectively  $m^2=\xi\Rcal$, and $\Rcal$ is 
negative for asymptotically hyperbolic spaces) lie in this region of two possible quantisations, however one should check 
whether both boundary conditions are consistent with the conformal coupling.   
Since unitarity in a CFT requires $\Dim\geq(d-1)/2$, and also  the fields in the bulk should have masses above
the Breitenlohner--Freedman bound of $-d^2/4$, it follows that the irregular modes can only exist in the range 
$0\leq d^2/4+\xi\Rcal<1$. However, the analysis of the irregular modes in \cite{MincesRivelles01} leads to a further restriction on the values 
of $\xi$: $\xi=\frac{1}{2d}\Dim_- $, where $\Dim_-=d/2-\sqrt{d^2/4+\xi\Rcal}$.
For massless scalars, there are two solutions for $\xi$ in $d+1=4$: $\xi=0$ (which is not interesting since
$\Dim_-$ takes the non--unitary value of zero) and $\xi=\frac{1}{6}$ (which gives $\Dim_-=1$). 
So the conformal value $\xi=1/6$ is allowed, and thus it is consistent to impose Dirichlet boundary conditions for conformal
coupling.

On the other hand, as dicussed e.g. in \cite{Solodukhin99,MincesRivelles01}, it is a consequence of 
the conformal coupling that we cannot consistently impose \emph{Neumann} boundary conditions for 
the action (\ref{Scalaraction}). Varying the action with respect to $\phi$ (and using the bulk
equations of motion) we have
\begin{equation}
\delta_\phi S=-\int_{\p \Mcal}\diff^d x\sqrt{h}\left(N^r\p_r\phi-2\xi\Kcal \phi\right)\delta\phi\;.
\end{equation}
So, while we are obviously free to impose the Dirichlet condition, the only other boundary condition that keeps the 
action invariant is the Robin--type condition:
\begin{equation} \label{Robincondition}
\left(N^r\p_r-\frac13\Kcal\right)\ra 0\; \text{as}\; r\ra 1/k\;.
\end{equation}
Here we have already substituted the conformal value of $\xi$. The precise scaling as we approach the boundary will be dicussed
below. In what follows we 
will sometimes call this condition \emph{conformal Neumann}, but will frequently revert to
calling it simply ``Robin'' for short.

\subsection{Green's functions on QTN} \label{Propagators}

Before writing down  the Green's functions on QTN constructed in \cite{Zoubos02}, we recall a few basic
facts about boundary problems and Green's functions in the asymptotically hyperbolic case. This is well known material in
AdS/CFT, but since the setting here is slightly more general we find it useful to retrace some steps.

Given an asymptotically hyperbolic manifold $\Mcal$ in $d+1$ dimensions, the massive laplacian 
$\nabla^2-m^2=\nabla^2+\Dim(d-\Dim)k^2$ admits two independent solutions which, in coordinates such that the boundary
is at $y=0$, scale as $y^\Dim$ and $y^{d-\Dim}$ (see e.g. \cite{MazzeoMelrose87}). So the general solution for a massive scalar 
behaves asymptotically as 
\begin{equation} \label{indepsolutions}
\phi(y,x)= y^{d-\Dim}f(y,\vec{x})+ y^{\Dim}h(y,\vec{x})\;.
\end{equation} 
The first terms $f_0=f(0,\vec{x})$ and $h_0=h(0,\vec{x})$ in an expansion of $f(y,\vec{x})$ and $h(y,\vec{x})$ are the only
independent quantities, with the higher orders being uniquely (and locally) specified by them \cite{KlebanovWitten99b,Skenderis02}. 
Now for large values of the conformal dimension (meaning $\Dim>d$), the first term in (\ref{indepsolutions}) 
is non--normalisable and corresponds to (modified) Dirichlet boundary conditions, while the second one is 
normalisable and corresponds to (modified) Neumann boundary conditions. So the 
Dirichlet problem consists of specifying $f_0$ on the boundary, while in the Neumann problem we specify $h_0$. In each case we can 
construct the solution for $\phi(y,\vx)$ in terms of a suitable Green's function. As discussed in \cite{Solodukhin99} 
the asymptotic behaviour of these Dirichlet and Neumann Green's functions is $G^\Dim_D\ra y^{\Dim}$ and $G^\Dim_N\ra y^{d-\Dim}$ 
respectively. The asymptotic behaviour of the Neumann Green's function comes from the requirement that it satisfy 
$(N^y\p_y+(d-\Dim)k)G^\Dim_N\ra y^{\Dim}$ asymptotically. 

Now we turn to our case of a conformally coupled scalar in $d+1=4$, where effectively $m^2=-2k^2$, resulting in 
the two solutions $\Dim_+=2$ and $\Dim_-=1$. Let us then write the asymptotic behaviour of $\phi(y,\vx)$ as
\begin{equation} \label{phitwo}
\phi(y,\vec{x})=y f(\vec{x})+y^2 h(\vec{x})+\cdots
\end{equation}
We have dropped the subscript of $f$ and $h$, which should cause no confusion since the higher order terms in each 
are contained in the ``$\cdots$''.  
As discussed in the previous section, in this case, where both modes are normalisable, we have a choice of boundary 
value problems, depending on which 
$\Dim$ we pick. Suppose we want to solve for $\phi(y,\vx)$ by specifying $h(\vx)$. If $\Dim=2$, this is clearly a Neumann problem. 
However, thinking of it from a $\Dim=1$ point of view, it is the Dirichlet problem. 
We recall that the (modified) Dirichlet problem in this case is the question of constructing a function $\phi(y,\vec{x})$ satisfying
\begin{equation}
\left\{\begin{array}{l}\left(\nabla^2+2k^2\right)\phi(y,\vec{x})=0 \\
\left[y^{\Dim-d}\phi(y,\vec{x})\right]_{\p\Mcal}=f(\vx)\end{array}\right.
\end{equation}
while the (modified) Neumann problem corresponds to finding a $\phi(y,\vec{x})$ such that\footnote{The extra term in the boundary
condition can be easily seen to follow from Green's third identity applied to the modified Neumann problem \cite{Solodukhin99}.}
\begin{equation}
\left\{\begin{array}{l}\left(\nabla^2+2k^2\right)\phi(y,\vec{x})=0\\
\left[y^{-\Dim}(N+(d-\Dim)k)\phi(y,\vec{x})\right]_{\p\Mcal}=-h(\vx)\end{array}\right.
\end{equation}
where $N=N^y\p_y$ is the outward unit normal, as before\footnote{It is basically given by $-y\p_y$ in the $(y,\vx)$ coordinates.}
. By $[\cdots]_{\p\Mcal}$ we mean the restriction of the point $y$ to 
the boundary, and $[\cdots]_{\p\Mcal'}$ below will mean the restriction of $y'$ to the boundary. Finally, if we know the 
Dirichlet and Neumann Green's functions, the Dirichlet problem can be solved through a double--layer potential (here 
$\diff\Omega(x)$ is the measure of the induced, finite boundary metric)
\begin{equation}
\phi(y,\vec{x})=-\int \left[(y')^{-\Dim}(N_{y'}+(d-\Dim)k) G^\Dim_D(y,\vec{x},y',\vec{x}')\right]_{\p\Mcal'}
\cdot f(\vec{x}')\diff\Omega(x')
\end{equation} 
while the Neumann problem is solved using a single--layer potential
\begin{equation}
\phi(y,\vec{x})=-\int [(y')^{\Dim-d}G^\Dim_N(y,\vec{x},y',\vec{x}')]_{\p\Mcal'} \cdot h(\vec{x}') \diff\Omega(x')\;.
\end{equation}
As we will see explicitly in sections \ref{HHfactorisation} and \ref{QTNfactorisation}, for $\Dim=2$ the factor of $(d-\Dim)k$ is 
precisely the restriction to the boundary of the extrinsic curvature term $-1/3\Kcal$ appearing in the Robin problem, so 
we can actually switch to calling $G^2_N(y,\vx,y',\vx')$ the Robin Green's function, and denote it $G^2_R(y,\vx,y',\vx')$. 

After these preliminaries, let us turn to the bulk--to--bulk Green's function for a conformally coupled scalar on QTN, which 
was found in \cite{Zoubos02} to be\footnote{Here we switch to the coordinates of (\ref{Pedersen}), where the boundary is at $r=1/k$,
and will use the notation $x_r=(r,x)$ and $x_s=(s,x')$ for the two bulk points, 
where $r,s$ are the corresponding radial coordinates, and $x,x'$ are shorthand for the angular coordinates 
$(\theta_r,\varphi_r, \psi_r)$ and $(\theta_s,\varphi_s,\psi_s)$ respectively.
 See section \ref{HHfactorisation} for $N^r$ in AdS and section \ref{QTNfactorisation} for $N^r$ in QTN.} 
\begin{equation} \label{Gone}
G^{(\mu,k)}(r,x,s,x')=\frac{(1-k^2r^2)(1-k^2s^2)}{\sqrt{D}}\frac{Q_+-Q_-}
{Q_++Q_--4rsC \vone}\;\;,
\end{equation}
where
\begin{equation}
Q_\pm=(r^2+s^2\pm\sqrt{D}+2\mu kr^2s^2\vtwo)(1\pm\mu k\sqrt{D}-\mu^2k^2r^2s^2(2\vtwo-1))
^{\half\left(1-\frac{\mu}{k}\right)}
\end{equation}
with $D=(r^2-s^2)^2-4r^2s^2(\vtwo-1)+4\mu^2k^2r^4s^4\vtwo(\vtwo-1)$ and 
\begin{equation}
C=\left[(1-\mu^2k^2r^4)(1-\mu^2k^2s^4)\right]^{\frac{1}{4}\left(1-\frac{\mu}{k}\right)}
\sqrt{(1+\mu k r^2)(1+\mu k s^2)}\;\;.
\end{equation}
The $\vone$ and $\vtwo$ are $\SU(2)\times\Urm(1)$--invariant combinations of the three Euler angles describing
the three--sphere (about which see appendix \ref{BergerSphere})\footnote{We found it convenient to rescale $\vone$ to 
half its value relative to \cite{Zoubos02}, which leads to slight changes in the formulas compared to 
that paper.}
\begin{equation} \label{vone}
\vone=\cos\frac{\tr}{2}\cos\frac{\ts}{2}\cos\half(\psi_r+\fr-\psi_s-\fs)
+\sin\frac{\tr}{2}\sin\frac{\ts}{2}\cos\half(\psi_r-\fr-\psi_s+\fs)
\end{equation}
and
\begin{equation} \label{vtwo}
\vtwo=\half\{1+\cos\theta_r\cos\theta_s+
\sin\theta_r\sin\theta_s\cos(\varphi_r-\varphi_s)\}\;\;.
\end{equation} 
Note that we have ignored the normalisation of (\ref{Gone}) and we will similarly ignore normalisation constants 
in the following propagators. $G^{(\mu,k)}$ satisfies $\sqrt{g}(\nabla^2+2k^2)G^{(\mu,k)}(r,x,s,x')=\delta^{(4)}(x_r,x_s)$ in
the bulk of QTN, and it scales as $(1-k^2r^2)$ at the boundary $r\ra1/k$. So, according to the discussion above,  
if we have $\Dim=1$ it is the Dirichlet Green's function $G^1_D$, while for $\Dim=2$ we should think of it as the 
Robin Green's function $G^2_R$. We can also verify for $\Dim=2$ that its Robin derivative scales as 
\begin{equation}
\left[(N+k)\Gmk(r,x,s,x')\right]_{\p\Mcal}\sim(1-k^2r^2)^2
\end{equation}
as it should. Now we can do two things starting from $\Gmk(r,x,s,x')$: Thinking in $\Dim=1$ terms, we can solve
the Dirichlet problem by writing
\begin{equation} \label{Dir}
\phi(r,x)=-\int_{\p\Mcal'}\left[\frac{1}{(1-k^2s^2)}(N^s\p_s+2k)\Gmk(r,x,s,x')\right]_{\p\Mcal'}\phi_0(x')\diff\Omega(x')\;.
\end{equation}
Note that $\phi_0(x)$ is really $h(x)$ in (\ref{phitwo}). Restricting $\phi(r,x)$ to the boundary will 
give $\phi_0(x)$. This is of course the standard way of doing things in AdS/CFT, and as is well--known (e.g. \cite{Giddings99}) 
leads us to the bulk--to--boundary propagator, 
\begin{equation}
\left[\frac{1}{(1-k^2s^2)}\cdot (N^s\p_s+2k)G^{(\mu,k)}(r,x,s,x')\right]_{\p\Mcal'} \sim 
\left[\frac{1}{(1-k^2s^2)} G^{(\mu,k)}(r,x,s,x')\right]_{\p\Mcal'}=:\Kmk(r,x,x')\;.
\end{equation} 
However we could also take a $\Dim=2$ point of view, and solve the Robin problem by directly writing a single--layer potential
\begin{equation} \label{Rob}
\phi(r,x)=\int_{\p\Mcal'}\left[\frac{1}{(1-k^2s^2)}\cdot \Gmk(r,x,s,x')\right]_{\p\Mcal'}\phi_0(x')\diff\Omega(x')
\end{equation}
where here we recover $\phi_0(x)$ by taking $(N^r\p_r+k)\phi(y,x)$ and restricting to the boundary. Note that in
both cases the same object, the bulk--to--boundary propagator (which is simply the scaled restriction of one point of 
$\Gmk(r,x,s,x')$ to the boundary), appears, but $\phi_0(x)$ can be recovered from it 
in two different ways. Let us now write it down:
\begin{equation} \label{bulktoboundary}
K^{(\mu,k)}(r,x,x')=
\lim_{s\ra1/k}\left\{\frac{G^{(\mu,k)}(r,x,s,x')}{(1-k^2s^2)}\right\}
=\frac{(1-k^2r^2)}{\sqrt{{\Delta}}}\frac{q_+-q_-}
{q_++q_--4krc\vone}
\end{equation}
where
\begin{equation}
q_{\pm}=(1+k^2r^2\pm\sqrt{{\Delta}}+2\mu k r^2\vtwo)
\left(1\pm\frac{\mu}{k}\sqrt{{\Delta}}-\mu^2r^2(2\vtwo-1)\right)^{\half\left(1-\frac{\mu}{k}\right)}\;,
\end{equation}
with 
\begin{equation}\label{Delta}
{\Delta}=(1-k^2r^2)^2-4k^2r^2(\vtwo-1)+4\mu^2k^2r^4\vtwo(\vtwo-1)
\end{equation}
 and
\begin{equation}
c=\left[(1-\mu^2k^2r^4)\left(1-\mu^2/k^2\right)\right]^{\frac{1}{4}\left(1-\frac{\mu}{k}\right)}
\sqrt{\left(1+\mu/k\right)(1+\mu k r^2)}\;.
\end{equation}
This Poisson--like kernel scales at the boundary as $(1-k^2r^2)^2\delta^{(3)}(x,x')$\footnote{By this we mean that
it behaves roughly like $\lim_{\epsilon\ra0}\epsilon/(\epsilon^2+x^2)\ra\epsilon^2\delta^{(3)}(x,x')$ 
close to the boundary. Properly normalising
(\ref{bulktoboundary}) would require integrating it over the Berger sphere, which is a rather involved calculation and
we have not performed it.}, which is the behaviour needed to obtain $\phi_0(x)$ from (\ref{Dir}). On the other hand, 
$(N^r\p_r+k)\Kmk(r,x,x')$ also scales as $(1-k^2r^2)^2\delta^{(3)}(x,x')$, showing that we could also  recover $\phi_0(x)$ from
(\ref{Rob})\footnote{The difference is that now one obtains 
$\epsilon^2\delta^{(3)}(x,x')$ from a distribution of the type $\lim_{\epsilon\ra0}\epsilon^3/(\epsilon^2+x^2)^2$, rather than 
the more ``exceptional'' case above.}. 
So indeed both ways of looking at this problem are equivalent. Of course, in both cases we 
obtain the same boundary behaviour, and thus the same conformal dimension for the dual operator. Given the scaling
of $\phi_0(x)$ it should be a density on the boundary of
conformal dimension $\Dim=2$, so (as standard in AdS/CFT) it will be a source for a dual operator $\Ocal$ of conformal 
dimension $\Dim=1$.  We will obtain the two--point function of $\Ocal$ in the next section, and will eventually give an 
independent verification
that it is conformally invariant and the dimension of $\Ocal$ is indeed 1 in section \ref{Bergersphere}.\footnote{Note 
that to solve the problem of prescribing 
$f(x)$ in (\ref{phitwo}), which would lead to correlation functions of a dual $\Dim=2$ operator, 
we would have to find a Green's function on QTN scaling as $(1-k^2r^2)^2$ on the boundary (which could of course then
be interpreted as the Dirichlet Green's function for $\Dim=2$ or the Neumann Green's function for $\Dim=1$). This could be 
constructed in a similar way to \cite{Zoubos02}, but has not yet been attempted.} 

\subsection{The boundary two--point function}

We can now follow standard AdS/CFT procedures \cite{Witten98} to obtain the two--point 
function on the boundary. We evaluate the action (\ref{Scalaraction}) by first integrating by parts 
and then imposing the bulk equations of motion ($(\nabla^2-1/6\Rcal)\phi=0$), which results in a pure boundary term:
\begin{equation}
S=-\frac{1}{2}\lim_{r\ra1/k}\int_{\p\Mcal}\diff^dx\sqrt{g}(N^r)\phi(r,x)\left(N^r\p_r-\frac{1}{3}\Kcal\right)\phi(r,x)\;.
\end{equation}
Now we can substitute the solution for $\phi(r,x)$ in terms of the bulk--to--boundary propagator
\begin{equation}
\phi(r,x)=\int_{\p\Mcal}  \Kmk(r,x,x')\phi_0(x') \diff\Omega(x')
\end{equation}
(where $\diff\Omega(x')$ is the measure with respect to the finite boundary metric) which gives 
\begin{equation}
(N^r\p_r-\frac{1}{3}\Kcal)\phi(r,x)=(1-k^2r^2)^2\phi_0(x)\;,
\end{equation}
and given the scaling of the induced measure ($\sqrt{h}=\sqrt{g}(N^r)\sim (1-k^2r^2)^{-3}$), we have 
\begin{equation} \label{Sfinal}
S\sim\int_{\p\Mcal}\int_{\p\Mcal'}\phi_0(x) \left(\lim_{r\ra1/k} \frac{\Kmk(r,x,x')}{(1-k^2r^2)}\right)\phi_0(x')\diff\Omega(x)\diff\Omega(x')\;.
\end{equation}
Taking the limit $r\ra1/k$, we find that the answer is finite, which would not have been the case had we forgotten 
the extrinsic curvature term\footnote{It is well known, of course, that this term is the first in a series of 
(covariant) regulators for the boundary theory in the case of scalars, derivable using holographic 
renormalisation \cite{deHaroetal01,Skenderis02}.
Since we are working at a rather naive level, being interested only in the functional form of the two--point function rather
than the precise normalisation, we will not perform a similar analysis here.}. 
Note that the way we obtained (\ref{Sfinal}) is exactly the opposite of the way it is usually done in AdS/CFT, i.e. it is the 
normal derivative term that gives us $\phi_0$, while the restriction
of $\phi$ to the boundary brings in the boundary value of $\Kmk(r,x,x')$. As discussed, we would have obtained the same 
result working in the standard way, but at various places we will find it best  to think of $\phi(x)$ being given by a 
Robin problem rather than a Dirichlet problem. 

Interpreting the boundary values $\phi_0$ as sources in the dual field theory, coupled in a conformally invariant 
way to an operator $\Ocal$ and, using the notation 
$M(x,x'):=\langle\Ocal(\theta_r,\varphi_r,\psi_r) \Ocal(\theta_s,\varphi_s,\psi_s)\rangle$  we arrive at the following proposal for the
two--point correlation function of the boundary operator $\Ocal$ dual to the bulk conformally coupled scalar $\phi$: 
\begin{equation} \label{correlator}
M(x,x') =
\frac{2}{\sqrt{\delta}}\frac{a_+-a_-}{a_++a_--2c_0 \vone}
\end{equation}
where $\delta=(1-\vtwo)(1-\mu^2/k^2\vtwo)$, and
\begin{equation}
\begin{split}
a_\pm&=(1\pm \sqrt{\delta}+\frac{\mu}{k}\vtwo)
(1\pm2\frac{\mu}{k}\sqrt{\delta}+\frac{\mu^2}{k^2}(1-2\vtwo))
^{\half\left(1-\frac{\mu}{k}\right)}\\
c_0&=(1-\frac{\mu^2}{k^2})^{\half(1-\mu/k)}(1+\frac{\mu}{k})\;.
\end{split}
\end{equation}
This slightly complicated object is what AdS/CFT tells us should arise on the boundary $\Squashedsphere$
as a conformal two--point correlation function of the scalar operator $\Ocal(x)$ dual to the bulk conformally
coupled scalar $\phi(y,x)$. In particular, as discussed extensively in the preceding section,
we expect $\Ocal(x)$ to have conformal dimension $\Dim=1$, and this is the statement that we would like to 
check in the following. Of course, in the absence so far of a concrete string/M theory embedding (in 
the sense of a decoupling limit of some brane configuration) of this model, it is not yet clear what 
the dual boundary theory actually is, and thus 
which fields constitute the gauge--invariant operator $\Ocal(x)$. It is expected \cite{Zoubos02} that one 
can think of it as some (non--supersymmetric) deformation of (the infrared limit of) $\Ncal=8$ SYM on round $\Srm^3$ 
\cite{Maldacena98} 
(here we are thinking of QTN$\times X^7$ (with $X$ some Einstein manifold) as a deformation of $\AdS_4\times \Srm^7$), and in
particular it is expected to at least contain scalar operators with $\Dim=1$ (since $\Ncal=8$ SYM does).
On the other hand, \emph{any} CFT on $\Squashedsphere$ (regardless of the fundamental degrees of 
freedom)  which contains $\Dim=1$ scalar operators \emph{could}, according to AdS/CFT, lead to 
correlators with the functional form of (\ref{correlator}) (in the same sense that any CFT in flat 
space will lead to powers of $1/(|x-x'|)$)\footnote{However, unlike the case of flat space, 
we cannot claim that $M(x,x')$ is the \emph{unique} $\Dim=1$ correlation
function. See the conclusions for a discussion of this point.}, so it is clearly important to make sure that this is
what is actually required by conformal invariance on $\Squashedsphere$. We will start developing the necessary ideas in
the next section. 
 
\subsection{The round--sphere limit} \label{Roundlimit}

Before we proceed, it is instructive to consider the limit $\mu=0$, which should bring us to completely 
familiar AdS/CFT territory. In this limit our correlator simplifies drastically, reducing to 
\begin{equation} \label{Mzero}
M^0(x,x')=\lim_{\mu\ra0}M(x,x')\sim\frac{1}{1-\vone}\;.
\end{equation}
This is a two--point function on the round sphere, and it can be easily seen to correspond to 
a boundary operator of conformal dimension $1$, by switching to the upper--half space (or Poincar\'e) notation.
The conversion can be found in appendix B of \cite{Zoubos02}, where the hyperbolic space 
chordal distance $u$ is written in two ways:\footnote{By ``upper--half space notation'' we mean 
the usual metric $\diff s^2=(\diff y^2+\diff\vec{x}^2)/(k^2y^2)$, where $y$ is the radial coordinate.} 
\begin{equation}
u=\left\{\begin{array}{l}  \frac{(y-y')^2+(\vec{x}-\vec{x}')^2}{2yy'} \quad\text{(upper--half space)}
\\ \\ \frac{2k^2(r^2+s^2-2rs\vone)}{(1-k^2r^2)(1-k^2s^2)} \quad\text{(polar--type)}\;.\\ 
\end{array}\right.
\end{equation}
The bulk--to--bulk propagator in this case is simply $1/u$, which, properly rescaled and 
restricted to the boundary
(which is $y,y'\ra0$ for the upper--half space) gives the two--point correlation function for the dual boundary 
operator
\begin{equation}
M^0(x,x')=\frac{1}{(x-x')^2}
\end{equation}
while in our polar case the corresponding limit ($r,s\ra 1/k$)  simply gives us (\ref{Mzero}).
From the upper--half--space picture it is obvious that the conformal dimension of the 
corresponding operator is $\Dim=1$. 

As discussed in the introduction, we have another way to check this result without ever leaving
the sphere, namely we can hit $M^0(x,x')$  with any of the four conformal Killing vectors available. 
Choosing e.g. the conformal Killing vector (denoted $k_1$ in appendix \ref{BergerSphere})
\begin{equation}
\upsilon=2\sin\frac{\theta}{2}\cos\left(\frac{\varphi+\psi}{2}\right)\p_\theta
+\frac{\sin\left(\frac{\varphi+\psi}{2}\right)}{\cos\frac{\theta}{2}}\left(\p_\varphi+\p_\psi\right)\\
\end{equation}
we can readily check that 
\begin{equation}
\left(\Lcal_{\upsilon(x_r)}+\Lcal_{\upsilon(x_s)}\right)\frac{1}{1-\vone}=
-(\sigma_\upsilon(x_r)+\sigma_\upsilon(x_s))\frac{1}{1-\vone}.
\end{equation}
(where $\sigma_\upsilon(x_r)=\cos(\frac{\theta_r}{2})\cos(\frac{\vf_r+\psi_r}{2})$ and
similarly for $\sigma_\upsilon(x_s)$). Recalling (\ref{conformaltransformation}), we conclude, this time directly on the sphere, 
that $M^0(x,x')$ corresponds to a $\Dim=1$ operator. Of course, due to the lack of conformal Killing vectors, this simple 
argument fails as soon as we leave the round sphere (by allowing $\mu\neq0$).

\section{Conformal Geometry} \label{Conformal}
 
In this section we give a brief overview of conformal geometry, which can roughly be thought of as an analogue of
riemannian geometry, but where only angles are important, not lengths. Reviews of conformal geometry, containing
references to the literature, can be found in \cite{Eastwood96,Branson97,Branson98}.

\subsection{Preliminaries} \label{Preliminaries}

A conformal manifold $\Mcal$ is a pair $(\Mcal,[g])$, where $\Mcal$ is a smooth $d$--dimensional 
manifold, and $[g]$ is a riemannian metric defined only up to scale, i.e. a conformal class of metrics.
That is, rather than dealing with smooth, positive sections of $\odot T^*\Mcal$, 
as in riemannian geometry, here we are considering positive sections of the ray bundle $R(\odot T^*\Mcal)$ (i.e. if $g$ is
a representative metric, the fibres are $t^2 g, t\in\Rset_+$) 
 Two metrics in the conformal class, say  $g_\mn$ and $\hat{g}_\mn$ that are related 
as $\hat{g}_\mn=\Omega^2 g_\mn$, with $\Omega$ a smooth function on the manifold, are equivalent in conformal geometry. 
 
 In practice, rather than implementing this equivalence directly, it is more convenient to choose a 
representative $g\in [g]$, and examine how other quantities on the manifold, defined relative to $g$,
 transform under conformal (Weyl) change of the metric: $g_\mn\ra \Omega^2 g_\mn$. To keep track of these transformations, we
introduce the bundle $\Ecal(w)$ of conformally weighted functions (conformal densities) of weight $w$: Under conformal change 
of the metric, a section $\phi\in\Gamma(\Ecal(w))$ transforms 
as\footnote{Note that the conformal weight $w$ as it is defined here corresponds to the length dimension of $\phi$. Often
in physics it is more convenient to use mass dimensions, which we have been calling $\Dim$, in which case
$\Dim=-w$. In this section we will use length dimensions which are more natural from a conformal geometry point of view. }
\begin{equation}
g_\mn\ra \Omega^2 g_\mn \quad \Longrightarrow \quad\phi\ra \Omega^w \phi
\end{equation}
One can also define tensors by tensoring $\Ecal(w)$ with the appropriate bundles. For instance the conformal metric
itself can be thought of as a section of $\Ecal_\mn(2)$, signifying that it is a two--tensor transforming 
with weight 2. We will only need to discuss  scalars, so we refer the reader to the reviews cited above for more
details.

The prototype example of conformal geometry is the sphere $\Srm^d=\SO(d+1)/\SO(d)$, which is acted on naturally
by the group of conformal diffeomorphisms $\SO(d+1,1)$. It plays a role similar to that of $\Rset^d$ in 
riemannian geometry. The high degree of symmetry leads to a beautiful interplay between differential--geometric
and group--theoretic ideas, and helps in gaining insight which is often helpful in the curved case. 

One of the main objectives in conformal geometry is to find conformal covariants, i.e. objects that transform 
in a simple way under conformal change. These could be tensors on the manifold, like the Weyl tensor $W^{\mu}_{\;\nu\rho\sigma}$ 
which (with this choice of indices) is conformally invariant. There also exist objects which, unlike the Weyl tensor, 
are only covariant in particular dimensions, as for example the Cotton tensor\footnote{Also sometimes referred to
as the \emph{Cotton--York} tensor \cite{Eastwood96} or the \emph{Weyl--Schouten} tensor \cite{SkenderisSolodukhin00}.}, 
\begin{equation} \label{Cotton}
C_{\mn\rho}=\nabla_\rho \mathrm{P}_{\mn}-\nabla_\nu \mathrm{P}_{\mu \rho}\;,\quad\text{where}\;\; 
\mathrm{P}_\mn=\frac{1}{d-2}\left(\Rcal_\mn-\frac{\Rcal}{2(d-1)}g_\mn\right) 
\end{equation}
which is conformally covariant only in $d=3$, where it is an indicator of conformal flatness (i.e. it
vanishes if the space is conformally flat).  
One of the most influential works in this field is of course the article by Fefferman and Graham 
\cite{FeffermanGraham85}, in which they construct scalar conformal invariants involving (derivatives of) a $d$--dimensional
riemannian metric by looking at differential invariants of an associated $d+2$ dimensional \emph{lorentzian} 
ambient metric, where the conformal group is linearly realised. These ideas have proved very useful in an AdS/CFT context, 
where, in the context of holographic renormalisation, these conformally invariant objects 
are used as counterterms in the $d+1$ dimensional gravitational action, leading to a regulated action for the dual 
$d$--dimensional field theory. See \cite{Skenderis02} for a review and references. 

As in riemannian geometry, an important question that arises when given a conformally covariant object on the sphere
(which, as indicated above, is the ``flat model'' of conformal geometry) is whether it can be extended to a conformally
covariant object in the curved case, that is on an arbitrary non--conformally flat manifold equipped with a conformal metric. 
This will be important for us, since the Berger sphere is not conformally flat (which can be easily seen from 
the fact that the Cotton tensor does not vanish). 

\subsection{Conformally invariant differential operators} \label{confdiff}

In conformal geometry it is also of great importance to construct conformally covariant \emph{differential} operators,
i.e. differential operators that transform nicely under conformal change of the metric. 
More precisely (e.g. \cite{Branson98}): 
A differential operator $\Dcal$, defined locally in terms of $g_\mn$, its inverse, 
and their partial derivatives, is called \emph{conformally covariant} if it transforms in the following way under 
conformal change:
\begin{equation}
g_\mn\ra\Omega^2 g_\mn \quad\Longrightarrow \quad\Dcal\ra \Omega^{-b}\Dcal\Omega^{a}\;.
\end{equation}
Note that $\Dcal$ will act on everything to its right, including the factor of $\Omega^a$. The quantity 
$(a,b)$ is often called the conformal biweight, or bidegree, of the operator $\Dcal$. To obtain a conformally 
\emph{invariant} operator, we have to also specify which density bundles the conformally covariant
operator acts between. If it acts between a section  $u\in\Gamma(\Ecal(-a))$ and a section $v\in\Gamma(\Ecal(-b))$, 
i.e. we have 
\begin{equation}
\Dcal u=v
\end{equation}
this equation is invariant under conformal change of the metric. 
The most fundamental conformally covariant operator is the conformal (or Yamabe) laplacian 
\begin{equation} \label{Yamabe}
Y=\Delta+\frac{d-2}{4(d-1)}\Rcal,
\end{equation}
where $\Rcal$ is the scalar curvature. Under the change $g_\mn\ra\Omega^2 g_\mn$, we have 
\begin{equation}
Y\ra \Omega^{-\frac{d+2}{2}}Y\Omega^{\frac{d-2}{2}}. 
\end{equation}
Thus, when acting between the density bundles $\Ecal(\frac{2-d}{2})$ and 
$\Ecal(-\frac{2+d}{2})$, the conformal laplacian is a conformally \emph{invariant} operator.
Note that, being a second order operator, it reduces conformal weight by two. This intuitive fact is a
consequence of working with length dimensions rather than mass dimensions (which was discussed in a
previous footnote).
There are other, perhaps less well known but also important, conformally covariant differential operators, as for instance the 
fourth--order Paneitz operator in $d=4$ (see e.g. \cite{Chang99} for a review with references) which acts on scalars and has biweight $(0,4)$. 
It is given by
\begin{equation}\label{Paneitz}
P_4=\sqrt{g}\left(\Delta^2+2\nabla_\mu\left(\Rcal^{\mn}-\frac13g^{\mn}\Rcal\right)\nabla_\nu\right)\;.
\end{equation}
This operator has principal part $\Delta^2$, where $\Delta$ is the positive laplacian. On $\Srm^d$ one can generalise
this construction to operators whose principal parts are higher powers of the laplacian (e.g. \cite{EastwoodRice87,Grahametal92}), 
called ``conformally invariant powers of the laplacian'' but the question of whether they extend to the curved case does
not always have a positive answer. For instance, the sixth--order operator with principal part $\Delta^3$ 
in $d=4$ does not have a curved space analogue \cite{Graham92}. In general these scalar operators $P_s$ have 
biweight $(\frac{d-s}{2},\frac{d+s}{2})$
and so will take $\Ecal(\frac{s-d}{2})\ra\Ecal(-\frac{s+d}{2})$. If $\phi$ has weight $w$, then the corresponding Green's function 
$M(x,x')=\langle\phi(x)\phi(x')\rangle$ will transform as $M(x,x')\ra\Omega^{w}(x)\Omega^{w}(x')M(x,x')$, and the equation 
\begin{equation}
\sqrt{g}P_sM(x,x')=\delta^{(d)}(x,x')
\end{equation}
will be conformally invariant if $w=\frac{s-d}{2}$. Starting from the invariance of this equation under both 
diffeomorphisms and conformal change, and assuming a nontrivial conformal Killing equation, 
one can now easily prove (\ref{conformaltransformation})
by adapting a discussion in \cite{Erdmenger97} (and also writing  $\Omega(x)=e^{\sigma(x)}$, and linearising). Note
of course that the mass conformal dimension of $\phi$ will be $\Dim=-w=\frac{d-s}{2}$. 

The operators $P_k$ act on scalar densities, but there are also generalisations that naturally act on tensor densities with
various symmetry properties, as well as on forms. See e.g. \cite{Erdmenger97,ErdmengerOsborn98,BransonGover02} for 
relatively recent discussions and references, as well as the reviews \cite{Branson97,Branson98}.

One important application of conformally invariant operators in physics has been the construction of non--local actions 
that generalise (to higher dimensions) Polyakov's two--dimensional action \cite{Polyakov81} in the sense that, upon 
variation, they generate the gravitational conformal anomaly (see e.g. \cite{Erdmenger97,Deser00}). 
In fact the action constructed by Riegert \cite{Riegert84} for this purpose contains a fourth--order conformally invariant 
operator which is the same as (\ref{Paneitz}), found by Paneitz at around the same time.  

In an AdS/CFT setting, conformally invariant operators of a slightly different kind played a crucial 
role in \cite{Dolanetal01}. There the goal was to construct the conformally invariant equations 
obeyed by partially conserved tensors in the boundary theory, which are dual to partially massless
fields in the bulk.  

So far we have been discussing \emph{differential}, i.e. local, operators. More relevant to 
our goal in this article is that one 
can also construct \emph{non--local} conformally invariant operators \cite{BransonGover01,Peterson00}. 
They will be discussed in due course, in section \ref{Nonlocal}, after we first introduce the Dirichlet--to--Robin operator
(which will turn out to be one of them) from a different perspective, that of inverse boundary problems.

\section{The Dirichlet--to--Robin operator} \label{DirichlettoNeumann}

In this section we first define the Dirichlet--to--Neumann operator, which plays a central
role in dealing with inverse boundary value problems (i.e. the general issue of obtaining 
information about the interior of a domain through boundary measurements). We sketch a way (detailed in 
\cite{LeeUhlmann89,Uhlmann92}) to explicitly construct the Dirichlet--to--Neumann operator for a given manifold with
boundary. However, the object we need is actually the slightly more general Dirichlet--to--Robin operator, and the
boundary in question actually lies at conformal infinity, so we will discuss some plausible guesses of how the construction 
might generalise to  this case.
Then, although the application we have in mind is very different from the usage of the Dirichlet--to--Neumann/Robin operator
in the inverse boundary  problems we mentioned, we give a short overview of the main ideas. After that we dicuss the crucial 
fact that the Dirichlet--to--Robin operator is \emph{conformally covariant}, and finally explain what all that has to do with
the proposed QTN/Berger sphere correspondence and the question of the conformal invariance of our correlation function
$M(x,x')$.

\subsection{Definition} \label{Definition}

Here, following \cite{LeeUhlmann89,Uhlmann92}, we will give a definition of the Dirichlet--to--Neumann operator which 
is natural from the viewpoint of riemannian geometry. These references also discuss slightly different definitions 
 which can be mapped to this one. 
So let us assume given a smooth riemannian metric $g_\mn$ on a manifold--with--boundary $\overline{\Mcal}$, and also some function
$f$ defined on the boundary $\p\Mcal$. 
The Dirichlet problem for the laplacian $\Delta$ requires finding a function $u$ such that:
\begin{equation} \label{Dirprob}
\hspace{1cm}\left\{\begin{array}{l} \Delta u=0 \quad \text{in}\;\; \Mcal \;\quad\text{($\Mcal$ is the interior of $\overline{\Mcal}$)}\\
u|_{\p \Mcal}=f\;.
\end{array}\right.
\end{equation}
In words, we need to find a harmonic function in the bulk, that restricts to the given data on the boundary. Similarly, 
the Neumann problem asks for a function $u$ which satisfies 
\begin{equation} \label{Neuprob}
\left\{\begin{array}{l}
\Delta u=0 \quad \text{in} \;\; \Mcal\hspace{2.3cm}\mbox{}\\ 
(N\cdot u)|_{\p\Mcal}=h\;,
\end{array}\right.
\end{equation}
where $h$ is a function on the boundary, and $N$ is the outward unit normal. Now, if we solve the Dirichlet problem as 
posed in (\ref{Dirprob}) to obtain $u$, it
is a valid question to ask what is the Neumann data $h$ \emph{defined} by $u$ through (\ref{Neuprob}). So we define the 
Dirichlet--to--Neumann operator $\Lambda$ as: 
\begin{equation} \label{DtNdef}
\Lambda f=[N\cdot u] |_{\p \Mcal}\;.
\end{equation}
That is,  the operator $\Lambda$ takes Dirichlet data to Neumann data.

One can thus think of the Dirichlet--to--Neumann operator as a composition of two operations: First one extends the
prescribed boundary data $f$ to the bulk in a way that satisfies (\ref{Dirprob}), and then restricts the normal 
derivative of the bulk solution to the boundary. 
The main point now is that $\Lambda$ can usefully be considered as an operator \emph{directly on the boundary}, 
which acts on the boundary function $f$ and 
gives another boundary function $h$ as the result. Note, however, that since the bulk equation admits
two \emph{independent} solutions, one of which gives the Dirichlet and the other one the Neumann data on the boundary, 
from a boundary perspective this data cannot be locally related. Thus we should expect that
the Dirichlet--to--Neumann operator, if viewed as a boundary operator, will be non--local.  
Indeed there are theorems (see e.g. \cite{SylvesterUhlmann88}) 
which assure us that this is a pseudodifferential operator on the boundary, of order one, and that in fact
its \emph{principal} symbol is the square root of the boundary laplacian\footnote{
See appendix \ref{Pseudodifferential} for some basic definitions on pseudodifferential operators.}. 
One (iterative) approach to obtaining its \emph{full} symbol for a smooth riemannian manifold 
has been outlined by Lee and Uhlmann in
\cite{LeeUhlmann89,Uhlmann92}: Start by writing the $d+1$--dimensional metric in boundary normal 
coordinates $\{x^1,..,x^{d},y\}$, where $y\geq0$ and the boundary is defined by $y=0$.
The remaining coordinates $\{x^1,...,x^d\}$ are local coordinates for $\p \Mcal$, and smoothly extend
to coordinates on a neighbourhood in $\Mcal$ of a point $p\in\p \Mcal$\footnote{See also \cite{Katchalovetal01} for a discussion 
of boundary normal coordinates in a similar context.}. In these coordinates the metric is written as 
\begin{equation}
g_\mn=\diff y^2+h_{ij}(y,x)\diff x^i\diff x^j
\end{equation}
and the laplacian takes the form 
\begin{equation}
\nabla^2=\p_{yy}+\half h^{ij}\p_y h_{ij}\p_y+{h}^{-\half}\p_i{h}^\half h^{ij}\p_j\;.
\end{equation}
The last term clearly reduces to the boundary laplacian (which we call $\Delta_B$) when $y=0$. 
Now the crucial step is the following \cite{LeeUhlmann89}: Suppose we can find a pseudodifferential, 
order one operator $\Acal$, depending smoothly on $y$, such that the Laplacian factorises in the following way:
\begin{equation} \label{factorization}
\nabla^2=(\p_y+E(y,x)+\Acal(y,x,\p_x))(\p_y-\Acal(y,x,\p_x))
\end{equation}
where $E(y,x)=\half h^{ij}\p_y h_{ij}$. Then $\Acal$, restricted to the boundary, 
is essentially the Dirichlet--to--Neumann operator\footnote{In 
\cite{LeeUhlmann89,Uhlmann92}, the Dirichlet--to--Neumann operator is defined as taking 
functions to $d$--forms, so the definition is slightly different.}. The proof (which we do not give here) is based on showing that 
$(\p_y-\Acal(y,x,\p_x))u$ is a smoothing operator (an operator ``of order $-\infty$'') on the boundary, 
so, modulo smoothing operators, the relation  $\p_y u|_{\p\Mcal}=\Acal(0,x,\p_x)\cdot f(x)$ holds on the boundary. See also 
\cite{Treves}, Ch. III for a related discussion.

As mentioned, this method allows us to construct the Dirichlet--to--Neumann operator in an \emph{iterative} manner.
We can start by  expanding the full symbol $a$ of the operator $\Acal$ as
\begin{equation} \label{asymbol}
a(x,\xi')=\sum_{i=0}^\infty a_{1-i} (x,\xi')
\end{equation}
where each term $a_{1-i}$ in this expansion is the symbol of an order--$(1-i)$ operator.  
Taking the symbol on both sides of (\ref{factorization}), expanding, and equating the highest degree terms, 
it follows easily that $a_1$, the principal symbol of $\Acal$, is given by
\begin{equation}
a_1=-\sqrt{\sigma_2(\Delta_B)},
\end{equation}
where $\sigma_2(\Delta_B)$ denotes the principal symbol of the boundary laplacian, and the choice of sign corresponds to 
the outward unit normal vector giving the positive square root, 
\begin{equation}\label{positive}
N\cdot u|_{\p\Mcal}=\grL_xf(x)\Rightarrow (-\p_y)u(y,x)|_{y=0}=(\sqrt{\Delta_B}+\cdots)f(x)\;.  
\end{equation}
Here the ``$\cdots$'' corresponds to the lower order terms in $\grL$. Substituting this result back into 
(\ref{factorization}) we can get a recursion formula for the $a_{1-i}$s, and so in
principle we can determine all the terms in the series (\ref{asymbol}) for $\Acal$. There will
obviously always be an ambiguity since we can arbitrarily add operators whose symbols are in $\mathbb S^{-\infty}$,
the smoothing operators alluded to above. 

For reasons to be fully explained in section \ref{Implications}, we will actually be interested in the 
Dirichlet--to--Robin operator. This can be defined by analogy with
(\ref{DtNdef}), as the operator taking Dirichlet data to Robin data:
\begin{equation} \label{DtR}
\Bcal f\mapsto \left[(N+\sigma)u\right]|_{\p \Mcal}
\end{equation}
where $\sigma$ is some function defined on the boundary, and we have named the operator $\Bcal$ to distinguish it
from the Dirichlet--to--Neumann operator. 

Motivated by the method we just outlined for the Dirichlet--to--Neumann operator, we can try to obtain the
Dirichlet--to-Robin operator for hyperbolic space and QTN by factorising the bulk laplacian.
There are several extrapolations we make that seem reasonable (and certainly work) but
which do not follow from the discussion above and perhaps lack rigorous justification. First, of course, in the
asymptotically hyperbolic case we are dealing with, the laplacian vanishes at the boundary. 
Therefore we will take the (perhaps) pragmatic approach of first finding $\Acal$ at some finite radius 
(i.e. the ``boundary'' will be a hypersurface corresponding to a fixed value of $r$--the radial coordinate on the ball) 
and then taking it to the boundary at infinity via some limiting process\footnote{Note here that the methods of \cite{Zoubos02} 
so far have only produced the solution of the Dirichlet problem at infinity of QTN, \emph{not} at some finite radius. Thus, 
unlike the AdS case where (thanks to the ability to Fourier transform) the equivalent of $M(x,x')$ is also available at 
some finite radius \cite{Freedmanetal98}, in the QTN case we know it solely at infinity. So, although presumably $\Acal$ at finite
radius contains useful information, we will need to take its limit at infinity to compare with $M(x,x')$.}.

Second, we will treat (\ref{factorization}) as an \emph{exact} factorisation. That is, we will look for an $\Acal$ that
satisfies (\ref{factorization}) precisely, without the need for smoothing operators. In practice, that means that
we will \emph{not} proceed iteratively, as in \cite{LeeUhlmann89}, but will treat (\ref{factorization}) as an actual
differential equation (in the radial variable). (See in this respect the discussion in \cite{Treves}, chapter III.) 

Third,  we would like to avoid boundary normal coordinates. The reason is that writing QTN in these
coordinates will turn the metric into a series expansion in the radial coordinate (which we have been denoting by 
$y$ here), while we find it easier to deal with the exact metric (\ref{Pedersen}) and exact laplacian. 
Ultimately this is probably a matter of taste, but a more serious issue is the fact (see appendix \ref{Pseudodifferential}) 
that the complete symbol of a pseudodifferential operator is not coordinate--invariant. Since we would eventually like to use
the operator we find in polar--type coordinates, we had better calculate it in polar--type coordinates.  
 
A final subtlety which arises is that, as we will see in section 
\ref{Implications}, the Dirichlet--to--Robin operator will turn out to be a Robin derivative of a double 
layer potential. As is well known in potential theory, (see e.g. \cite{Folland, Mikhlin,McLean}), 
although the single layer potential is continuous at the boundary, the double layer potential typically exhibits a 
discontinuity. This means that the boundary data of the double layer potential is given by the (appropriately normalised) 
difference of the limits of the solutions to the interior and the exterior problems. The \emph{Neumann} derivative 
of a double layer has no jump, but since we are taking the Robin derivative, i.e. we are adding a constant term, it
is not unreasonable to expect that the Dirichlet--to--Robin operator
we are looking for will also have to exhibit this discontinuity. To cover this possibility, the final part of the 
practical recipe we give is to find $\Acal$ by factorising the laplacian on the inside and on the ``outside'' of 
QTN, i.e. the regions $r<1/k$ and $r>1/k$, and then taking the difference of the limits $r\ra1/k$ on both sides. 

Needless to say,  the fact that we might need to examine the exterior boundary problem in order to properly 
define the Dirichlet--to--Robin operator, although very natural if we think in terms of boundary value
problems on the ball in $\Rset^{d}$, is rather unusual from an AdS/CFT point of view, and potentially raises
 some conceptual issues. Hyperbolic and asymptotically
hyperbolic metrics on the ball blow up on the boundary, which we thus interpret as the ``conformal infinity'' of the
manifold. It is thus very natural  to discuss holography in terms of only what is happening in the interior.
We will see later that this distinction is actually irrelevant for hyperbolic space, which possibly explains why no need for
it has been seen so far in discussions of AdS/CFT. It could be that some new principle is at play in the asymptotically 
hyperbolic case, or
(more likely perhaps) that our rough analysis above can be refined to avoid having to look at the exterior. For now 
we will just blindly accept this recipe and see that it gives the expected answer. 

Another unpleasant feature is that QTN is generically incomplete for $r>1/k$, containing a conical singularity. 
A brief analysis can be found in \cite{Hitchin95}, where it is shown that the metric can be extended to a smooth one
only for certain \emph{discrete} values of $\mu$. (See also \cite{CalderbankSinger02} and \cite{BehrndtDallAgata02} for 
related matters.). This is probably an important issue, but for the time being we will simply ignore it, by assuming that
since the singularity is at $r\ra\infty$, and we are only interested in some region close to the boundary $r=1/k$ on both 
sides, it will not affect the limiting procedure. 

It should be clear that the prescription we gave above, although perhaps reasonable, is rather \emph{ad hoc}. In this respect 
we should mention that there \emph{exists} a formalism which has been explicitly constructed to deal with boundary
value problems on asymptotically hyperbolic manifolds, namely the $0$--pseudodifferential calculus of Mazzeo and Melrose
\cite{MazzeoMelrose87,Mazzeo88,Mazzeo91,Melrose95}. Some recent applications (in the context of 
scattering theory) to asymptotically hyperbolic manifolds are \cite{Borthwick97,JoshiSaBarreto00, GrahamZworski01,SaBarreto03}. 
Part of
the idea is to treat operators that vanish on the boundary, e.g. the laplacian, by introducing a so--called ``stretched 
product'' which is a blow up of the diagonal in $\Mcal\times\Mcal$ (the space that Green's functions naturally inhabit)
at the boundary $\p\Mcal\times\p\Mcal$. We expect that working within this formalism will resolve, or at least clarify, 
the various problematic issues we cited. Regrettably, we have nothing more to say on these matters here.

\subsection{Inverse boundary value problems} \label{InverseProblems}

Dirichlet--to--Neumann and Dirichlet--to--Robin maps are extensively used in the discussion of inverse boundary
value problems, a field that has its roots in a very influential paper of Calder\'on \cite{Calderon80} and, 
as mentioned, concerns obtaining information about the interior of some domain from suitable
measurements made at its boundary. 
In particular, one could ask the question: Assuming knowledge of voltages and current fluxes at the boundary of an object,
can we infer the (generically anisotropic) conductivity of the bulk?  Although the 
main motivation for these problems (which come under the name of ``Electrical impedance tomography'')
arose in geophysics (understanding properties of 
the earth's interior by measuring at the surface), and also in medical physics (non--invasively 
determining the density function of e.g. a human head by measuring the response to an applied current on the
skin), they can be precisely mapped to questions on riemannian manifolds that are interesting
in their own right. For more information and references, we suggest the reviews \cite{Uhlmann92, Uhlmann99} 
and the book \cite{Katchalovetal01}. One of the main results in this field \cite{LeeUhlmann89} is that
in $d\geq 3$, and for $(\Mcal, g)$ a compact, strongly convex, real--analytic  riemannian manifold with boundary, 
the Dirichlet--to--Neumann
operator uniquely determines the metric in the bulk, in the sense that if there is another metric $g'$ such
that $\Lambda_g=\Lambda_{g'}$, these two metrics are diffeomorphic, with the diffeomorphism equal to the identity
on the boundary\footnote{The proof of this is based on the expansion we discussed in section
\ref{Definition}. Knowledge of the  coefficients $a_{1-i}$ is enough to find all terms in the Taylor expansion of the metric in 
boundary normal coordinates in a local neighbourhood, and (given the convexity requirement) this can be extended to 
the full metric \cite{LeeUhlmann89}.}.

It goes without saying that inverse problems of this kind become especially relevant when embedded in a 
holographic context, since they are related to the question (e.g. \cite{HenningsonSkenderis98,Balasubramanianetal98})
of how information about the bulk is encoded in the boundary theory, and how to extract it (``decode the hologram'') 
from boundary data (like correlation functions).  The recent article \cite{PorratiRabadan04} surveys many of the known
results and to what extent they apply to manifolds that might be interesting from an AdS/CFT point of view. We direct the
reader to that article for discussion and references. 

A related field that seems to be a natural setting for discussing holography is that of 
geometric scattering theory (see e.g. \cite{Melrose95} for a survey). In geometric scattering (and the subfield of
inverse scattering) the main object of study is the scattering matrix $\Scal(\Dim)$. In the context of scattering on hyperbolic 
(e.g. \cite{Perry89}) and asymptotically hyperbolic manifolds (see e.g. \cite{Borthwick97,JoshiSaBarreto00,GrahamZworski01,SaBarreto03}), on 
which we recall that the solution of $\Delta-\grl(d-\grl)$ is given by 
\begin{equation}
u\sim f(y,\vec{x}) y^{d-\Dim}+h(y,\vec{x})y^\Dim
\end{equation}
the scattering matrix is defined as the operator taking $f$ to $h$. It thus clearly generalises the Dirichlet--to--Robin
operator, being a pseudodifferential operator of order $\mathrm{Re}(\Dim)-d/2$. Note that usually one discusses scattering in
the ``physical scattering'' region, where $\mathrm{Re}(\Dim)=d/2$ but $\Dim$ has an imaginary part (in other words, in 
the region $m^2<-d^2/4$, where the continuous spectrum of the laplacian lies), but it can be analytically continued to
a large part of the complex $\Dim$--plane, and thus describe ``boundary value'' type problems. As noted at the end
of the previous section, in the context of geometric scattering there is a well--developed formalism for dealing
with differential operators and Green's functions on asymptotically hyperbolic metrics.

Returning to holography, the interpretation of the Dirichlet--to--Robin operator (and, more generally, the scattering matrix) 
in this context (but from a direct, rather than inverse, viewpoint) is easy to understand: In the dual gauge theory, taking 
Dirichlet data to Neumann/Robin data has the effect of going from sources to vacuum expectation values. So the action of the
Dirichlet--to--Robin operator corresponds to ``taking the path integral'' in the gauge theory. 
Although perhaps there is no real content in the above statement, it is fair to say that there exist several
unexplored connections between the mathematical field of scattering theory and the AdS/CFT correspondence
(and its generalisations to more general manifolds). However in this article  we have in mind a very different
application of the Dirichlet--to--Robin operator, which we finally start discussing in the next subsection.

\subsection{The Dirichlet--to--Robin operator as a conformally invariant operator} \label{Nonlocal}

The discussion of the Dirichlet--to--Neumann operator in section \ref{Definition} can be embedded in 
the framework of conformal geometry (which was the topic of section \ref{Conformal}), based on the
following observation (e.g. \cite{Branson97}): Let $(\Mcal,g)$ be a riemannian manifold, and consider boundary
value problems for the conformal laplacian $Y$ (defined in (\ref{Yamabe}) above). Let us take the dimension
of the manifold to be $n$ in this section. It is often 
convenient to define a boundary value problem as a pair $(\Dcal,\Bcal)$, where $\Dcal$ is the bulk
operator under study, and $\Bcal$ is the operator imposing the boundary condition. The boundary
value problem is conformally covariant if both $\Dcal$ and $\Bcal$ transform covariantly under
conformal change, and also if the initial weights of both operators match. For the conformal
laplacian, there are two well--known possibilities for $\Bcal$ that respect conformal covariance \cite{Branson97}:
The Dirichlet operator
\begin{equation}
\Bcal_D:\quad u\mapsto u|_{\p \Mcal}
\end{equation}
and the conformal Neumann, or Robin, operator\footnote{As always, we use ``Robin'' as short 
for ``conformal Neumann''.} 
\begin{equation}
\Bcal_R:\quad u\mapsto \left(N-\frac{n-2}{2(n-1)}\Kcal\right)u\;|_{\p \Mcal}\;.
\end{equation}
The conformal biweights of these two operators are $(a,a)$, where $a$ is arbitrary (restriction to the boundary
does not affect the conformal weight of a density), and $(\frac{n-2}{2},\frac{n}{2})$ respectively. Since
the conformal laplacian has biweight $(\frac{n-2}{2},\frac{n+2}{2})$, both $(Y,\Bcal_D)$ (choosing $a=\frac{n-2}{2}$) 
and $(Y,\Bcal_R)$ are conformally covariant.  

If the Dirichlet and Robin problems for $Y$ 
are conformally covariant, it follows that the Dirichlet--to--Robin operator that takes Dirichlet
data to Robin data should also be covariant. Following Branson \cite{Branson97}, we can show this explicitly
by starting with Dirichlet data, given by a smooth function $f$ in $\p \Mcal$, such that its extension
to the bulk $\Bcal_D^{-1}f$ is smooth and satisfies $Y\Bcal_D^{-1}f=0$. ($\Bcal_D^{-1}$ can be thought of as the 
operator taking $f$ to its double layer potential, and is clearly unique.) Since the expression $Y\Bcal_D^{-1}f=0$ is conformally
invariant, the covariance of $Y$ implies that $\Bcal_D^{-1}f\ra \Omega^{-(n-2)/2}\Bcal_D^{-1}f$ under conformal change, and thus 
\begin{equation}
\hat{\Bcal}_D^{-1}(\Omega^{-\frac{n-2}{2}}f)=\Omega^{-\frac{n-2}{2}}(\Bcal_D^{-1}f)
\end{equation}
which means $\Bcal_D^{-1}$ has biweight $(\frac{n-2}{2},\frac{n-2}{2})$, which is exactly what is 
needed for the Dirichlet--to--Robin
operator $\Bcal=\Bcal_R\cdot \Bcal_D^{-1}$ to be conformally invariant. 
It also follows that $\Bcal$ is an operator of conformal biweight $(\frac{n-2}{2},\frac{n}{2})$. So it acts
invariantly between the bundles $\Ecal_{\p \Mcal}(1-n/2)$ and $\Ecal_{\p \Mcal}(-n/2)$, reducing the conformal weight
by one, as should be the case for a first order operator.

\subsection{Implications for holography} \label{Implications}

Let us now return to the problem of checking our proposal (\ref{correlator}) for the boundary two--point function on the
Berger sphere. We saw that the Dirichlet--to--Robin operator defined as above on a boundary manifold of 
dimension $d=n-1$ acts naturally 
and in a conformally invariant way on boundary densities of conformal weight $1-n/2$. Restricting
to $n=4$, we see that it acts on densities of conformal weight $-1$ on the three--dimensional boundary.

The next crucial bit of information we will need is the fact that, on the boundary, the Green's
function of $\Bcal$ is nothing but the (restriction to the boundary of the) Robin Green's function.
We will show this by following, step--by--step, a very simple argument in \cite{BarvinskyNesterov03}, 
adapting the notation in that article to suit our needs.   
Let us start by thinking of the solutions of the Dirichlet and Robin problems (for an arbitrary smooth
manifold $\Mcal$ with boundary $\p \Mcal$) in terms of single and double layer potentials. In this approach
we assume known the corresponding Green's functions $G_D(y,\vec{x},y',\vec{x}')$ and
 $G_R(y,\vec{x},y',\vec{x}')$ in the bulk\footnote{In this section we use coordinates $(y,\vec{x})$,
where $y$ is the radial variable and $\vec{x}=(x^1,\ldots, x^d)$ parametrise the $d$--dimensional boundary of a 
conformally compact manifold $\Mcal$. If $\Mcal$ is $\AdS$, the metric would be the usual ``upper--half--space''
metric. For simplicity, we set $k=1$. 
Note finally that the outward normal vector in these coordinates is roughly $N=-y\p_y$ (and exactly that for $\AdS$).}, 
which behave as (see the discussion in section 2, and \cite{Solodukhin99})
\begin{equation} \label{Greensasympt}
G_D(y,\vec{x},y',\vec{x}')\ra y^\Dim\;, G_R(y,\vec{x},y',\vec{x}')\ra y^{d-\Dim}\;,[N+(d-\Dim)]G_R(y,\vx,y',\vx')\ra O(y^\Dim)\;,
\end{equation}
where $N$ is the outward unit normal vector. As in section \ref{Propagators}, we write for a function $u(y,\vec{x})$, satisfying 
the laplacian in the bulk, the expression 
\begin{equation} \label{usol}
u(y,\vec{x})\ra f(\vec{x}) y^{d-\Dim} +h(\vec{x}) y^\Dim+\cdots
\end{equation}
In writing these formulas we are mainly thinking of the cases $\Dim=1$ and $\Dim=2$, where the two independent solutions
are found at contiguous powers of $y$. It is convenient to introduce the boundary operators
\begin{equation}
\Bcal_D^\Dim: u(y,\vec{x})\mapsto \left[y^{\Dim-d} u(y,\vec{x})\right]_{\p\Mcal} \quad \text{and} \quad 
\Bcal_R^\Dim: u(y,\vec{x})\mapsto \left[y^{-\Dim} \left(N+(d-\Dim)\right)u(y,\vec{x})\right]_{\p\Mcal}\;. 
\end{equation}
They are designed to either restrict to the boundary, or to restrict to the boundary after applying the Robin 
derivative, respectively. Note that we have included the correct scaling factors which will give a finite result acting
on $u$ as in (\ref{usol}), so $\Bcal_D^\Dim\cdot u(y,\vec{x})=f(\vec{x})$ and likewise 
$\Bcal_R^\Dim\cdot u(y,\vec{x})=(d-2\grl)h(\vec{x})$. Now let us restrict to $d=3,\Dim=2$, and thus write simply $\Bcal_D$ and $\Bcal_R$ 
from now on. In that case, as discussed in section \ref{Propagators}, the bulk Green's function (\ref{Gone}) corresponds
to $G_R(y,\vx,y',\vx')$. 

As usual, the Robin problem (specifying $h(\vec{x})$ on the boundary) can be solved via a single--layer potential 
\begin{equation} \label{singlelayer}
u(y,\vec{x})=-\int_{\p\Mcal'} (\Bcal_D)_{y'}G_R(y,\vec{x},y',\vec{x}') h(\vec{x}') \diff\Omega(x')
\end{equation}
which clearly, given (\ref{Greensasympt}), gives $(\Bcal_R)_y\cdot u(y,\vec{x})=-h(\vec{x})$. However, we can do something 
different: We can restrict $u(y,\vec{x})$, given by (\ref{singlelayer}), to the boundary,  and thus obtain the 
Dirichlet data $f(\vec{x})$ as 
\begin{equation} \label{ffromg}
f(\vec{x})=-\int_{\p\Mcal'} (\Bcal_D)_y(\Bcal_D)_{y'} G_R(y,\vec{x},y',\vec{x}') h(\vec{x}') \diff\Omega(x')\;.
\end{equation}
This is a boundary operator, relating $f(\vec{x})$ and $h(\vec{x})$. Note that the kernel in (\ref{ffromg}) 
is simply the appropriately scaled restriction of (both points of) the Robin Green's function to the boundary, 
which is precisely what we have been calling $M(x,x')$.
On the other hand, we can definitely find $f(\vec{x})$ in the usual way from the Dirichlet function 
$G_D(y,\vec{x},y',\vec{x}')$, using a double layer potential:
\begin{equation}
u(y,\vec{x})=-\int_{\p\Mcal'}(\Bcal_R)_{y'}\cdot G_D(y,\vec{x},y',\vec{x}')f(\vec{x}')\diff\Omega(x')\;.
\end{equation}
But taking the Robin derivative of \emph{this} expression for $u(y,\vec{x})$ and restricting it 
to the boundary, we are assured (by the uniqueness of the Cauchy problem) to get back $h(\vec{x})$:
\begin{equation}\label{gfromf}
h(\vec{x})=-(\Bcal_R)_y\cdot u(y,\vec{x})=
\int_{\p\Mcal'}(\Bcal_R)_y(\Bcal_R)_{y'}\cdot G_D(y,\vec{x},y',\vec{x}') f(\vec{x}')\diff\Omega(x')\;.
\end{equation}
So we have found a function on the boundary giving $-h(\vec{x})$ from $f(\vec{x})$. Of course, by definition,  
this is nothing but the Dirichlet--to--Robin operator $\Bcal$, so we can write
\begin{equation}
-\int_{\p\Mcal'}(\Bcal_R)_y(\Bcal_R)_{y'} G_D(y,\vec{x},y',\vec{x}')
(\cdots)\diff\Omega(x')=\Bcal_x (\cdots)
\end{equation}
as an operator\footnote{See also \cite{Katchalovetal01}, chapter 4,  for a different proof of this.}. 
But now substituting (\ref{ffromg}) in (\ref{gfromf}) gives
\begin{equation}
\begin{split}
h(\vec{x})&=-\int_{\p\Mcal''} \int_{\p\Mcal'} (\Bcal_R)_y(\Bcal_R)_{y'}G_D(y,\vec{x},y',\vec{x}')\cdot
(\Bcal_D)_{y'}(\Bcal_D)_{y''}G_R(y',\vec{x}',y'',\vec{x}'') h(\vec{x}'')\diff\Omega(x')\diff\Omega(x'')\\
&=-\int_{\p\Mcal''} \left\{\int_{\p\Mcal'}\left[(\Bcal_R)_y(\Bcal_R)_{y'}G_D(y,\vec{x},y',\vec{x}')\right]
M(\vec{x}',\vec{x}'')\diff\Omega(x')\right\}h(\vec{x}'')\diff\Omega(x'')\\
&=\int_{\p\Mcal''} \Bcal_{x}M(\vec{x},\vec{x}'')h(\vec{x}'')\diff\Omega(x'')
\end{split}
\end{equation}
in other words, 
\begin{equation} \label{BMproof}
\Bcal_{x}M(\vx,\vx') =\frac{1}{\sqrt{g}}\delta^{(d)}(\vx,\vx')\;.
\end{equation}
This rather general argument shows that the Dirichlet--to--Robin operator is the inverse to the 
\emph{restriction} of the Robin Green's function to the boundary. This is the main result that, 
combined with the conformal covariance of the Dirichlet--to--Robin operator, will be used to check 
whether $M(x,x')$ has the form that is required by conformal invariance on the Berger sphere.


Let us also point out that in the hyperbolic case $\Bcal$ can be thought of in a 
group--theoretical way, as an intertwining operator acting between representations of the conformal group $\SO(d+1,1)$ 
\cite{Branson87,Branson97}. In fact this aspect is treated in the AdS/CFT context by Dobrev \cite{Dobrev99}, where it is 
shown that the operator interpolating between $f$ and $h$ (called $\phi_0$ and $\Ocal$ there) is the boundary 
correlator $G^b_{\Dim}(x-x')\sim 1/(x-x')^{2\Dim}$: 
\begin{equation} \label{hfromf}
h(x)\sim\int G^b_\Dim(x,x')f(x') \diff^d x'\;.
\end{equation}
Let us actually check this using the definition of the Dirichlet--to--Robin operator, in upper--half--space coordinates, where the boundary is $\Rset^d$:
In this case, as we will show towards the end of section \ref{HHfactorisation}, the Dirichlet--to--Robin operator for
hyperbolic space is simply 
\begin{equation}
\Bcal=\sqrt{\Delta_0}
\end{equation}
where $\Delta_0$ denotes the positive boundary laplacian. We can now apply the definition of the square root of the 
laplacian, which, according to appendix \ref{Pseudodifferential}, can be given (in $\Rset^d$) through the Fourier transform:
\begin{equation}
(\sqrt{\Dzero})_xf(x)=\frac{1}{(2\pi)^d}\int_{\Rset^d} |\xi|\hat{f}(x)e^{ix\xi}\diff^d\xi
=\frac{1}{(2\pi)^d}\int_{\Rset^3}\left[\int_{\Rset^d}|\xi|e^{i\xi(x-x')}\diff^d\xi\right]f(x')\diff^d x'\;.
\end{equation}
Now we can use the formula \cite{Dobrev99} (see also \cite{Dobrevetal77}, p. 67)
\begin{equation}
\int\frac{e^{-i\xi x}}{x^{2\Dim}}\diff^d x=\frac{(2\pi)^{\frac{d}{2}}\Gamma(\frac{d}{2}-\Dim)}{2^{2\Dim-\frac{d}{2}}
\Gamma(\Dim)}|\xi|^{2\Dim-d}\;.
\end{equation}
Inverting (by multiplying with $e^{-iy\xi}$ and integrating over $\diff^d y$), and specialising to $d=3,\Dim=2$ we find
\begin{equation}
\int|\xi|e^{i\xi (x-x')}\diff^3\xi\sim \frac{1}{[(x-x')^2]^2}\;.
\end{equation}
This is the same as the kernel $G_\Dim$ in (\ref{hfromf}), so we have verified that the intertwining operator relating the 
two independent boundary values in the group--theoretical approach of \cite{Dobrev99} is simply  the Dirichlet--to--Robin
operator. Note that a very similar calculation can be found in \cite{ArefevaVolovich98}.

After these remarks, let us now 
specialise to QTN and conformal coupling. We now know that the (pseudo)differential operator, whose Green's function
is our correlation function $M(x,x')$ (\ref{correlator}), is the Dirichlet--to--Robin operator $\Bcal$. 
Given the discussion above on the conformal invariance of $\Bcal$, we see that if we can verify (\ref{BMproof}) 
for $M(x,x')$, we can immediately conclude that it is a two--point function of a boundary
operator of conformal dimension $1$. In this case, this is simply a nice (and, we believe, necessary) check of our results, 
since $M(x,x')$ \emph{should} satisfy (\ref{BMproof}) by construction. However perhaps cases will arise where one 
is given a boundary correlation function without knowing its origins as a restriction of a bulk Green's function, and 
then (\ref{BMproof}) could provide a way of checking whether it corresponds to a $\Dim=1$ operator.  

Having discussed the background ideas we need, we will now move on to the calculations.

\section{The Round Sphere} \label{Roundsphere}

Before applying the methods we discussed previously to the Berger sphere, we believe it is instructive to 
do so for the simplest possible case, namely the round three--sphere, which has $\HH$ as infilling manifold.
This case, which of course corresponds to the standard AdS/CFT situation on $\AdS_4$, has many simplifying features 
which allow us to demonstrate the main idea of the calculation without going into details. 

\subsection{Factorisation of the hyperbolic laplacian} \label{HHfactorisation}

So let us start by applying the ideas of section \ref{DirichlettoNeumann}, which boil down to 
trying to find a suitable factorisation of the hyperbolic space laplacian. We begin by recalling  the polar--type metric on 
global $\HH$:
\begin{equation}
\diff s^2=\frac{4}{(1-k^2r^2)^2}\left(\diff r^2+r^2(\sone^2+\stwo^2+\sthree^2)\right)\;.
\end{equation}
The conformal Laplace--Beltrami operator is 
\begin{equation} \label{HHlap}
\nabla^2+2k^2=\frac{(1-k^2r^2)^2}{4}\left[\p_{rr}+\frac{3+k^2r^2}{r}\p_{r}
+\frac{4}{r^2}(-\Delta_0)\right]+2k^2\;.
\end{equation}
Here $\Delta_0$ denotes the positive laplacian on the boundary three--sphere (see also appendix \ref{BergerSphere}) , 
\begin{equation}
-\Delta_0=\nabla^2_{S^3}=\p_{\theta\theta}+\cot\theta\p_{\theta}+\csc^2\theta(\p_{\varphi\varphi}
-2\cos\theta\p_{\varphi\psi}+\p_{\psi\psi})\;.
\end{equation}
The right--hand--side of (\ref{HHlap}) can be easily factorised to give
\begin{equation} \label{hyperbolicfactorisation}
\nabla^2+2k^2=\left(N^r\p_r-\frac{2}{3}\Kcal+\Acal\right)\left(N^r\p_r-\frac{1}{3}\Kcal-\Acal\right)
\end{equation}
where $N^r=(1-k^2r^2)/2$ is the $r$--component of the contravariant unit normal vector 
$n^\mu=(N^r,0,0,0)$ to a surface defined by $r=r_0$ (where $r_0$ is some constant), $\Kcal$ is the trace of the extrinsic 
curvature, 
\begin{equation}
\Kcal=-\nabla_r N^r=-\frac{3(1+k^2r^2)}{2r}\;,
\end{equation}
and the pseudodifferential first order operator $\Acal$ is given by\footnote{As mentioned in the introduction, the 
factor $\frac{1}{4}$ could be thought of as \emph{four} dimensional conformal coupling $\frac{1}{6}\Rcal_0$ on the 
three--sphere, where $\Rcal_0=3(1-k^2r^2)^2/(2r^2)$.}
\begin{equation} \label{AcalAdS}
\Acal=\frac{1-k^2r^2}{r}\sqrt{\Delta_0+1/4}\;.
\end{equation}
This can be easily checked by expanding out the right-hand side of (\ref{hyperbolicfactorisation}) 
and equating with the left--hand  side. Notice that the linear in $\Acal$ terms on the right--hand side 
\begin{equation}
-N^r\p_r\Acal+\frac{1}{3}\Kcal\Acal=\left(-\frac{1-k^2r^2}{2}\p_r\left(\frac{1-k^2r^2}{r}\right)-\frac{1-k^4r^4}{2r^2}\right)
\sqrt{\Dzero+1/4}
\end{equation}
cancel on their own. This means we could have taken $\Acal$ to have the opposite sign and it would still be a solution. 
So we have made a choice of sign here, which is dictated by requiring that $\Acal$ be a positive operator. This can be traced
back to the choice of the positive square root for the outer normal vector in (\ref{positive})\footnote{The sign
of $\Acal$ also depends on the sign in (\ref{hyperbolicfactorisation}) but that is purely conventional, and indeed here it
is opposite from that in section \ref{Definition}. Even with the opposite sign, we would still choose
 $\Acal$ as in (\ref{AcalAdS}) based on positivity.}. 
Now following the arguments in section \ref{DirichlettoNeumann}, we conclude that the 
Dirichlet--to--Robin operator at the inner boundary ($r=r_0<1/k$) is
\begin{equation}
\left[N^r\p_r-\frac{1}{3}\Kcal\right]|_{r=r_0}=\Acal|_{r=r_0}\;.
\end{equation}
The last step is to rescale $\Acal$ to obtain an operator on the boundary at infinity. The most obvious
way is to simply scale to the Dirichlet--to--Robin operator $\Bcal$ on this conformal boundary as: 
\begin{equation} \label{BAdS}
\Bcal=\lim_{r\ra1/k} \frac{1}{1-k^2r^2}\Acal=k\sqrt{\Delta_0+1/4}\;.
\end{equation}
So here we use the same defining function we used to define the boundary laplacian (see appendix \ref{BergerSphere}). 
This operator is the same as the one appearing (as an example of a conformally invariant operator) in Branson's 
work \cite{Branson87,BransonGover01} (where the bulk manifold of study is not hyperbolic space but $\Rset\times \Srm^{d}$).\footnote{In \cite{Branson87}, for $d=3$ the expression for $\Bcal$ is $\sqrt{\Delta_B+1}$, which comes about 
because there the definition of the laplacian on $\Srm^3$ is such that its eigenvalues are $j(j+2)$. To compare
 (since we are allowing half-integral $l$) we need to take $j=2l$, in which case (setting also $k=2$ which corresponds to the unit
sphere---see appendix \ref{BergerSphere}) the expressions match precisely. The \emph{reason} they match is possibly
related to that $\HH$ is conformal to flat space, and $\Bcal$ is conformally invariant.}

Although it is not needed in the following, it is instructive to  also derive the 
Dirichlet--to--Robin operator for $\AdS_4$ in the Poincar\'e or upper--half--space parametrisation. 
Here the laplacian is simply (see e.g. \cite{DHokerFreedman02})
\begin{equation}
\nabla^2_P=y^2\p_{yy}-2y\p_y-y^2\Delta_{B}
\end{equation}
where here $\Delta_{B}=-\sum_{i=1}^{3}\p_{ii}$ is simply the positive laplacian on $\Rset^3$, the boundary
of $\AdS_4$ in upper--half space coordinates. Now the conformally coupled laplacian can be easily factorised as 
(we set $k=1$ here)
\begin{equation}
\nabla^2_P+2=\left(-y\p_y+2+y\sqrt{\Delta_{B}}\right)\left(-y\p_y+1-y\sqrt{\Delta_{B}}\right)\;.
\end{equation}
Note that the outward unit normal vector is $-y\p_y$, and $\Kcal=-3$. Taking the limit $y\ra 0$ we find 
\begin{equation} \label{BPoincare}
\Bcal_P=\sqrt{\Delta_B}\;.
\end{equation}
Note the absence of the $1/4$ factor, which is a good illustration of the fact that only the \emph{principal} symbol 
of a pseudodifferential operator needs to 
be coordinate--invariant. Lower order terms \emph{do} depend on the coordinate system used. 
The form (\ref{BPoincare}) of the Dirichlet--to--Robin operator was used in section 
\ref{Implications} in comparing with the results of \cite{Dobrev99}.

\subsection{Eigenfunction expansion of the correlation function} \label{Roundexpansion}

Now we need to act with $\Bcal$ on the two-point function 
\begin{equation} \label{MHH}
M^0(x,x')=C\cdot\frac{1}{1-\vone}
\end{equation}
where $C$ is some normalisation constant. Hopefully we will obtain a delta function, demonstrating that, 
as claimed in the introduction,  $M^0(x,x')$ is the Green's function of $\Bcal$.
The problem of course is that $\Bcal$ is a pseudodifferential operator and it is not immediately clear
(to this author!) how to act with it on a function on $\Srm^3$. We find it most 
convenient to expand $M^0(x,x')$ in eigenfunctions\footnote{This is, after all, a physics paper.}, on which
(as discussed at the end of appendix \ref{Pseudodifferential}) it is 
trivial to act with $\Bcal$ since we can easily find its eigenvalues.  
 
So first we need to expand (\ref{MHH}) in eigenfunctions on the round sphere. Since the round sphere is the 
symmetric space $\SO(4)/\SO(3)$, 
there are various ways to do this. For instance we could
use directly the results of Drummond \cite{Drummond75}, who gives a formula for the expansion 
 of arbitrary powers of $|x-x'|$ in terms of spherical harmonics\footnote{Our expression 
$1/(1-\vone)$ is really just $1/|x-x'|$ if we define $x,x'$ in terms of the embedding coordinates
of $\Srm^3$ in $\Rset^4$ as in \cite{Drummond75}.}. However, this expansion depends on the bi--invariance of the metric, 
so keeping in mind the equivalent calculation for the Berger sphere (where we have only left--invariance) that awaits us, we will 
choose a different route. 

To begin, we need to retrace our steps a little bit. In section \ref{Roundlimit}, we found $M^0(x,x')$ as the 
limit of $M(x,x')$ as 
$\mu=0$. We could instead have taken the limits differently, that is \emph{first} take $\mu=0$ in 
$K^{(\mu,k)}(r,x,x')$, and \emph{then} obtain $M^0(x,x')$ as the restriction to the boundary. In that case, we write
\begin{equation}
M^0(x,x')=\lim_{r\ra1/k} C'\cdot\frac{1}{(1-k^2r^2)} K^{(0,k)}(r,x,x')
=C'\cdot\lim_{r\ra1/k} \frac{1}{1+k^2r^2-2kr \vone}
\end{equation}
where $C'=2C$. Now, for reasons soon to become clear, let us decide \emph{not} to take the 
limit $r\ra1/k$ until later in the calculation. Then, the expansion we are looking for is:
\begin{equation}
M^0(x,x')=C'\cdot\sum_{lmn}\alpha_{lmn}\Ylmn(x)\overline{\mathrm{Y}}^{l}_{\;mn}(x')\;,
\end{equation}
where $l$ takes all integer and half--integer values from $0$ to $\infty$, while $m,n$ range from $-l,-l+1,\ldots l$. 
As discussed in  appendix \ref{BergerSphere}, 
$\Ylmn(x)$ (appropriately normalised) equals $\sqrt{2l+1}e^{-im\varphi}e^{-in\psi}\Plmn(\cos\theta)$.
To simplify matters, we can choose particular values for the coordinates of the point $x'$, and 
reinstate them at the end. We will use $\theta'=\varphi'=\psi'=0$, which gives $\overline{\mathrm{Y}}^{l}_{mn}(0)=\sqrt{2l+1}$. 
Also, $\vone$ in (\ref{vone}) simplifies to $\vone=\cos(\theta/2)\cos((\varphi+\psi)/2)$. 
Now we are looking at the expansion
\begin{equation}
M^0(x,0)=C'\cdot\sum_{lmn}\sqrt{2l+1}\alpha_{lmn}\Ylmn(x)=
C'\cdot\sum_{lmn}(2l+1)\alpha_{lmn}e^{-im\varphi}e^{-in\psi}\Plmn(\cos\theta)
\end{equation}
where the coefficients $\alpha_{lmn}$ are given by (see e.g. \cite{Vilenkin,Varshalovichetal})
\begin{equation}
\alpha_{lmn}=\lim_{r\ra1/k}\int_0^{\pi}\int_{0}^{2\pi}\int_0^{4\pi} \frac{b}{1+k^2r^2-2kr\cos\half\theta\cos\half(\varphi+\psi)} 
e^{im\varphi}e^{in\psi}\Plmn(\cos\theta) 
\diff\psi\diff\varphi\sin\theta\diff\theta\;.
\end{equation}
Here $b=(-1)^{m-n}/(16\pi^2)$. First we do the $\psi$ integration, by converting to a contour integral, and end up with:
\begin{equation}
\alpha_{lmn}=4\pi b\lim_{r\ra1/k} \int_{0}^{\pi}\int_{0}^{2\pi} \frac{1}{\sqrt{\Delta}}
\left(\frac{1+k^2r^2-\sqrt{\Delta}}{2kr\cos\half\theta}\right)^{2n}e^{i(m-n)\varphi}\Plmn(\cos\theta)\diff\varphi
\sin\theta\diff\theta
\end{equation}
with $\Delta$ given by (\ref{Delta}) with $\mu=0,\theta'=0,\varphi'=0$, i.e. $\Delta=1+k^4r^4-2k^2r^2\cos\theta$. 
Performing the (trivial) $\varphi$ integration, we get:
\begin{equation} \label{integralround}
\alpha_{lmn}=8\pi^2b\lim_{r\ra 1/k} \int_0^\pi \frac{1}{\sqrt{\Delta}}
\left(\frac{1+k^2r^2-\sqrt{\Delta}}{2kr\cos\half\theta}\right)^{2n}
P^{l}_{\;nn}(\cos\theta)\delta_{mn}\sin\theta\diff\theta
\end{equation}
This is a relatively nontrivial integral, and we can now see the value of not taking the 
limit $r\ra1/k$ previously. Defining $h=k^2r^2$, we see that the integrand is 
\begin{equation} \label{integrandround}
\frac{1}{\sqrt{1+h^2-2h\cos\theta}}
\left(\frac{1+h-\sqrt{1+h^2-2h\cos\theta}}{2\sqrt{h}\cos\half\theta}\right)^{2n}\;.
\end{equation}
A look at, for example, \cite{Vilenkin} or \cite{Varshalovichetal} reveals the close similarity of
(\ref{integrandround}) to the generating function of the $\Plmn(\cos\theta)$, which we reproduce here 
for $h<1$:
\begin{equation} \label{Plmngenerating}
\frac{t^{m-n}\left(it\sin(\theta/2)+\cos(\theta/2)\right)^{2n}}{\sqrt{1-2 h\cos\theta+h^2}}=
\sum_{l=n}^{\infty}\sqrt{\frac{(l-n)!(l+n)!}{(l-m)!(l+m)!}}\Plmn(\cos\theta)h^{l-n}
\end{equation}
with
\begin{equation}
t=\frac{1-h\cos\theta+\sqrt{1-2h\cos\theta+h^2}}{ih\sin\theta}\;.
\end{equation}
Note that this holds for $m,n$ (and thus $l$) half--integer as well as integer. Setting $m=n$ in  (\ref{Plmngenerating}) 
we can easily conclude that our integrand (\ref{integrandround}) can be expanded in $h$ as 
\begin{equation} \label{expansionround}
\frac{1}{\sqrt{\Delta}}\left(\frac{1+h-\sqrt{\Delta}}{2\sqrt{h}\cos\half\theta}\right)^{2n}
=h^n\frac{1}{\sqrt{\Delta}}\left(\frac{1+h-\sqrt{\Delta}}{2h\cos\half\theta}\right)^{2n}
=\sum_{s=n}^{\infty} P^{s}_{\;nn}(\cos\theta)h^s\;.
\end{equation}
Substituting in the integral (\ref{integralround}), we have
\begin{equation}
\alpha_{lmn}=8\pi^2b\lim_{h\ra1}\sum_{s=n}^{\infty}\delta_{mn}h^s\int_0^\pi 
P^{s}_{\;nn}(\cos\theta)P^{l}_{\;nn}(\cos\theta)\;.
\sin\theta\diff\theta
\end{equation}
Using the well--known orthonormality property of the $\Plmn(\cos\theta)$ (for any $m,n$, and thus also for $m=n$):
\begin{equation}
\int_0^\pi P^{s}_{\;mn}(\cos\theta)P^{l}_{\;mn}(\cos\theta)\sin\theta\diff\theta
=\int_{-1}^{1}P^{s}_{\;mn}(z)P^{l}_{\;mn}(z)\diff z=\frac{2}{2l+1}\delta_{sl}
\end{equation}
(where $z=\cos\theta$) and summing over $s$, we conclude
\begin{equation}
\alpha_{lmn}=16\pi^2b\lim_{h\ra1}\delta_{mn}\frac{h^l}{2l+1}\;.
\end{equation}
We are now free to take the limit $h\ra1$ \footnote{Of course, one can question the validity of
taking $h\ra1$ when (\ref{expansionround}) is really only valid for $h<1$. But since
orthonormality leaves only one term in the $h$--expansion for each $l$, namely the one proportional 
to $h^s$ for $s=l$, questions of convergence should not arise. Of course one can also check this by first taking $h\ra1$ in 
(\ref{integralround}) and explicitly doing the integral, say for particular values of $l,n$, using e.g. the tables for
the $\Plmn(z)$ given in \cite{Varshalovichetal}.} and obtain the final formula for the
$\alpha_{lmn}$: 
\begin{equation}
\alpha_{lmn}=16\pi^2\frac{(-1)^{m-n}}{16\pi^2}\frac{\delta_{mn}}{2l+1}=\frac{\delta_{mn}}{2l+1}
\end{equation}
(where we substituted $b$). From this we obtain
\begin{equation}
M^0(x,0)=C\cdot\frac{1}{1-\vone}=C'\cdot\sum_{lmn} \delta_{mn} e^{-i(m\varphi+n\psi)}\Plmn(\cos\theta)\;.
\end{equation}
Reinstating $\theta',\varphi',\psi'$ using (an adaptation of) the composition theorem for the $\Plmn(\cos\theta)$ 
(see e.g. \cite{Vilenkin},
III.4)\footnote{In comparing with the formulas in \cite{Vilenkin}, note that the sign of $\theta'$ is opposite to ours,
so one must use the property $t^{l}_{mn}(-\theta)=\bar{t}^{l}_{nm}(\theta)$ given on p.123 of that reference.}  
\begin{equation}
e^{in(\varphi+\psi)}P^l_{nn}(\cos\theta)\ra\sum_{k=-l}^{l}e^{-in(\psi-\psi')}e^{-ik(\varphi-\varphi')}P^{l}_{nk}(\cos\theta)P^{l}_{nk}(\cos\theta')
\end{equation}
and reverting to the normalised eigenfunctions $\Ylmn(x)$, we finally conclude that
\begin{equation}
M^0(x,x')=C\cdot \sum_{lmn} \frac{2}{2l+1} \Ylmn(x)\overline{\mathrm{Y}}^l_{\;mn}(x')\;.
\end{equation}
It is worth stressing once more that the calculation we just went through is \emph{not} the recommended way
of expanding a function into harmonics on the round sphere! We could have arrived at this simple result
with less effort. However the Berger sphere calculation to follow (in section \ref{Bergerexpansion})
will be modelled, step--by--step, on the previous calculation, which will thus serve as a guide. 

\subsection{Checking the correlation function}

Now recall that, on the round sphere, the eigenfunctions $\Ylmn(x)$  satisfy (see appendix \ref{BergerSphere})
\begin{equation}
\Delta_0 \Ylmn(x)=l(l+1) \Ylmn(x)\;.
\end{equation}
We then find that the $\Ylmn(x)$ are eigenfunctions of the operator $\Delta_0+1/4$ with
eigenvalues $(2l+1)^2/4$
\begin{equation}
(\Delta_0+1/4)\Ylmn(x)=\Bcal^2\Ylmn(x)=\frac{1}{4}(2l+1)^2 \Ylmn(x)
\end{equation}
and thus they are also eigenfunctions of $\Bcal=\sqrt{\Delta_0+1/4}$ \footnote{We set $k=1$ here for simplicity.} with 
eigenvalues $(2l+1)/2$. 
\begin{equation}
\Bcal_x\Ylmn(x)=\frac{(2l+1)}{2}\Ylmn(x)\;.
\end{equation}
 We are finally ready to act with $\Bcal$ on $M^0(x,x')$ (ignoring the unknown normalisation constant):
\begin{equation}
\begin{split}
\Bcal_x M^0(x,x')=& \Bcal_x \left(\sum_{lmn} \frac{2}{2l+1}\Ylmn(x)\overline{\mathrm{Y}}^l_{mn}(x')\right)
=\sum_{lmn} \frac{2}{2l+1}(\Bcal_x \Ylmn(x))\overline{\mathrm{Y}}^{l}_{mn}(x')\\
=&\sum_{lmn}\Ylmn(x)\overline{\mathrm{Y}}^l_{mn}(x')=\frac{1}{\sqrt{g}}\delta^{(3)}(x,x') \;.
\end{split}
\end{equation}
We just rederived the well--known result that the Green's function $M^0(x,x')$ on the boundary of 
hyperbolic space corresponds to a boundary operator $\Ocal$ of conformal dimension $\Dim_\Ocal=1$. Of
course there are easier ways to see this (two of them have been demonstrated in section \ref{Roundlimit})
but we believe it was worthwhile to go through all the steps before tackling the case
of QTN in the next section.

\section{The Berger Sphere}   \label{Bergersphere}

In the previous section we demonstrated that the Dirichlet--to--Robin operator is indeed 
the pseudodifferential operator whose Green's function is our two--point function $M^0(x,x')$ 
in the case of round $\Srm^3$. Now we turn to the Berger sphere, and try to follow the calculations in section
\ref{Roundsphere} as closely as possible. Of course things get rather complicated, but remarkably we can 
obtain exact answers that generalise those for hyperbolic space and the round sphere.

\subsection{Factorisation of the QTN laplacian} \label{QTNfactorisation}

In this section we find the Dirichlet--to--Robin operator for the boundary of QTN. We will follow the same 
procedure as in the previous section, i.e. we will look for a factorisation of the bulk conformally coupled
laplacian of the form:
\begin{equation} \label{QTNansatz}
\nabla^2+2k^2=\left(N^r\p_r-\frac{2}{3}\Kcal+\Acal\right)\left(N^r\p_r-\frac{1}{3}\Kcal-\Acal\right)\;.
\end{equation}
Here the outward unit normal vector $N=N^r\p_r$ and the extrinsic curvature trace $\Kcal=-\nabla_rN^r$ at some radial distance $r$ are 
given by
\begin{equation}
N^r=\frac{(1-k^2r^2)}{2}\sqrt{\frac{1-\mu^2k^2r^4}{1-\mu^2r^2}}
\end{equation}
and 
\begin{equation}
\Kcal=-\sqrt{\frac{1-\mu^2r^2}{1-\mu^2k^2r^4}}\frac{(3-4\mu^2r^2+3k^2r^2-7\mu^2k^2r^4-\mu^2k^4r^6+6\mu^4k^2r^6)}{2r(1-\mu^2r^2)^2}\;.
\end{equation}
As in section \ref{HHfactorisation}, the goal is to determine the nonlocal operator $\Acal$. We proceed by expanding out 
the various terms:
\begin{equation} \label{QTNexpansion}
\nabla^2+2k^2=(N^r)^2\p_{rr}+(N^r\p_rN^r-\Kcal N^r)\p_r+(2/9\Kcal^2-1/3 N^r\p_r \Kcal)
-(N^r\p_r\Acal-1/3\Kcal\Acal)-\Acal^2\;.
\end{equation}
Comparing with (\ref{Pedlaplacian}) one sees immediately that the first two terms on the right--hand side 
match the corresponding ones in $\nabla^2+2k^2$, i.e.
\begin{equation}
(N^r)^2\p_{rr}=\frac{(1-k^2r^2)^2(1-\mu^2k^2r^4)}{4(1-\mu^2r^2)}\p_{rr}\;,
\end{equation}
and
\begin{equation}
(N^r\p_rN^r-\Kcal N^r)\p_r=\frac{(1-k^2r^2)(3+k^2r^2-7\mu^2k^2r^4+3\mu^2k^4r^6)}{4r(1-\mu^2r^2)}\p_r\;.
\end{equation}
The remaining terms on the left hand side of (\ref{QTNexpansion}) are 
\begin{equation}
-\frac{(1-k^2r^2)^2}{(1-\mu^2r^2)r^2}\left[\Dzero-\frac{\mu^2r^2(2-\mu^2r^2-k^2r^2)}{(1-\mu^2k^2r^4)}\Done\right]+2k^2
\end{equation}
where the \emph{positive} operators $\Dzero$ and $\Done$ are defined in appendix \ref{BergerSphere}. Requiring that 
the left--hand side of 
(\ref{QTNexpansion}) equals the right--hand side leads us to the following Ricatti--type equation 
for $\Acal$:
\begin{equation} \label{Ricattione}
N^r\p_r\Acal=-\Acal^2+\frac{1}{3}\Kcal\Acal
+\frac{(1-k^2r^2)^2}{(1-\mu^2r^2)r^2}\left(\Dzero-\frac{\mu^2r^2(2-\mu^2r^2-k^2r^2)}{(1-\mu^2k^2r^4)}\Done\right)+X(r)
\end{equation}
where the zeroth--order part $X(r)$ is given by 
\begin{equation}
X(r)=2/9\Kcal^2-1/3 N^r\p_r \Kcal-2k^2=-\frac{\mu^4r^2(1-k^2r^2)^4}{9(1-\mu^2r^2)^3(1-\mu^2k^2r^4)}
-\frac16\Rcal_{r}\;,
\end{equation}
$\Rcal_r$ being the scalar curvature of the induced metric at the (inner) boundary, which is (see
appendix \ref{BergerSphere})
\begin{equation}
\Rcal_r=\frac{(1-k^2r^2)^2}{2r^2(1-\mu^2r^2)^3}(3-8\mu^2r^2+\mu^2k^2r^4+4\mu^4r^4)\;.
\end{equation}
The nonlinear, first--order equation (\ref{Ricattione}) can be solved for $\Acal$ in the usual manner, 
by transforming it into a \emph{linear}, second--order equation. We will outline the main steps: First, 
we bring it to the standard Ricatti form:
\begin{equation}
\p_r\Acal=a \Acal^2+b\Acal+c
\end{equation}
with 
\begin{equation}
a=-\frac{2}{(1-k^2r^2)}\sqrt{\frac{1-\mu^2r^2}{1-\mu^2k^2r^4}}\;,
\end{equation}
\begin{equation}
b=\frac{3+3k^2r^2-4\mu^2r^2-7\mu^2k^2r^4-\mu^2k^4r^6+6\mu^4k^2r^6}{3r(1-k^2r^2)(1-\mu^2r^2)(1-\mu^2k^2r^4)}
\end{equation}
and 
\begin{equation}
\begin{split}
c=&-\frac{1}{a}\left[\frac{4\Dzero}{r^2(1-\mu^2k^2r^4)}-\frac{4\mu^2(2-k^2r^2-\mu^2r^2)\Done}{(1-\mu^2k^2r^4)^2}
\right.\\
&\left.+\frac{9-24\mu^2r^2-6\mu^2k^2r^4+8\mu^4r^4+32\mu^4k^2r^6-7\mu^4k^4r^8-12\mu^6k^2r^8}
{9r^2(1-\mu^2r^2)^2(1-\mu^2k^2r^4)^2}\right]\;.
\end{split}
\end{equation}
The next step is to choose $\Acal=-\frac{1}{a}\frac{\p_rw(r)}{w(r)}$, which gives us the equation
\begin{equation} \label{secondorderw}
\p_{rr}w(r)-\left(\frac{\p_r a}{a}+b\right)\p_r w(r)+ac w(r)=0\;.
\end{equation}
In order to solve this, we must make a convenient rescaling of $w(r)$, to reduce (\ref{secondorderw}) 
to a known second--order equation. It turns out that a particularily nice equation results if we rescale
$w(r)=f(r)v(r)$, where\footnote{The {\sc maple} algebraic manipulation program was instrumental in
finding this expression.} 
\begin{equation}
f(r)=(1-\mu^2r^2)^{\frac16}(1-\mu^2k^2r^4)^{\frac16-\frac{\mu}{2k}\sqrt{\Done}}
\left(\frac{1-\mu k r^2}{1+\mu k r^2}\right)^{\half\sqrt{\Done}} r^{-1+\frac{2\mu}{k}\sqrt{\Done}}\;.
\end{equation}
The resulting equation for $v(r)$ is much simpler:
\begin{equation}
\p_{rr}v(r)-\left[\frac{k+3\mu^2k^3r^4-4\mu(1-k^2r^2)\sqrt{\Done}}{kr(1-\mu^2k^2r^4)}\right]\p_r v(r)
-\frac{4(k^2\Dzero+\mu k\sqrt{\Done}-\mu^2\Done)}{k^2r^2(1-\mu^2k^2r^4)} v(r)=0\;.
\end{equation}
The final step is to perform a change of variables: $r\ra u=\frac{1+\mu k r^2}{2\mu k r^2}$ (where we
temporarily assume $\mu$ is nonzero), after
which we have (note that $u(1-u)=-(1-\mu^2k^2r^4)/(4\mu^2k^2r^4)$)
\begin{equation} \label{Fequation}
u(1-u)\p_{uu}v(u)+\left[1-(1+\frac{\mu}{k})\sqrt{\Done}-2\left(1-\frac{\mu}{k}\sqrt{\Done}\right)u\right]\p_u v(u)
+\frac{k^2\Dzero+k\mu\sqrt{\Done}-\mu^2\Done}{k^2} v(u)=0.
\end{equation}
Here we have omitted an overall factor of $4/[(1-\mu^2k^2r^4)r^2]$. We can recognise (\ref{Fequation}) 
as the hypergeometric equation 
\begin{equation} \label{hypergeom}
\left(u(1-u)\p_{uu}+[\gamma-(1+\alpha+\beta)u]\p_u-\alpha\beta\right)v(u)=0,
\end{equation} 
with 
\begin{equation} \label{params}
\begin{split}
\alpha=&\half-\sqrt{\Dzero+1/4}-\frac{\mu}{k}\sqrt{\Done} \\
\beta=&\half+\sqrt{\Dzero+1/4}-\frac{\mu}{k}\sqrt{\Done} \\
\gamma=&1-\left(1+\frac{\mu}{k}\right)\sqrt{\Done},
\end{split}
\end{equation}
whose solution is generically given by (e.g. \cite{Lebedev})
\begin{equation} \label{hypergeomsolution}
v(u)=c_1 F(\alpha,\beta;\gamma;u)+c_2 u^{1-\gamma}F(1-\gamma+\alpha,1-\gamma+\beta;2-\gamma;u)\;.
\end{equation}
However, before we proceed to solve (\ref{Fequation}), we have to backtrack a little and consider the range of validity
of (\ref{hypergeomsolution}): It is really valid when $|u|<1$, and that is clearly not always the case, as can be seen from the
definition of $u$ above. So we will need to perform a fractional linear transformation  to bring it into the 
required domain. We will distinguish two cases, the interior ($r<1/k$) and the exterior ($r>1/k$) of QTN, which will be 
dealt with separately. 

\paragraph{The interior of QTN}

 Here $r<1/k$, and since also $\mu<k$ it is clear that $u=(1+\mu k r^2)/(2\mu k r^2)\geq1$, with equality for $r=1/k$ and only
if $\mu=k$ also. So the relevant coordinate transformation \cite{Lebedev} is $u=1/x$. In terms of this 
$x=2\mu k r^2/(1+\mu k r^2)$, and after rescaling $v(x)=x^\alpha \tilde{v}(x)$, equation (\ref{Fequation}) becomes 
\begin{equation} \label{hyperinterior}
x(1-x)\p_{xx}\tilde{v}(x)+\left(1-\beta+\alpha+(\gamma-2-2\alpha)x\right)\p_x\tilde{v}(x)-\alpha(1+\alpha-\gamma)\tilde{v}(x)\;.
\end{equation}
This new equation has $\alpha'=\alpha$, $\beta'=1+\alpha-\gamma$ and $\gamma'=1-\beta+\alpha$. There is a slight subtlety
here: Recall that we should really be thinking of our operator as acting on eigenfunctions, in which case $\Dzero$ is 
replaced by $l(l+1)$, and $\Done$ by $n^2$. But then $\gamma'=1-(2l+1)=-2l$, i.e. it is a (negative) integer and the usual 
solution (\ref{hypergeomsolution}) does not hold (i.e. does not lead to two independent solutions). 
Looking at, for instance, \cite{ErdelyiI} we can find the correct solution for this  degenerate case: 
\begin{equation}
\tilde{v}(x)=c_1x^{1-\gamma'}F(\alpha'-\gamma'+1,\beta'-\gamma'+1;2-\gamma';x)+c_2F(\alpha',\beta';1+\alpha'+\beta'-\gamma';1-x)\;
\end{equation}
Replacing $x$ with the original variable  $r$, we thus have  
\begin{equation}
\tilde{v}(x)=c_1\left(\frac{2\mu k r^2}{1+\mu k r^2}\right)^{\beta-\alpha} \cdot F_1(r)+c_2 F_2(r)
\end{equation}
(notice that here we use the original $\alpha$ and $\beta$ from (\ref{params})), where 
\begin{equation}
F_1(r)=F\left(\half+\sqrt{\Dzero+1/4}+\sqrt{\Done},\half+\sqrt{\Dzero+1/4}-\frac{\mu}{k}\sqrt{\Done};1+2\sqrt{\Dzero+1/4};
\frac{2\mu kr^2}{1+\mu k r^2}\right)
\end{equation}
and
\begin{equation}
F_2(r)=F\left(\half-\sqrt{\Dzero+1/4}+\sqrt{\Delta1},\half-\sqrt{\Dzero+1/4}-\frac{\mu}{k}\sqrt{\Done};1+(1-\frac{\mu}{k})
\sqrt{\Done};
\frac{1-\mu k r^2}{1+\mu k r^2}\right)\;.
\end{equation}
Putting everything together (not forgetting the extra factor of $x^{\alpha}$ from $v(x)=x^\alpha\tilde{v}(x)$), 
the solution  for $\Acal^i$ is 
\begin{equation} \label{Aintfirstsolution}
\Acal^i=\frac{(1-k^2r^2)}{2}\sqrt{\frac{1-\mu^2 k^2 r^4}{1-\mu^2 r^2}}\frac{\p_r\left[f(r)\left(
\left(\frac{2\mu k r^2}{1+\mu k r^2}\right)^\beta F_1(r)+c\cdot\left(\frac{2\mu k r^2}{1+\mu k r^2}\right)^\alpha F_2(r)\right)\right]}{f(r)\left(
\left(\frac{2\mu k r^2}{1+\mu k r^2}\right)^\beta F_1(r)+c\cdot\left(\frac{2\mu k r^2}{1+\mu k r^2}\right)^\alpha F_2(r)\right)}
\end{equation}
where the constant $c=c_1/c_2$. To fix this constant, we can check the behaviour for $r\ll 1/k, 1/\mu$, where the QTN metric 
resembles flat space:
\begin{equation}
\diff s^2=\diff r^2+r^2(\sone^2+\stwo^2+\sthree^2)+O(r^2\diff r^2,r^4)\;.
\end{equation}
By factorising the corresponding laplacian we conclude that $\Acal\ra 1/r\sqrt{\Dzero+1/4}$, for small $r$, 
where the important piece of information is the positive sign of $\Acal$. Looking at the two extreme values of
$c$ in (\ref{Aintfirstsolution}), we can check that
\begin{equation}
\Acal^i(c=0)\ra \frac{\sqrt{\Dzero+1/4}}{r}, \quad \Acal^i(c=\infty)\ra -\frac{\sqrt{\Dzero+1/4}}{r}\;.
\end{equation}
Thus it is clear that taking $c=0$ gives the full answer. So we will take this as our solution for the interior of
QTN. Explicitly, 
\begin{equation} \label{Aint}
\Acal^i=\frac{(1-k^2r^2)}{2}\sqrt{\frac{1-\mu^2 k^2 r^4}{1-\mu^2 r^2}}\p_r\ln\left[f(r)
\left(\frac{2\mu k r^2}{1+\mu k r^2}\right)^{\half+\sqrt{\Dzero+1/4}-\frac{\mu}{k}\sqrt{\Done}}F_1(r)\right]\;.
\end{equation}

\paragraph{The exterior of QTN}
We turn to the case $r>1/k$. First note that in this case the factorisation equation (\ref{QTNansatz}) should be modified 
slightly, since by convention we are defining $\Acal$ with respect to the outward pointing normal vector. If we are solving 
the exterior problem, the sign of the normal vector will change, and so will the sign of the extrinsic curvature of the boundary
(seen as the sphere embedded in the 4--dimensional manifold that is the ``whole'' of QTN). So essentially we need to solve
(\ref{QTNansatz}) with the opposite sign in front of $\Acal$. This clearly changes nothing in the calculations, we can
effectively simply flip the sign of $\Acal^e$ relative to $\Acal^i$.  

On the other hand, we  should check whether the solution of (\ref{Fequation}) changes if $r>1/k$. This condition translates
to $u>1/2$  and we should transform as $u=1/(1-y)$, so that $y=(u-1)/u \leq 1$. The limiting case $y=1$ occurs when $\mu=0$ 
(opposite to the interior problem). 
Substituting in (\ref{Fequation}), and this time rescaling $v(y)=(1-y)^{\alpha} \tilde{v}(y)$, we end up with
\begin{equation} \label{hyperexterior}
y(1-y)\p_{yy}\tilde{v}(y)+\left(1+\beta+\alpha-\gamma-(2-\gamma+2\alpha)y\right)\p_y\tilde{v}(y)
-\alpha(1+\alpha-\gamma)\tilde{v}(y)=0\;.
\end{equation}
This equation is also degenerate, in a similar way to (\ref{hyperinterior}). In this case $\alpha''=\alpha$, 
$\beta''=1+\alpha-\gamma$, and $\gamma''=1+\beta+\alpha-\gamma$, so $\gamma''$ is not an integer, but $\beta''$ is a 
negative integer, being equal to $\half-\sqrt{\Dzero+1/4}+\sqrt{\Done}=-l+n$ acting on eigenfunctions. (We take $n$ 
always less than $l$, see appendix \ref{BergerSphere}). Noticing that also $\gamma''-\alpha''-\beta''$ is integer, 
we can find the two independent solutions by looking at \cite{ErdelyiI}. So we solve (\ref{hyperexterior}) as
\begin{equation}
\tilde{v}(y)= F_3(r)+
\left(\frac{1+\mu k r^2}{1-\mu k r^2}\right)^{(1-\frac{\mu}{k})\sqrt{\Done}}
\left(\frac{2\mu k r^2}{1+\mu k r^2}\right)^{2\sqrt{\Delta_0+1/4}} \cdot F_4(r)\;.
\end{equation}
Here $\alpha$ is as in (\ref{params}), and the two hypergeometric functions are given by
\begin{equation}
F_3(r)= F\left(\half-\sqrt{\Dzero+1/4}+\sqrt{\Done},\half-\sqrt{\Dzero+1/4}-\frac{\mu}{k}\sqrt{\Done};1+(1-\frac{\mu}{k})
\sqrt{\Done};
\frac{1-\mu k r^2}{1+\mu k r^2}\right)
\end{equation}
and
\begin{equation}
F_4(r)= F\left(\half+\sqrt{\Dzero+1/4}-\sqrt{\Done},\half+\sqrt{\Dzero+1/4}+\frac{\mu}{k}\sqrt{\Done};1-(1-\frac{\mu}{k})
\sqrt{\Done};
\frac{1-\mu k r^2}{1+\mu k r^2}\right)\;.
\end{equation}
The important thing to note is that $F_3(r)$ is actually the same as the \emph{second} solution $F_2(r)$ of the interior problem, 
which was discarded there.  
We can identify $F_3(r)$ as the correct one to take for the exterior problem by the following reasoning: The transition 
between $u>1$ and $u<1$ is at the line $r=1/(\mu k)$ which always lies outside the boundary (apart from the point $\mu=k$). 
So for $1/k<r<1/(\mu k)$ we could still use one of the solutions to the interior problem. Assuming that the only discontinuity 
where the solution could change is at the boundary $r=1/k$, clearly the solution for $1/k<r<1/(\mu k)$ should be the same as 
that for $r>1/(\mu k)$, and (since $F_1(r)$ presumably should not be used for $r>1/(\mu k)$) that leaves us with the 
solution which is common in both regions, that is $F_3(r)=F_2(r)$. Notice that this is actually a polynomial, which is
in fact the reason it shows up in both cases. So we write for the solution in the exterior: 
\begin{equation} \label{Aext}
\Acal^e=-\frac{(1-k^2r^2)}{2}\sqrt{\frac{1-\mu^2 k^2 r^4}{1-\mu^2 r^2}}\p_r\ln\left[f(r)
\left(\frac{2\mu k r^2}{1+\mu k r^2}\right)^{\half-\sqrt{\Dzero+1/4}-\frac{\mu}{k}\sqrt{\Done}}F_3(r)\right]\;.
\end{equation}
Note the extra minus sign, which comes from the definition of the exterior problem.

As discussed in section \ref{Definition}, we can now find the Dirichlet--to--Robin operator on the boundary of QTN
by taking the sum of the boundary limits of $\Acal^i$ and $\Acal^e$.  
\begin{equation}
\Bcal=\lim_{r\ra1/k}\half\frac{1}{(1-k^2r^2)}\left[\Acal^i+\Acal^e\right]
\end{equation}
where the limit is understood as coming from lower values for $\Acal^i$ and higher values for $\Acal^e$. This expression
corresponds to the \emph{difference} of the inner and outer normal derivatives at the boundary, and the reason 
we get a plus sign is of course that we decided to flip the sign of $\Acal^e$, so that it matches the outward normal 
vector of the exterior. We will give the final expression for $\Bcal$ in section \ref{QTNcheck}, after we have discussed the 
eigenfunction expansion of $M(x,x')$. Now we turn to some interesting special cases. 

\paragraph{Special Cases}

One special case to consider is the situation where the operators $\Acal^i$ and $\Acal^e$ act on functions that have no 
$\psi$ dependence. Then we can effectively set $\sqrt{\Done}=\sqrt{-\p_{\psi\psi}}=0$ everywhere, and we
obtain\footnote{Here we use the following formulas from \cite{Lebedev}, p. 200, relating the hypergeometric function to
the Legendre functions of the first and second kind : 
\begin{equation} \nonumber
P_l(z)=\left(\frac{z+1}{2}\right)^{l}F\left(-l,-l;1;\frac{z-1}{z+1}\right),\quad 
\text{and}\quad 
Q_l(z)=\frac{\sqrt{\pi}\Gamma(l+1)}{2^{l+1}\Gamma(l+3/2)}(z+1)^{-l-1}
F\left(1+l,1+l;2+2l;\frac{2}{1+z}\right)
\end{equation}
where in both cases $z=1/(\mu k r^2)$.}
\begin{equation}
\Acal^i_0=\frac{(1-k^2r^2)}{2}\sqrt{\frac{1-\mu^2 k^2 r^4}{1-\mu^2 r^2}}\p_r \ln\left[f_0(r)
Q_{\sqrt{\Dzero+1/4}-\half}\left(\frac{1}{\mu k r^2}\right)\right]\;,
\end{equation}
\begin{equation}
\Acal^e_0=-\frac{(1-k^2r^2)}{2}\sqrt{\frac{1-\mu^2 k^2 r^4}{1-\mu^2 r^2}}\p_r \ln\left[f_0(r)
P_{\sqrt{\Dzero+1/4}-\half}\left(\frac{1}{\mu k r^2}\right)\right]
\end{equation}
where $f_0=(1-\mu^2 r^2)^{\frac{1}{6}}(1-\mu^2 k^2 r^4)^{\frac16}/r$. Of course we could also retrieve this 
result by working directly with the hypergeometric equation (\ref{Fequation}) and demonstrating that
it reduces to the Legendre equation when $\Done$ is ignored. 

Another special case arises when we consider the $\mu=0$ limit, which should make contact with the results in 
section \ref{HHfactorisation}. Taking the limit $\mu\ra 0$ of (\ref{Aint}) and (\ref{Aext}) we find 
\begin{equation}
\Acal_0^i=\Acal_0^e=\frac{(1-k^2r^2)}{2r}\sqrt{\Dzero+1/4}\;.
\end{equation}
Thus we have a formula for $\Acal$ that holds everywhere, and precisely matches what we found for hyperbolic space (equation
(\ref{AcalAdS})). This
shows why in that case we could get away with being naive and not paying attention to the difference between
the interior and exterior problem. Note that the opposite sign of $\Acal^e$ was crucial in obtaining this result.

\subsection{Eigenfunction expansion of the correlation function} \label{Bergerexpansion}

In the previous section we obtained an expression for the Dirichlet--to--Robin operator, with which
we have to act on our Green's function. Now we face the task of expanding the two--point function $M(x,x')$
in (\ref{correlator}) into eigenfunctions on the squashed sphere. As discussed in appendix \ref{BergerSphere}, 
separation of variables works in exactly the same way as on the round sphere, which was the topic of section
\ref{Roundexpansion}. Let us then write formally:
\begin{equation} \label{Mexpansionformal}
M(x,x')=\sum_{lmn} \alpha_{lmn} \Ylmn(x)\overline{\mathrm{Y}}^l_{mn}(x')
\end{equation} 
where $\Ylmn(x)=\sqrt{2l+1}e^{-im\varphi}e^{-in\psi}\Plmn(\cos\theta)$ for $x=\{\theta,\varphi,\psi\}$, and similarly for
$\overline{\mathrm{Y}}^l_{mn}(x')$. Recall that the factor of $\sqrt{2l+1}$ is needed for the $\Ylmn(x)$ to be 
orthonormal, and also that
$l=0,\half,1,\ldots$, while $m$ and $n$ range from $-l,-l+1,\ldots l$. 
We can now proceed in exactly the same way as in section \ref{Roundexpansion}, by picking special 
values for the coordinates of the point $x'$, $\theta'=0,\varphi'=0,\psi'=0$, which can be restored at the end. 
Then $\overline{\mathrm{Y}}^l_{mn}(0)=\sqrt{2l+1}$. We can now expand $M(\theta,\varphi,\psi)$ in the $\Ylmn(\theta,\varphi,\psi)$ as follows:
\begin{equation} \label{MexpansionBerger}
M(\theta,\varphi,\psi)=\sum_{lmn}\sqrt{2l+1}\alpha_{lmn}\Ylmn(\theta,\varphi,\psi)=
\sum_{lmn}(2l+1)\alpha_{lmn}e^{-im\varphi}e^{-in\psi}\Plmn(\cos\theta)
\end{equation}
where the coefficients $\alpha_{lmn}$ are given by\footnote{The factor $b$ is $(-1)^{m-n}/(16\pi^2)$, as 
in section \ref{Roundexpansion}.} 
\begin{equation} \label{alphafirst}
\alpha_{lmn}=b\int_0^{\pi}\int_{0}^{2\pi}\int_0^{4\pi} M(\theta,\varphi,\psi) e^{im\varphi}e^{in\psi}
\Plmn(\cos\theta)\diff\psi\diff\varphi\sin\theta\diff\theta\;.
\end{equation}
We thus need to calculate a triple integral to find the $\alpha_{lmn}$. As in section \ref{Roundexpansion},
it will prove convenient not to take the limit $r\ra 1/k$ until the very end of the calculation. That means that 
we will actually replace $M(x,x')$ in (\ref{Mexpansionformal}) by $K(r,x,x')=K^{(\mu,k)}(r,x,x')$ from (\ref{bulktoboundary}) 
and then evaluate it on the boundary at the end of the calculation. 
Let us first do the integration over $\psi$. Converting to a contour integral by defining  $w=e^{i\psi/2}$, we find
\begin{equation}
\lim_{r\ra1/k}\int_0^{4\pi} K(r,\theta,\varphi,\psi) e^{in\psi} \diff\psi=\lim_{r\ra1/k}\frac{4\pi}{\sqrt{\Delta}}
\left(\frac{q_-}{2krc\cos(\theta/2) 
e^{i\varphi/2}}\right)^{2n}
\end{equation} 
where we recall from section \ref{Propagators} 
that $\Delta=(1-k^2r^2)^2-4k^2r^2(\vtwo-1)+4\mu^2k^2r^4\vtwo(\vtwo-1)$,  $q_-$ is given by 
\begin{equation}
q_{-}=(1+k^2r^2-\sqrt{{\Delta}}+2\mu k r^2\vtwo)
\left(1-\frac{\mu}{k}\sqrt{{\Delta}}-\mu^2r^2(2\vtwo-1)\right)^{\half\left(1-\frac{\mu}{k}\right)}\;,
\end{equation}
and 
\begin{equation}
c=\left[(1-\mu^2k^2r^4)\left(1-\mu^2/k^2\right)\right]^{\frac{1}{4}\left(1-\frac{\mu}{k}\right)}
\sqrt{\left(1+\mu/k\right)(1+\mu k r^2)}\;.
\end{equation}
Notice that with our choice of $\theta',\varphi',\psi'=0$, we have $\vtwo=\half(1+\cos\theta)$, so (defining $z=\cos\theta$) 
we get
\begin{equation}
\Delta=1+k^4r^4-2(k^2r^2)z-\frac{\mu^2}{k^2}(k^4r^4)(1-z^2)
\end{equation}
and
\begin{equation}
q_-=\left(1+k^2r^2-\sqrt{\Delta}
+\frac{\mu}{k} k^2r^2(1-z)\right)\left(1-\frac{\mu}{k}\sqrt{\Delta}-\frac{\mu^2}{k^2}k^2r^2z\right)^{\half(1-\frac{\mu}{k})}\;.
\end{equation}
Since $\mu,k$, and $r$ (which is now viewed just as a regularising parameter) appear only in the combinations 
$\mu/k$ and $k^2r^2$, it is convenient to rename them as $\mu/k\ra \mu$, and $k^2r^2\ra h$ (as in 
\ref{Roundexpansion}). After these cosmetic changes, we can write 
\begin{equation}
\alpha_{lmn}=4\pi b\lim_{h\ra1}\int_0^\pi\int_0^{2\pi}\frac{1}{\sqrt{\Delta}}
\left(\frac{q_-}{2c\sqrt{h}\sqrt{(1+z)/2}}\right)^{2n} e^{i(m-n)\varphi}\Plmn(z)\diff\varphi\diff z\\
\end{equation}
where now 
\begin{equation} \label{DeltaPlmn}
\Delta=1+h^2-2hz-\mu^2 h^2(1-z^2)\;,
\end{equation}
\begin{equation}
q_-=\left(1+h-\sqrt{\Delta}+\mu h(1-z)\right)\left(1-\mu\sqrt{\Delta}-\mu^2 hz\right)^{\half(1-\mu)}
\end{equation}
and
\begin{equation}
c=\left((1-\mu^2h^2)(1-\mu^2)\right)^{\frac{1}{4}(1-\mu)}\sqrt{(1+\mu)(1+\mu h)}\;.
\end{equation}
Note that the range of $\mu$ now is from 0 to 1. Noticing that $\varphi$ appears only in the exponential, it is trivial to 
do the integration over $\varphi$:
\begin{equation} \label{almnintegral}
\alpha_{lmn}=8\pi^2b\delta_{mn}\lim_{h\ra1}\int_{-1}^1\frac{1}{\sqrt{\Delta}}
\left(\frac{q_-}{2c\sqrt{h}\sqrt{(1+z)/2}}\right)^{2n}\Plnn(z)\diff z\;.
\end{equation}
We are now left with the nontrivial integral over $z$. Although there may be much more elegant ways 
of calculating
it, motivated by the resemblance of the integrand to the generating function for the $\Plnn(z)$ (\ref{expansionround}) 
we chose to expand in the parameter $h$ and write each factor in the expansion in terms of the $\Plnn(z)$, so that we can
use their orthonormality properties to perform the integration. This is the tactic we employed in section \ref{Roundexpansion}, 
but recall that there the integrand was \emph{exactly} the same as the generating function. Here things will turn out to
be somewhat more complicated.

\paragraph{A simple case: n=0}

Let us first look at the case where $n=0$, in other words we want to calculate the coefficient $\alpha_{l00}$ for 
any $l$, where of course now $l$ has to be integer (since $n$ is). If we set $n=0$, the integrand 
(apart from the $P^l_{nn}(z)$ factor and the $8\pi^2b$) is just (from (\ref{DeltaPlmn}))
\begin{equation}
I:=\frac{1}{\sqrt{1+h^2-2hz-\mu^2h^2(1-z^2)}}
\end{equation}
and we expand in $h$ as follows:
\begin{equation}
I=\sum_{s=0}^{\infty}h^s G_{s}(z)\;.
\end{equation}
To be concrete, let us write down the first few $G_s(z)$s: 
\begin{equation}
\begin{split}
G_0=1\; ;\; G_1=z\; ; \; G_2=-\half(1-\mu^2+\mu^2z^2-3z^2)\; ;\; G_3=-\frac{z}{2}(3-5z^2-3\mu^2+3\mu^2z^2)\;.
\end{split}
\end{equation} 
We would like to express the $G_s(z)$s in terms of eigenfunctions of the boundary laplacian, i.e. functions
$H_s(z)$ that satisfy (see appendix \ref{BergerSphere})
\begin{equation}
\Delta_B H_s(z)=s(s+1)H_s(z)
\end{equation}
(since $n=0$ for the moment, the term proportional to $n^2$ coming from the extra $\p_{\psi\psi}$ part 
in $\Delta_B$ will not play a role. That is, for now $\Delta_B$=$\Delta_0$.) Clearly the $H_s(z)$ will have
to be proportional to the Legendre functions $P_s(z)$, and we write $H_s(z)=N_s P_s(z)$, with
$N_s$ a normalisation factor to be determined. First we find linear combinations of the $G$s that \emph{are} 
eigenfunctions, and call these combinations $H_s(z)$: 
\begin{equation}
\begin{split}
H_0(z)&=G_0(z)\;;\;H_1(z)=G_1(z)\;;\;H_2(z)=G_2(z)-\frac13\mu^2G_0(z)\;;\;H_3(z)=G_3(z)-\frac35\mu^2G_1(z)\\
H_4(z)&=G_4(z)-\frac67\mu^2G_2(z)+\frac{3}{35}G_0(z)\;;\;H_5(z)=G_5(z)-\frac{10}{9}\mu^2G_3(z)+\frac{5}{21}\mu^4G_1(z)
\;;\;\ldots
\end{split}
\end{equation}
and so on. It is now easy to find expressions for the $G$s in terms of the $H$s at each (fixed, low) order in $h$ 
by inverting these relations. Here are the first few:
\begin{equation}
\begin{split}
G_0&=H_0\; ;\; G_1=H_1\; ;\; G_2=H_2+\frac13\mu^2 H_0\;;\; G_3=H_3+\frac35\mu^2H_1\\
G_4&=H_4+\frac67\mu^2H_2+\frac15 H_0\;;\;G_5=H_5+\frac{10}{9}\mu^2H_3+\frac37\mu^4H_1\;;\;\ldots \\
\end{split}
\end{equation} 
As discussed, the $H_s(z)$ are completely specified apart from their normalisations $N_s$. Let us
write the first few of those:
\begin{equation} \label{Nnzero}
\begin{split}
N_0=N_1=1\;;\;N_2=1-\frac13\mu^2\;;\;N_3=1-\frac35\mu^2\;;\;N_4=1-\frac67\mu^2+\frac{3}{35}\mu^4\;;\;
N_5=1-\frac{10}{9}\mu^2+\frac{5}{21}\mu^4\;;\; \ldots
\end{split}
\end{equation}
We now have to start summing up these series to end up with a closed form. By examining the terms that occur 
for higher $s$, we eventually find that the series for $N_s$ has a simple generating function:
\begin{equation}
N_s=\frac{s!\mu^sP_s(\frac{1}{\mu})}{(2s+1)!!}\;.
\end{equation}
As for the $G_s(z)$, it turns out that it is most convenient to treat the $s$ even and odd cases 
slightly differently in the beginning. We eventually find (this is probably the most nontrivial 
part of the paper from a calculational point of view, but as there is little insight to be gained
from the intermediate steps, we spare the reader the details.) 
\begin{equation}
G_{2r}(z)=\sum_{i=0}^{r}T_{ri}^{(e)}\mu^{2i} H_{2r-2i}(z)
\end{equation}
and
\begin{equation}
G_{2r-1}(z)=\sum_{i=0}^{r-1}T_{ri}^{(o)}\mu^{2i} H_{2r-2i-1}(z)
\end{equation}
where the coefficients $T_r^e$ and $T_r^o$ of the even and odd terms respectively are given by
\begin{equation}
T_{r}^{(e)}=\frac{1}{i!}\frac{(4r+1-4i)!!}{(4r+1-2i)!!}\frac{(2r-1)!!}{(2r-1-2i)!!}\frac{r!}{(r-i)!}
\end{equation}
and 
\begin{equation}
T_{r}^{(o)}=\frac{1}{i!}\frac{(4r-1-4i)!!}{(4r-1-2i)!!}\frac{(2r-1)!!}{(2r-1-2i)!!}\frac{r!}{(r-i-1)!}\;.
\end{equation}
So we have succeded in writing the integrand as an expansion in the Legendre functions:
\begin{equation}
I=\sum_{s=0}^\infty h^s\left(\sum_{i=0}^{r}T_{ri}^{(e)}\mu^{2i} N_{2r-2i} P_{2r-2i}(z)
+\sum_{i=0}^{r-1}T_{ri}^{(o)}\mu^{2i}N_{2r-2i-1} P_{2r-2i-1}(z)\right)\;.
\end{equation}
Now we can proceed by first doing the integral over $z$ in (\ref{almnintegral}). We get
\begin{equation} \label{bigexpression}
\begin{split}
\int_{-1}^1 I \cdot P_l(z)\diff z&=\sum_rh^{2r}\sum_{i=0}^{r} T_{ri}^{(e)}\mu^{2i}N_{2r-2i}\int_{-1}^{1}
P_{2r-2i}(z)P_l(z)\diff z+\\
&\quad\quad+\sum_r h^{2r-1}\sum_{i=0}^{r-1} T_{ri}^{(o)}\mu^{2i} N_{2r-2i-1} \int_{-1}^{1}
P_{2r-2i-1}(z)P_l(z)\diff z=\\
&=\sum_r h^{2r}\mu^{2r}\left(\frac{(2r-1)!!r!}{(2r+l+1)!!(r-l/2)!}\right)
\left(\frac{2\cdot l!P_l(1/\mu)}{(l/2)!(l-1)!!}\right)+\\
&\quad\quad+\sum_r h^{2r-1}\mu^{2r-1}\left(\frac{(2r-1)!!(r-1)!}{(2r+l)!!(r-(l+1)/2)!}\right)
\left(\frac{2\cdot l!P_l(1/\mu)}{((l-1)/2)!l!!}\right)\;.
\end{split}
\end{equation}
Here we first used the orthonormality property of the Legendre functions, 
\begin{equation}
\int_{-1}^1P_s(z)P_l(z)\diff z=\frac{2\delta_{sl}}{2l+1}
\end{equation}
which gave $2r-2i=l$ and $2r-2i-1=l$ in the first and second sums, respectively. Summing over the
delta functions then gives the final result. Clearly the first sum in (\ref{bigexpression}) will
be there only for $l$ even, and the second one only for $l$ odd, otherwise they should be thought of
as being zero. Now we have to put these two sums (\ref{bigexpression}) together again.  

Manipulating the series expansion, using relations such as $(2n)!!=2^n n!$ (for which we first need 
to convert the double factorials of odd numbers to ones of even numbers, through 
$(2n+1)!!=(2n+1)!/(2n)!!$) and $\Gamma(2n)=2^{2n-1}/\sqrt{\pi}\Gamma(n)\Gamma(n+1/2)$ (e.g.\cite{Lebedev}), 
and substituting $r=k+l/2$ for $l$ even, we find for the even $l$ series in (\ref{bigexpression}) the result
\begin{equation} \label{summed}
P_l\left(\frac{1}{\mu}\right)\frac{2}{\mu}
\left(\frac{\sqrt{\pi}\Gamma(l+1)\mu^{l+1}}{\Gamma(l+3/2)2^{l+1}}\right)
\sum_{k=0}^{\infty}\frac{\Gamma(k+l/2+1)\Gamma(k+l/2+1/2)\Gamma(l+3/2)}
{\Gamma(l/2+1)\Gamma(l/2+1/2)\Gamma(k+l+3/2)} \mu^{2k} h^{2k+l}\;.
\end{equation}
Notice that now the range of the summation variable $k$ is from zero to infinity. Similar manipulations
of the odd $l$ part of (\ref{bigexpression}), substituting this time $r=k+(l+1)/2$, give the \emph{same} result. 
This means that (\ref{summed}) is actually the full solution to our integral. What's more, the 
series over $k$ turns out to actually be a rather simple function: Noticing the similarity to the
hypergeometric generating function (which we have tried to exhibit by writing the series in this way) 
and recalling that the Legendre function of the \emph{second} kind can be expressed as (e.g \cite{Lebedev})
\begin{equation}
Q_l(z)=\frac{\sqrt{\pi}\Gamma(l+1)}{\Gamma(l+3/2)(2z)^{l+1}}\cdot
F\left(\frac{l}{2}+1,\frac{l}{2}+\half;l+\frac32;\frac{1}{z^2}\right)
\end{equation}
for $|z|>1$ (i.e. $\mu<1$ in which certainly is the case here), we have shown that  
\begin{equation} \label{Iresult}
\lim_{h\ra1}\int_{-1}^{1} \frac{1}{\sqrt{1+h^2-2h z-\mu^2h^2(1-z^2)}} P_l(z)\diff z=
\sum_l \frac{2}{\mu}P_l\left(\frac1\mu\right)Q_l\left(\frac1\mu\right).
\end{equation}
This final formula is refreshingly simple, so much so in fact that it cries out for a smarter 
(and less onerous) way to obtain it\footnote{There is one, of course: We can first take the limit
$h\ra 1$ in $I$ and  \emph{then} integrate it with a Legendre function, which can be done for
 particular values of $l$ since the $P_l(z)$s are polynomials. Then we can try to find a general formula 
for the result, which again leads us to (\ref{Iresult}).}. However recall that the purpose of doing things 
the way we did was to prepare us for the more demanding calculation for $n\neq 0$. As we will see, much of the work
for that case has already been done. 
 
To summarise, so far we have found one coefficient in the expansion (\ref{Mexpansionformal}) of $M(x,x')$ in spherical harmonics
(\ref{MexpansionBerger}) for each (integer) $l$:
\begin{equation} \label{alzerozero}
\alpha_{l00}=8\pi^2b\frac2\mu P_l\left(\frac1\mu\right) Q_l\left(\frac1\mu\right)\;.
\end{equation}

\paragraph{The general case}

Now let $n$ be arbitrary. Let us again expand the integrand of (\ref{almnintegral}) as a series in $h$:
\begin{equation}
I=\frac{1}{\sqrt{\Delta}}\left(\frac{q_-}{2c\sqrt{h}\sqrt{(1+z)/2}}\;\right)^{2n}=\sum_{s=0}^{\infty} h^s G^{s+n}_{nn}(z)\;.
\end{equation}
The reason for the notation $G^{s+n}_{nn}(z)$ is that, as we will soon see,  
these functions are related in a simple way to the Wigner functions $P^{s+n}_{nn}(z)$.
The first few terms in the series are\footnote{Note that this expansion holds for both integer and half--integer values
of $n$.} 
\begin{equation}
\begin{split}
&s=0:\; G^{n}_{nn}(z)=[(1+\mu)^{1+\mu}(1-\mu)^{1-\mu}]^{\frac{n}{2}}
\cdot\left(\frac{1+z}{2}\right)^{n}\\
&s=1:\; G^{n+1}_{nn}(z)=-(n\mu^2z-nz-z+n)\cdot G^{n}_{nn}(z)\\
&s=2:\; G^{n+2}_{nn}(z)=\frac14(2\mu^4z^2n^2-2\mu^2z^2-7n\mu^2z^2-4n^2\mu^2z^2+6z^2+7nz^2
+2n^2z^2+ \\
\quad\quad &\quad\quad\quad\quad\quad\quad\quad+4n^2\mu^2z+2n\mu^2z-6zn-4zn^2+3n\mu^2+2\mu^2
-2+2n^2-n)\cdot G^{n}_{nn}(z)\;.\\ 
\end{split}
\end{equation}
The expressions rapidly become awkward, so we have only written out the first three terms.  
We can easily verify that $G^{n}_{nn}(z)$ satisfies the $\SU(2)$ differential equation, with
$m=n$, and eigenvalue $-n(n+1)$\footnote{In (\ref{SUtwoplus}) we have added the extra factor of 1/4 which makes the 
eigenvalues complete squares.}:
\begin{equation} \label{SUtwoplus}
\left[(1-z^2)\p_{zz}-2z\p_z-\frac{2n^2}{1+z}-\frac{1}{4}\right] G^{n}_{nn}(z)=
-\frac{(2n+1)^2}{4} G^{n}_{nn}(z)\;.
\end{equation}
So $G^{n}_{nn}(z)$ has to be proportional to $P^{n}_{nn}(z)$. On the other hand, we find that $G^{n+1}_{nn}(z)$
 is \emph{not} an eigenfunction of (\ref{SUtwo}) with $m=n$. As in the previous calculation for $n=0$, we 
should consider linear combinations of the $G$'s that \emph{are} eigenfunctions. In this case we define
\begin{equation}
H^{n+1}_{nn}(z)=G^{n+1}_{nn}(z)+\frac{\mu^2n^2}{n+1} G^{n}_{nn}(z)
\end{equation}
which turns out to be an eigenfunction, with eigenvalue $-(n+1)(n+2)$:
\begin{equation}
\left[(1-z^2)\p_{zz}-2z\p_z-\frac{2n^2}{1+z}-\frac{1}{4}\right] H^{n+1}_{nn}(z)=
-\frac{(2(n+1)+1)^2}{4} H^{n+1}_{nn}(z)\;.
\end{equation}
Proceeding in this way, we can find linear combinations of the $G^{n+l}_{nn}(z)$ 
that are (unnormalised) eigenfunctions of the $\SU(2)$ differential equation, with 
eigenvalues $(n+l)(n+l+1)$, and thus have to be proportional to the $P^{n+l}_{nn}(z)$.
Let us write down the first few:
\begin{equation} 
\begin{split}
&H^{n}_{nn}(z)=G^{n}_{nn}(z)\\
&H^{n+1}_{nn}(z)=G^{n+1}_{nn}(z)+\frac{\mu^2n^2}{n+1} G^{n}_{nn}(z)\\
&H^{n+2}_{nn}(z)=G^{n+2}_{nn}(z)+\frac{\mu^2n^2}{n+2} G^{n+1}_{nn}(z)
-\frac{4-2n^4\mu^2+2n^3+7n^2+n^3\mu^2+2n^2\mu^2+8n}{2(n+2)(n+3)} G^{n}_{nn}(z)\;.\\
\end{split}
\end{equation}
The relations rapidly become unwieldy as $l$ increases, so we will not write 
any more of them out explicitly. In any case, what we are more interested in is the 
\emph{inverse} relation: How to write the $G$'s in terms of the $H$'s. Let us
again write down only a few terms:
\begin{equation} \label{GH}
\begin{split}
&G^{n}_{nn}(z)=H^n_{nn}(z)\\
&G^{n+1}_{nn}(z)=H^{n+1}_{nn}(z)-\frac{\mu^2n^2}{n+1} H^n_{nn}(z)\\
&G^{n+2}_{nn}(z)=H^{n+2}_{nn}(z)-\frac{\mu^2n^2}{n+2} H^{n+1}_{nn}(z)
+\mu^2\frac{2n^4\mu^2+2n^3+n^3\mu^2+n^2\mu^2+5n^2+5n+2}{2(2n+3)(n+1)} H^n_{nn}(z)\\
&G^{n+3}_{nn}(z)=H^{n+3}_{nn}(z)-\frac{\mu^2n^2}{n+3}H^{n+2}_{nn}(z)
+\mu^2\frac{12+16n+9n^2+2n^3+\mu^2(2n^2+n^3+2n^4)}{2(n+2)(2n+5)}H^{n+1}_{nn}(z)\\
&\qquad\qquad\quad-\frac{\mu^4n^2(26+42n+25n^2+4\mu^2n^2+6n^3+3\mu^2n^3+2\mu^2n^4)}{6(n+1)(n+2)(2n+3)}H^{n}_{nn}(z)\;.
\end{split}
\end{equation}
We have thus managed to write the first few terms in the expansion 
in terms of the $H^{n+s}_{nn}(z)$, which are just unnormalised versions of
the $P^{n+s}_{nn}(z)$. The normalisation factors can be obtained by dividing them, $N_{snn}$=$H^{n+s}_{nn}(z)/P^{n+s}_{nn}(z)$ 
(where we can find the $P^{n+s}_{nn}(z)$ either by looking at tables, or by expanding the generating function
(\ref{Plmngenerating}) for $m=n$). We have 
\begin{equation}
\begin{split}
N_{0nn}&=\left((1+\mu)^{\half(1+\mu)}(1-\mu)^{\half(1-\mu)}\right)^n\\
N_{1nn}&=\frac{1+n-\mu^2n}{n+1}\cdot N_{0nn}\\
N_{2nn}&=\frac{6+7n+2n^2-2\mu^2-7\mu^2n-4\mu^2n^2+2\mu^4n^2}{(n+2)(2n+3)}\cdot N_{0nn}\\
N_{3nn}&=\frac{30+37n+15n^2+2n^3-\mu^2(18+45n+30n^2+6n^3)+\mu^4(8n+15n^2+6n^3)-2\mu^6n^3}{(n+2)(n+3)(2n+5)}
\cdot N_{0nn}.
\end{split}
\end{equation}

Let us now write the integral (\ref{almnintegral}) as
\begin{equation} \label{inttry}
\int_{-1}^{1}I\cdot \Plmn(z)\diff z=\sum_{s} h^s\int_{-1}^1 G^{n+s}_{nn}(z)\Plmn(z)\diff z
=\sum_{s} h^s N_{lnn}\sum_{i=0}^{s} \alpha^{sl}(\mu) \int H^{n+s-i}_{nn}(z) \Plmn(z)\diff z\;,
\end{equation}
where $\alpha^{sl}(\mu)$ symbolises the various coefficients in (\ref{GH}). This integral receives contributions from all $s>l-n$, 
since then for $i=n+s-l$ we have a nonzero contribution. 
Now notice in (\ref{GH})  that the terms to the right (higher $i$ in (\ref{inttry})) have higher powers of $\mu^2$, and that
the power increases by $\mu^2$ every two terms or so (this is really only evident in $G^{n+3}_{nn}(z)$, but can be checked 
to hold for the higher terms). So, truncating the sum in $s$ should give us the answer to some order in $\mu$.
For example, we can check the case $l=n+1$, to order $\mu^4$:\footnote{Taking also $h\ra1$, as in all expressions that follow.}
\begin{equation} \label{intexample}
\begin{split}
\int_{-1}^{1} I \cdot &P^{n+1}_{\;nn}(z)\diff z=\int_{-1}^{1}\left(G^{n+1}_{nn}(z)+G^{n+2}_{nn}(z)+G^{n+3}_{nn}(z)+
\cdots\right)P^{n+1}_{nn}(z)\diff z\\
&=N_{1nn}\frac{2}{2(n+1)+1}\left(1-\frac{\mu^2 n^2}{n+2}+\mu^2\frac{12+16n+9n^2+2n^3+\mu^2(2n^2+n^3+2n^4)}{2(n+2)(2n+5)}+\cdots\right) \\
&=\frac{2}{2n+3}+\frac{2(2n^3+8n^2+9n+6)}{(2n+5)(2n+3)(n+2)(n+1)}\mu^2+O(\mu^4)\;.
\end{split}
\end{equation}
So this is a contribution to $\alpha_{lnn}$ when $l=n+1$. 
It can be easily checked that it reduces to the $\alpha_{100}$ above when $n=0$, as is the case for all the previous formulas.
Now the question is whether we can extend these relations to arbitrary $l$, 
i.e. write the $\alpha_{lnn}$ in closed form. This is where the special
case we considered earlier comes in especially handy. Recall that we found
that $\alpha_{l00}=2/\mu P_l(1/\mu)Q_l(1/\mu)$. Clearly the formula for 
$n\neq 0$ must be a generalisation of this. We can cheat a bit by looking at the expressions in section \ref{QTNfactorisation},
where, as we saw, there appeared some hypergeometric functions that reduced to Legendre functions in a special case. 
Motivated by this, we can make a guess as to how to generalise $P_l(1/\mu)$, $Q_l(1/\mu)$. The guess is (see a previous footnote):
\begin{equation}
\begin{split}
P_l\left(\frac{1}{\mu}\right)&\ra \left(\frac{1+\mu}{2\mu}\right)^{l+\mu n}
F\left(-l+n, -l-\mu n;1+\left(1-\mu\right);
\frac{1-\mu}{1+\mu}\right)\\
Q_l\left(\frac{1}{\mu}\right)&\ra \left(\frac{2\mu}{1+\mu}\right)^{1+l-\mu n}F\left(1+l+n,1+l-\mu n;2l+2;\frac{2\mu}{1+\mu}\right)\;.
\end{split}
\end{equation}
Trying this out, and fixing coefficients to match with the lower order terms for the $\alpha_{lmn}$ (that we can calculate
as in (\ref{intexample})) we end up with:
\begin{equation} \label{alnnfinal}
\begin{split}
\alpha_{lmn}=&\half\delta_{mn}\frac{2}{n(1-\mu^2)}\left(\frac{1-\mu}{1+\mu}\right)^{(1-\mu)n}
\frac{\Gamma(1+l+n)\Gamma(1+l-\mu n)}{\Gamma(2l+2)\Gamma((1-\mu) n)}\times\\
&\quad\times F\left(-l+n, -l-\mu n; 1+n-\mu n;\frac{1-\mu}{1+\mu}\right)
\times F\left(1+l+n,1+l-\mu n;2l+2; \frac{2\mu}{1+\mu}\right)
\end{split}
\end{equation}
where we have also substituted the factor $b$ from (\ref{almnintegral}). 
This proposal can be checked to give the correct answer for the coefficients $\alpha_{lmn}$ that we can calculate
as in (\ref{intexample}) by expanding in $\mu$. Note that despite the explicit appearance of $n$ to the first power, 
its expansion in $\mu$ contains only powers of $n^2$, as it should by looking at e.g. (\ref{GH}). 
Although one could probably perform further checks of this result, we take the matching with the low--order terms
as sufficient evidence to take it as our final answer\footnote{To check that we recover (\ref{alzerozero}) for $n=0$, note 
that $n\Gamma((1-\mu)n)\ra1/(1-\mu)$ for $n\ra0$.}.

We can now proceed (as in (\ref{Roundexpansion})) to restore the angles $\theta',\varphi',\psi'$ that we had set to 
zero, and conclude that the expansion of $M(x,x')$ is given by (\ref{Mexpansionformal}), with the $\alpha_{lmn}$ 
given by (\ref{alnnfinal}) without the $\delta_{mn}$.

\subsection{Checking the correlation function} \label{QTNcheck}

In section \ref{QTNfactorisation} we found two different solutions for the Dirichlet--to--Robin operator, one that should
hold ``inside''  the manifold (i.e. for $r<1/k$) and another that should be used ``outside''. We also argued that 
the correct operator to consider \emph{on} the boundary $r=1/k$ is the sum of these two:
\begin{equation} \label{Acaltotal}
\Acal=\half(\Acal^i+\Acal^e)\;.
\end{equation}
Here we finally check the results of \ref{QTNfactorisation} against the eigenfunction expansion for $M(x,x')$ that we 
obtained in the previous section. First we focus on the special case where $\Acal$ is acting on a
function that does not depend on the angle $\psi$. That means we can effectively set $\sqrt{\Done}=0$ everywhere, and we
obtain (see the end of section \ref{QTNfactorisation})
\begin{equation}
\Acal^i_0=\frac{(1-k^2r^2)}{2}\sqrt{\frac{1-\mu^2 k^2 r^4}{1-\mu^2 r^2}}\p_r \ln\left[f_0(r)
Q_{\sqrt{\Dzero+1/4}-\half}\left(\frac{1}{\mu k r^2}\right)\right]
\end{equation}
and
\begin{equation}
\Acal^e_0=-\frac{(1-k^2r^2)}{2}\sqrt{\frac{1-\mu^2 k^2 r^4}{1-\mu^2 r^2}}\p_r \ln\left[f_0(r)
P_{\sqrt{\Dzero+1/4}-\half}\left(\frac{1}{\mu k r^2}\right)\right]
\end{equation}
where $f_0=(1-\mu^2 r^2)^{\frac{1}{6}}(1-\mu^2 k^2 r^4)^{\frac16}/r$. Notice that although these are supposed to be operators
on the boundary, where $r=1/k$, we have kept $r$ free since it will prove useful in the calculations below\footnote{This is just
a formal trick, and could easily be dispensed with.}. From (\ref{Acaltotal}) we obtain
\begin{equation} 
\Acal=\frac{(1-k^2r^2)}{4}\sqrt{\frac{1-\mu^2 k^2 r^4}{1-\mu^2 r^2}}
\p_r\ln\left(\frac{Q_{\sqrt{\Dzero+1/4}-\half}\left(\frac{1}{\mu k r^2}\right)}{P_{\sqrt{\Dzero+1/4}-\half}\left(\frac{1}{\mu k r^2}\right)}\right)\;.
\end{equation}
From this expression, using the property $\p_z\ln(x/y)=(y\p_z x -x\p_z y)/(xy)$, and the identity\footnote{This is of
course simply the Wronskian of the two linearly independent Legendre functions, given e.g. in \cite{Lebedev}.} 
\begin{equation} \label{LegendreID}
P_{l}\left(z\right)\p_zQ_{l}\left(z\right)
-Q_{l}\left(z\right)
\p_zP_l\left(z\right)=\frac{1}{1-z^2}\;,
\end{equation}
we finally obtain 
\begin{equation}
\Acal=\frac{(1-k^2r^2)}{(1-\mu^2 r^2)}\frac{\mu k r^2}{2}\frac{1}
{P_{\sqrt{\Dzero+1/4}-\half}\left(\frac{1}{\mu k r^2}\right)Q_{\sqrt{\Dzero+1/4}-\half}\left(\frac{1}{\mu k r^2}\right)}\;.
\end{equation}
Now we can take $r\ra 1/k$, with the usual defining function
\begin{equation} \label{Bcalzero}
\Bcal^{(n=0)}=\lim_{r\ra1/k}\frac{1}{(1-k^2r^2)}\Acal=\frac{\mu}{2k(1-\mu^2/k^2)}\frac{1}
{P_{\sqrt{\Dzero+1/4}-\half}\left(\frac{k}{\mu}\right)Q_{\sqrt{\Dzero+1/4}-\half}\left(\frac{k}{\mu}\right)}\;.
\end{equation}
Recall that $\Bcal^{(n=0)}$ only makes sense when acting on functions on $\Srm^3$ that do not depend on $\psi$. Looking
at the expansion (\ref{MexpansionBerger}) of $M(x,x')$, we see that this is the case when $n=0$. So $\Bcal^{(n=0)}$ can 
only be allowed to act on the terms that have coefficients $\alpha_{l00}$. We get\footnote{We could have written here
$\alpha_{lm0}$, but since they are all equal to $\alpha_{l00}$ we just write the latter.}
\begin{equation}
\Bcal_x^{(n=0)}M^{(\psi=0)}(x,x')=\Bcal_x^{(n=0)}\sum_{lm}\alpha_{l00} \mathrm{Y}^l_{m0}(x)\overline{\mathrm{Y}}^l_{m0}(x')
=\sum_{lm}\alpha_{l00} (\Bcal_x^{(n=0)}\mathrm{Y}^l_{m0}(x))\overline{\mathrm{Y}}^l_{m0}(x')\;.
\end{equation} 
From  (\ref{alzerozero}), restoring $\mu\ra\mu/k$, we see that $\Bcal_x\Ylmn(x)=(1-\mu^2/k^2)/\alpha_{l00}\Ylmn(x)$, and it immediately follows 
that $\Bcal$ is indeed the inverse of $M(x,x')$ on this
restricted space\footnote{The extra factor $(1-\mu^2/k^2)$ clearly has to do with the normalisation of $M(x,x')$ which we have
left arbitrary.}. Before we proceed to the general case, let us perform a consistency check: If $\Bcal^{(n=0)}$ really
is (part of) the Dirichlet--to--Robin operator, it has to be first order, and in particular it should have the square
root of the laplacian as its principal symbol. One (perhaps not very rigorous) way to check this is to examine the limit
of (\ref{Bcalzero}) as $\Delta_0$ becomes large. This is easy to do, using the formulas \cite{Lebedev} for the 
asymptotic expansion of the Legendre functions for $l$ large:
\begin{equation}
P_l(\cosh\alpha)=\frac{e^{(l+1/2)\alpha}}{(2l\pi\sinh\alpha)^{\half}}\left[1+O\left(\frac{1}{l}\right)\right], \quad
Q_l(\cosh\alpha)=\frac{\pi^{\half}e^{-(l+1/2)\alpha}}{(2l\sinh\alpha)^{\half}}\left[1+O\left(\frac{1}{l}\right)\right]\;.
\end{equation}
So clearly, for $l$ large
\begin{equation}
\frac{2k}{\mu}\frac{1}{P_l(k/\mu)Q_l(k/\mu)}= k l\left[1+O\left(\frac{1}{l}\right)\right]
\end{equation}
where we used $\cosh\alpha=k/\mu$, and so $\sinh\alpha=1/\mu\sqrt{1-\mu^2/k^2}$. Thus (noting that $\Delta_0\sim l^2$ for $l$ large) 
we can conclude that $\Bcal^{(n=0)}$ does indeed scale as 
\begin{equation} \label{Basymptotic}
\Bcal^{(n=0)}=k\sqrt{\Dzero}\left[1+O\left(\frac{1}{\sqrt{\Dzero}}\right)\right]
\end{equation}
and thus has the right principal symbol to be the Dirichlet--to--Robin operator.

After this warm--up, we now turn to the $n\neq0$ case. Again we have the sum (\ref{Acaltotal}) of
the interior and exterior Dirichlet--to--Robin operators, which are  given in (\ref{Aint}) and (\ref{Aext}). Adding them gives:
\begin{equation}
\begin{split}
\Acal&=\half(\Acal^{i}+\Acal^{e})=\frac{(1-k^2r^2)}{4}\sqrt{\frac{1-\mu^2k^2r^4}{1-\mu^2r^2}}\p_r\ln
\frac{\left(\frac{2\mu k r^2}{1+\mu k r^2}\right)^{\sqrt{\Dzero+1/4}}F_1(r)}
{\left(\frac{2\mu k r^2}{1+\mu k r^2}\right)^{-\sqrt{\Dzero+1/4}}F_3(r)}\\
&=\frac{(1-k^2r^2)}{4}\sqrt{\frac{1-\mu^2k^2r^4}{1-\mu^2r^2}}\frac{1}{F_1(r)F_3(r)}\times\\
&\left[u^{\sqrt{\Dzero+1/4}} \left\{F_3(r)\p_r\left(u^{-\sqrt{\Dzero+1/4}}F_1(r)\right)-u^{-\sqrt{\Dzero+1/4}}F_1(r)
\p_r\left(u^{\sqrt{\Dzero+1/4}}F_3(r)\right)\right\}\right]\;.
\end{split}
\end{equation}
Recall here that we have defined $u=(1+\mu k r^2)/(2\mu k r^2)$. To proceed, we use the identity\footnote{Which we can verify
by inserting specific values of $l$ and $n$.}
\begin{equation} \label{hypergeomID}
\begin{split}
&u^{\sqrt{\Dzero+1/4}}F_3(r)\p_r\left(u^{-\sqrt{\Dzero+1/4}}F_1(r)\right)-u^{-\sqrt{\Dzero+1/4}}F_1(r)
\p_r\left(u^{\sqrt{\Dzero+1/4}}F_3(r)\right)=\\
&\quad \quad \frac{2n(1-\mu/k)}{r(1-\mu k r^2)}
\left(\frac{1+\mu k r^2}{1-\mu k r^2}\right)^{\left(1-\frac{\mu}{k}\right)n}
\frac{\Gamma(2l+2)\Gamma((1-\mu/k)n)}{\Gamma(1+l+n)\Gamma(1+l-\mu/k\cdot n)}
\end{split}
\end{equation}
which generalises (\ref{LegendreID}). Calling $T$ the right--hand side of (\ref{hypergeomID}), we write
\begin{equation}
\Acal=\frac{(1-k^2r^2)}{4}\sqrt{\frac{1-\mu^2k^2r^4}{1-\mu^2r^2}}\frac{T}{F_1(r)F_3(r)}\;.
\end{equation}
Now we finally take the limit $r\ra 1/k$, scaling $\Acal$ as in (\ref{Bcalzero}) to obtain the complete Dirichlet--to--Robin 
operator for QTN:
\small
\begin{equation} \label{Bcalfinal}
\begin{split}
\Bcal&=\frac{\sqrt{\Done}}{2}\left(\frac{1+\mu/k}{1-\mu/ k}\right)^{\left(1-\frac{\mu}{k}\right)\sqrt{\Done}}
\frac{\Gamma(1+2\sqrt{\Dzero+1/4})\Gamma((1-\mu/k)\sqrt{\Done})}
{\Gamma(\half+\sqrt{\Dzero+1/4}+\sqrt{\Done})\Gamma(-\half+\sqrt{\Dzero+1/4}-\mu/k\cdot \sqrt{\Done})}\times \\
&\left[F\left(\half-\sqrt{\Dzero+1/4}+\sqrt{\Done},\half-\sqrt{\Dzero+1/4}-\frac{\mu}{k}\sqrt{\Done};1+(1-\frac{\mu}{k})
\sqrt{\Done};\frac{1-\mu/k}{1+\mu/k}\right) \right.\times\\ 
&\left.\quad F\left(\half+\sqrt{\Dzero+1/4}+\sqrt{\Done},\half+\sqrt{\Dzero+1/4}-\frac{\mu}{k}\sqrt{\Done};1+2\sqrt{\Dzero+1/4};
\frac{2\mu/k}{1+\mu/k}\right)\right]^{-1}\;.
\end{split}
\end{equation}
\normalsize
At this point we really should repeat the check (that we previously performed for $\Bcal^{(n=0)}$) 
of whether the principal symbol of $\Bcal$ given
by (\ref{Bcalfinal}) really is the square root of the boundary laplacian $\Delta_B$ (\ref{DeltaB}). Clearly in the
case $\Dzero\gg\Done$ (\ref{Bcalfinal}) reduces to (\ref{Bcalzero}) so we can apply the previous result,
 but taking other scaling limits, like for instance both $\Delta_0$ and $\Delta_1$ large, seems a rather daunting task. 
Perhaps it is possible to use more formal methods to prove that $\Bcal$ is first--order, but for the moment we  
content ourselves with the simple consistency check we just discussed, and proceed to the final step. 

The expression for $\Bcal$ in (\ref{Bcalfinal}) should look familiar. If we act with $\Bcal$ on an eigenfunction 
$\Ylmn(x)$ on the Berger sphere, which really means substituting $\Dzero=l(l+1)$ and $\Done=n^2$ everywhere, we get 
precisely\footnote{Up to a $(1-\mu^2/k^2)$ term and other possible normalisation factors that do not depend on $l$ and $n$.} 
\begin{equation}
\Bcal_x \Ylmn(x)=\frac{1}{\alpha_{lnn}}\Ylmn(x)
\end{equation}
with $\alpha_{lnn}$ as defined in (\ref{alnnfinal}). So we finally have all the necessary ingredients to  show that
\begin{equation}
\Bcal_x M(x,x')=\sum_{lmn}\Bcal_x \alpha_{lnn}\Ylmn(x)\overline{\mathrm{Y}}^l_{mn}(x')
=\sum_{lmn}\Ylmn(x)\overline{\mathrm{Y}}^l_{mn}(x')=\frac{1}{\sqrt{g}}\delta^{(3)}(x,x')
\end{equation}
which was the main calculational objective of this article. 

Given the discussion in section \ref{DirichlettoNeumann}, we conclude that $M(x,x')$ (as defined in (\ref{correlator})) is
indeed a conformally invariant correlation function on the Berger sphere, corresponding to an operator of conformal 
dimension one.

\section{Conclusions} \label{Conclusions}

In this article we demonstrated that our correlation function $M(x,x')$ (see (\ref{correlator})) on the Berger
sphere is conformally invariant, and is a two--point function of an operator of conformal dimension $\Dim=1$. 
 We did this by explicitly constructing a conformally covariant pseudodifferential operator, the Dirichlet--to--Robin 
operator $\Bcal$, which we 
expected to be the inverse of $M(x,x')$ on general grounds, and then verifying that this was indeed the case. Note
that the methods that led to these two objects were (at least superficially) very different from each other:
$M(x,x')$ was found using twistor methods \cite{Zoubos02}, namely Penrose's nonlinear graviton construction, 
while $\Bcal$ was found by factorising the QTN laplacian, which in the end came down to solving a Ricatti--type equation. Thus
we believe we have performed an independent verification of the conformal invariance of our correlation function. On the other
hand, as discussed in section \ref{Definition}, to obtain this result we have made a number of  assumptions and generalisations 
which should eventually be better justified. We believe that this  will probably require some input from geometric scattering 
theory.

We should emphasise here that we are not claiming  \emph{uniqueness} of the correlation function. There could very
well be other two--point functions on the Berger sphere that correspond to $\Dim=1$ scalar operators. In fact, the 
very method we used to prove conformal invariance of $M(x,x')$ also seems to hint that it is not unique! Recall 
\cite{Chamblinetal99,Hawkingetal99} that there are parameter ranges where, apart from QTN, the Berger sphere is the 
conformal infinity of another 
asymptotically hyperbolic 4--manifold, known as AdS--Taub--Bolt. Now, as discussed in (\ref{InverseProblems})  the main feature 
of the Dirichlet--to--Robin operator is that it is different for non--diffeomorphic manifolds. So, although the  calculation
of the corresponding $M(x,x')$ for AdS--Taub--Bolt seems out of reach at the moment, it is likely that it will have 
a different functional form from the one for QTN we have been investigating (although both will have the same conformally
flat limit, since $M(x,x')$ is unique there). It is intriguing to wonder whether that is all, or whether there are 
several other  conformally invariant correlation functions on the Berger sphere. 

Although we focused narrowly on the very particular problem of $\Dim=1$ and the Berger sphere, 
we believe that the techniques we employed should be applicable to a wider range of questions that might arise in applications 
of holographic ideas to asymptotically hyperbolic bulk manifolds. A drawback is that calculability in our case hinged
on the existence and relative simplicity of an eigenfunction expansion, which is beyond reach for many manifolds. But
perhaps the formal aspects of the Dirichlet--to--Robin operator (and related nonlocal boundary operators) viewed as 
\emph{conformally invariant} operators will eventually turn out to be useful in better understanding conformal field theory, 
especially on curved manifolds. Of course, as the dicussion in section \ref{InverseProblems}
tries to make clear, the Dirichlet--to--Robin operator should also play an important role in exploring the
connections between holography in an AdS/CFT context and the mathematical fields of geometric scattering and inverse 
boundary problems. 

One could conceive looking for the nonlocal boundary conformally invariant operators corresponding to correlation functions
of higher conformal dimension. Branson and Gover \cite{BransonGover01} construct higher order, and also \emph{negative}
order, generalisations of the Dirichlet--to--Robin operator, which thus have principal parts $(\Delta_B)^{\frac{s}{2}}$ 
(in \cite{Peterson00} this is extended further to arbitrary real order). (We put emphasis on the negative order
cases, because by the relation $2\Dim=d-s$ (\ref{conformaldimension}), the order of the pseudodifferential operators 
corresponding to  CFT operators with the usual conformal dimensions one encounters (i.e. $\Dim\geq d/2$) will be negative.)
They do this by considering conformally invariant boundary value problems for the (bulk) higher order conformally invariant
powers of the laplacian we briefly discussed in section \ref{confdiff} (of which the  Paneitz operator (\ref{Paneitz}) is 
an example). There are restrictions to doing this on an arbitrary curved conformal manifold (related to the restrictions
in defining the powers of the laplacian), but supposing that one \emph{can} construct conformally invariant boundary problems 
in the bulk, which result in boundary pseudodifferential operators associated to a given correlation function, one could ask how 
this (conformal, nonlocal if the bulk operators are negative order) description of the bulk degrees of freedom differs 
from the usual (non--conformal, local) one based on the massive bulk laplacian. 

 After these comments, let us  return to the Berger sphere. 
In the end, are we any closer to understanding the nature of the field theory that might be dual
to gravity on QTN? In an important sense we are, since we have demonstrated, in an explicit (but very special) 
example, that the usual AdS/CFT dictionary between bulk fields and boundary operators holds 
unmodified in our more general case. However, a lot more needs to be done to fully understand how
holography works for QTN. The best one could wish for, of course, is a string/M theory realisation, which should give
a better picture of the gauge theory side (of course, given the apparent lack of supersymmetry, such a configuration 
would be interesting on its own, were it to be stable). One could also try to focus on the region close to $\mu=k$ (where the 
manifold is almost $\Bergman$) which we have been largely ignoring in this article: Despite the 
discussions in \cite{Britto--Pacumioetal99,Taylor--Robinson00}, holography in the CR case is still a bit mysterious and 
deserves to be better studied.

It will also be important to obtain correlation functions of higher $\Dim$, since they might
be able to give us insight into how the conformal field theory makes the transition from the ``stable'' region 
($\mu^2<3/4k^2$) where the boundary scalar curvature is positive, to the (presumably) unstable region ($\mu^2>3/4k^2$) 
where it becomes negative. Recall \cite{Zoubos02} that one of the main motivations for studying this 
model is understanding this instability (implied by the results of \cite{SeibergWitten99,WittenYau99}), 
and how/whether it is reflected in the geometry of the bulk. As an indication of the issues that need to 
be addressed, we mention a comment in \cite{Lee95} that, as the boundary scalar curvature shifts to 
negative values, numerically it seems that the bulk laplacian for Quaternionic Taub--NUT begins, as expected perhaps, 
to develop normalisable modes 
above the Breitenlohner--Freedman bound of $-9/4$. It is plausible that these modes could render the usual AdS/CFT correspondence 
ambiguous. What is \emph{more} interesting however, is that there seems to be a range in which the boundary scalar curvature is 
negative, but \emph{no} such modes have yet appeared in the bulk spectrum. Resolving these and other issues connected
to holography on QTN promises to be a worthwhile endeavor.

\section*{Acknowledgments}
I wish to thank M. Anderson and M. Ro\v{c}ek for useful discussions. I am especially thankful to M. Kulaxizi
for very fruitful discussions and help with some of the calculations. I am also 
grateful to the Physics Department at Stony Brook University for financial support. 

\appendix

\section{More on the Berger sphere} \label{BergerSphere}
In this appendix we include a few details on the homogeneously squashed three--sphere, or Berger sphere, which we 
sometimes denote by $\Squashedsphere$. 
Its metric is simply 

\begin{equation} \label{SquashedMetric}
g_{ij}\;:\quad \diff s^2=4\left[a^2(r)(\sigma_1^2+\sigma_2^2)+b(r)^2\sigma_3^2\right]
\end{equation}

where the sigmas are the usual left--invariant $\SU(2)$ one--forms:
\begin{equation}
\begin{split}
\sone&=\half (\cos\psi\diff\theta+\sin\psi\sin\theta\diff\varphi)\;,\\
\stwo&=\half (-\sin\psi\diff\theta+\cos\psi\sin\theta\diff\varphi)\;,\\
\sthree&=\half(\diff\psi+\cos\theta\diff\varphi)\;.
\end{split}
\end{equation}
normalised to satisfy $\diff \sone=-2\stwo\wedge \sthree$, and cyclic permutations thereof. The 
ranges of $\theta,\varphi$ and $\psi$ are $[0,\pi), [0,2\pi)$ and $[0,4\pi)$ respectively.

The squashing is given by the ratio $\grl(r)=b(r)^2/a(r)^2$. $\grl(r)=1$ corresponds to the round
three--sphere, $0<\grl<1$ is known as oblate squashing, while $\grl>1$ is called prolate squashing. For
definiteness we only discuss the prolate case in this article, but our results should hold for the oblate
case too, by taking $\mu\ra -i\mu$ everywhere \cite{Zoubos02}.  

 Any constant--$r$ hypersurface in QTN defines a squashed three--sphere, but the squashing depends on
the radial distance. Also, the overall scale of the metric is important in taking various limits. 
So we will mostly keep $a(r)$ and $b(r)$ arbitrary, and write down special cases at the end. 

The squashed sphere was famously considered by Hitchin \cite{Hitchin73} in his discussion of the
space of harmonic spinors (the null space of the Dirac operator) on a manifold:
$\Squashedsphere$ is a good illustration of the fact that the number of harmonic spinors is 
not a topological invariant of the manifold, but depends on the particular metric too.  On the physics 
side, 
scalar quantum field theory on the homogeneously squashed sphere has been considered by several authors 
\cite{Hu73,Huetal73,CritchleyDowker81}, mainly due to its appearance as (a particular case of) 
the spatial section of the mixmaster cosmological 
model \cite{Misner69}. Rather remarkably, the $\SU(2)\times\Urm(1)$ symmetry means that the problem
can be mapped to that of the symmetric quantum--mechanical rotor (see e.g. \cite{Wigner}, chapter 19, or 
\cite{LandauLifshitz}, sections 82 and 103) which 
(like that of the spherical rotor) is integrable. (As we will see, this means that it is relatively
trivial to find the eigenfunctions of the laplacian on $\Squashedsphere$.) There has also been
some work on $\Squashedsphere$ in the context of Kaluza--Klein compactification 
\cite{Okada86,ShenSobczyk87} (see also a discussion in \cite{Duffetal86}), and 
the asymptotic expansion of the heat kernel for the laplacian 
has been computed in \cite{ShtykovVassilevich95}. More recently, Dowker and collaborators
\cite{Dowker99,DeFranciaetal01} have considered the calculation of effective actions for both
scalars and fermions on $\Squashedsphere$, aiming at a comparison with the AdS/CFT results of
\cite{Chamblinetal99, Hawkingetal99,Emparanetal99}. The relation is far from obvious, since 
the regimes where the results are expected to apply are vastly different (see \cite{Taylor--Robinson00} for
a discussion).

\subsection{Geometry}

Written in terms of the coordinates $\theta,\varphi,\psi$, the metric (\ref{SquashedMetric}) is
\begin{equation}
g_{ij}=a(r)^2 \threebythree{1}{0}{0}{0}{\sin^2\theta+\grl\cos^2\theta}{\grl\cos\theta}{0}{\grl\cos\theta}{\grl}\;.
\end{equation}
The nonzero Christoffell symbols are 
\begin{equation}
\begin{split}
\Gamma^{\theta}_{\;\varphi\varphi}=(\grl-1)\sin\theta\cos\theta\;,\quad
\Gamma^{\theta}_{\;\varphi\psi}=\frac{\grl}{2}\sin\theta\;, \\
\Gamma^{\varphi}_{\;\theta\varphi}=\left(1-\frac{\grl}{2}\right)\frac{\cos\theta}{\sin\theta}\;,
\quad 
\Gamma^{\varphi}_{\;\theta\psi}=-\frac{\grl}{2}\frac{1}{\sin\theta}\;, \\
\Gamma^{\psi}_{\;\theta\varphi}=-\frac{\sin^2\theta+(2-\grl)\cos^2\theta}{2\sin\theta}\;,\quad
\Gamma^{\psi}_{\;\theta\psi}=\frac {\grl}{2}\frac{\cos\theta}{\sin\theta}\;.
\end{split}
\end{equation}
The Berger sphere is not an Einstein manifold, but it comes quite close: It satisfies the \emph{Einstein--Weyl} equation
$\Rcal_{[ij]}\sim g_{ij}$, where $\Rcal_{ij}$ is computed using the Weyl (non--Levi--Civita) connection. 
For a recent survey of Einstein--Weyl geometry, 
see \cite{CalderbankPedersen99}. On the physics side, it was shown in \cite{CritchleyDowker81} that (at least for 
$\mu$ close to zero), the spacetime can be self--consistently supported by the stress--tensor of a conformally coupled scalar.

The scalar curvature is given by 
\begin{equation}
\Rcal_h=\frac{1}{a(r)^2}\frac{4-\grl(r)}{2}
\end{equation}
which for the round sphere ($\grl=1$) of radius $1/k$ results in $\Rcal_h=3/2 k^2$. Notice that the unit sphere (which 
has $\Rcal_h=6$) corresponds to choosing $k=2$.  
There are two main cases to consider: If we are interested in the squashed sphere at a particular
radial distance in QTN, we have
\begin{equation} \label{abinner}
a(r)^2=\frac{r^2(1-\mu^2r^2)}{(1-k^2r^2)^2}\; , 
\quad b(r)^2=\frac{r^2(1-\mu^2k^2r^4)}{(1-k^2r^2)^2(1-\mu^2r^2)}\;.
\end{equation}
Then the scalar curvature at a radial distance $r$ is given by
\begin{equation}
\Rcal_r=\frac{(1-k^2r^2)^2}{2r^2(1-\mu^2r^2)^3}(3-8\mu^2r^2+4\mu^4r^4+\mu^2k^2r^4)
\end{equation}
This is the expression that is used when factorising the laplacian on QTN in section \ref{QTNfactorisation}. 
If, on the other hand, we want to look at the squashed sphere at the boundary at infinity, 
we have to choose a defining function. Choosing  $(1-k^2r^2)$ for this purpose, we end up with the
metric (\ref{SquashedMetric}) with 
\begin{equation} \label{abboundary}
a^2=\frac{1}{k^2}(1-\mu^2/k^2)\; , \quad b^2=\frac{1}{k^2}
\end{equation}
then the scalar curvature is 
\begin{equation}
\Rcal_B=\frac{k^2}{2}\left(\frac{3-4\mu^2/k^2}{(1-\mu^2/k^2)^2}\right)
\end{equation}
As mentioned in the main text (and also in \cite{Zoubos02}), the fact that $\Rcal_B$ becomes negative for $\mu^2>3/4 k^2$ should
have very interesting implications for the gauge theory side.

Before moving on to discuss the symmetries, we can quickly calculate the Cotton tensor (\ref{Cotton}), which, as discussed in 
section \ref{Conformal}, will tell us whether the space is conformally flat. Of course, it turns out to be nonzero apart
from the case of the round sphere. 

\subsection{Isometries and conformal isometries}

The squashed three--sphere has 4 Killing vectors, corresponding to its $\SU(2)\times\Urm(1)$
isometry group. We use the basis \cite{Britto--Pacumioetal99}
\begin{equation}
\begin{split}
l_1&=i\p_\varphi\;,\\
l_2&=i\{\sin\varphi\p_\theta+\frac{\cos\varphi}{\sin\theta}(\cos\theta\p_\varphi-\p_\psi)\}\;,\\
l_3&=i\{\cos\varphi\p_\theta-\frac{\sin\varphi}{\sin\theta}(\cos\theta\p_\varphi-\p_\psi)\}\;,\\
r_1&=i\p_\psi\;.
\end{split}
\end{equation}
The $l_i$ satisfy the commutation relations of $\SU(2)$, $[l_1,l_2]=i l_3$ etc., while
$r_1$ commutes with all of them. The notation comes from the fact that the $l_i$ 
correspond to the unbroken (left) part of the bi--invariant $\SU(2)_L\times\SU(2)_R$ 
isometry group, and $r_1$ is the remaining $\Urm(1)$ subgroup of $\SU(2)_R$.

For completeness, we include here the Killing vectors for the case of the round sphere, 
i.e. for the case that $\grl=1$ above. Then the four vectors above can be supplemented with
two more, 
\begin{equation}
\begin{split}
r_2&=i\{\sin\psi\p_\theta+\frac{\cos\psi}{\sin\theta}(\cos\theta\p_\psi-\p_\varphi)\}\;,\\
r_3&=i\{\cos\psi\p_\theta-\frac{\sin\psi}{\sin\theta}(\cos\theta\p_\psi-\p_\varphi)\}\;,\\
\end{split}
\end{equation}
restoring the full $\SU(2)_L\times\SU(2)_R$. Of course, in this case we also have a nontrivial 
conformal group. The conformal Killing vectors can be easily found by restricting the Killing
vectors of the $\AdS_4$ bulk, and can be taken to be
\begin{equation} \label{roundCKVs}
\begin{split}
k_1&=2i\sin\frac{\theta}{2}\cos\left(\frac{\varphi+\psi}{2}\right)\p_\theta
+i\frac{\sin\left(\frac{\varphi+\psi}{2}\right)}{\cos\frac{\theta}{2}}\left(\p_\varphi+\p_\psi\right)\\
k_2&=2i\sin\frac{\theta}{2}\sin\left(\frac{\varphi+\psi}{2}\right)\p_\theta
-i\frac{\cos\left(\frac{\varphi+\psi}{2}\right)}{\cos\frac{\theta}{2}}\left(\p_\varphi+\p_\psi\right)\\
k_3&=2i\cos\frac{\theta}{2}\cos\left(\frac{\psi-\varphi}{2}\right)\p_\theta
+i\frac{\sin\left(\frac{\psi-\varphi}{2}\right)}{\sin\frac{\theta}{2}}\left(\p_\varphi-\p_\psi\right)\\
k_4&=2i\cos\frac{\theta}{2}\sin\left(\frac{\psi-\varphi}{2}\right)\p_\theta
-i\frac{\cos\left(\frac{\psi-\varphi}{2}\right)}{\sin\frac{\theta}{2}}\left(\p_\varphi-\p_\psi\right)\\
\end{split}
\end{equation}
All told, we have ten Killing and conformal Killing vectors, corresponding to the conformal group
$\SO(1,4)$, which is of course the same as the isometry group of $\AdS_4$. We can easily check the
conformal Killing equation for the $k_n$s:
\begin{equation} \label{Killingeqk}
\Lcal_{k_n(x)}g_{ij}=\nabla_i k_{n,j}+\nabla_j k_{n,i}=2 \sigma_{k_n}(x) g_{ij}
\end{equation}
where $\sigma_{k_n}(x)=\nabla \cdot k_n(x)/3$. The coefficient of $\sigma_k$ is of course uniquely determined to
be 2, as we can see by acting with the inverse metric on (\ref{Killingeqk}).  

There is another special value of the squashing parameter $\grl$: When $\grl\ra \infty$ ($\mu\ra k$),
which corresponds to extreme prolate squashing, the three--sphere degenerates to a CR structure, 
the boundary of the unit sphere in $\Cset^2$. This is the boundary at infinity 
of the Bergman space $\Bergman$, which is the coset space $\SU(2,1)/\Urm(2)$. One would expect that, 
despite the degeneracy of the metric, 
the Bergman isometry group $\SU(2,1)$ will manifest itself as the conformal isometry group of the
boundary, as in the AdS case. In fact this was shown to be true by the authors of \cite{Britto--Pacumioetal99}, who
explicitly constructed the requisite conformal Killing vectors:
\begin{equation}
\begin{split}
b_1&=2i\sin\frac{\theta}{2}\cos\left(\frac{\varphi+\psi}{2}\right)\p_\theta
+ i\frac{\sin\left(\frac{\varphi+\psi}{2}\right)}{\cos\frac{\theta}{2}}\left(\p_\varphi+(2+\cos\theta)\p_\psi\right)\\
b_2&=2i\sin\frac{\theta}{2}\sin\left(\frac{\varphi+\psi}{2}\right)\p_\theta
- i\frac{\cos\left(\frac{\varphi+\psi}{2}\right)}{\cos\frac{\theta}{2}}\left(\p_\varphi+(2+\cos\theta)\p_\psi\right)\\
b_3&=2i\cos\frac{\theta}{2}\cos\left(\frac{\psi-\varphi}{2}\right)\p_\theta
+ i\frac{\sin\left(\frac{\psi-\varphi}{2}\right)}{\sin\frac{\theta}{2}}\left(\p_\varphi-(2+\cos\theta)\p_\psi\right)\\
b_4&=2i\cos\frac{\theta}{2}\sin\left(\frac{\psi-\varphi}{2}\right)\p_\theta
- i\frac{\cos\left(\frac{\psi-\varphi}{2}\right)}{\sin\frac{\theta}{2}}\left(\p_\varphi-(2+\cos\theta)\p_\psi\right)\;.\\
\end{split}
\end{equation}
The interesting point about these conformal Killing vectors is that they satisfy the conformal Killing equation with a 
different coefficient than the canonical one in (\ref{Killingeqk}): If we again 
define $\sigma_{b_n}(x)=(\nabla \cdot b_n)/3$, we find that
\begin{equation}
\Lcal_{b_n(x)} g_{ij}=\frac32 \sigma_{b_n}(x)g_{ij}\;.
\end{equation}
As discussed in \cite{Britto--Pacumioetal99}, there is nothing wrong with this since
here there is no inverse metric so it need not impose a particular coefficient.

After this discussion of the two interesting limiting cases of the metric, we will now assume we are at a generic 
value of $\grl$. 
In discussing two--point functions on the squashed sphere, we find it extremely 
convenient to work with variables invariant under the $\SU(2)\times\Urm(1)$ symmetry group, rather
than the Euler angles of the two points ($\theta_r,\varphi_r,\psi_r$ and $\theta_s,\varphi_s,\psi_s$ 
respectively). We would also, of course, like to work with variables that are symmetric under
the interchange $x_r\leftrightarrow x_s$, since then the symmetry of the two--point function is
manifest. These two requirements lead us to define the two combinations 
\begin{equation}
\vone=\cos\frac{\tr}{2}\cos\frac{\ts}{2}\cos\half(\psi_r+\fr-\psi_s-\fs)
+\sin\frac{\tr}{2}\sin\frac{\ts}{2}\cos\half(\psi_r-\fr-\psi_s+\fs)
\end{equation}
and
\begin{equation} 
\vtwo=\half\{1+\cos\theta_r\cos\theta_s+
\sin\theta_r\sin\theta_s\cos(\varphi_r-\varphi_s)\}\;\;.
\end{equation} 
Any Green's function on $\Squashedsphere$ can be written as a function of $\vone$ and $\vtwo$ only
(six Euler angles minus four constraints from $\SU(2)\times\Urm(1)$ equals two independent variables).
One can check that both $\vone$ and $\vtwo$ are $\SU(2)\times\Urm(1)$ invariant, by which 
we mean that for any Killing vector $v\in{\{l_i,r_1\}}$, we have
\begin{equation}
(\Lcal_{v(x_r)}+\Lcal_{v(x_s)})\vone=0\;,\quad (\Lcal_{v(x_r)}+\Lcal_{v(x_s)})\vtwo=0\;.
\end{equation}
However, if we want to impose invariance under the full $\SU(2)_L\times\SU(2)_R$ of the round sphere, 
i.e. allow $v$ to be $r_2$ or $r_3$, we find that $\vtwo$ is not invariant, while $\vone$ remains so. This means that 
on the round sphere two--point functions will depend only on $\vone$, as is indeed the case for
$M^0(x,x')$ in section \ref{Twopoint}.

\subsection{The laplacian}

From the metric (\ref{SquashedMetric}) we readily obtain the laplacian 

\begin{equation} \label{LaplacianSquashed}
\begin{split}
\nabla_B^2&=\frac{1}{\sqrt{g}}\p_i\sqrt{g}g^{ij}\p_j\\
&=\frac{1}{a^2}\left[\p_{\theta\theta}+\cot\theta\p_\theta
+\frac{1}{\sin^2\theta}(\p_{\varphi\varphi}-2\cos\theta\p_{\varphi\psi}+\p_{\psi\psi})\right]
+\left[\frac{1}{b^2}-\frac{1}{a^2}\right]\p_{\psi\psi}
\end{split}
\end{equation}
Here the subscript $B$ can be taken to stand for ``Berger'' or ``Boundary''. It is useful to define the operators
\begin{equation}
-\Delta_0=\p_{\theta\theta}+\cot\theta\p_\theta
+\frac{1}{\sin^2\theta}(\p_{\varphi\varphi}-2\cos\theta\p_{\varphi\psi}+\p_{\psi\psi})
\end{equation}
and
\begin{equation} 
-\Delta_1=\p_{\psi\psi}
\end{equation}
We will also find it convenient to work with a laplacian that is positive definite, and which depends
only on the squashing (and not on the overall scale). So we define an operator $\Delta_{(r)}$ as
\begin{equation} \label{Deltar}
\Delta_{(r)}=-a^2\nabla_B^2\;.
\end{equation} 
Given the definitions of $\Delta_0$ and $\Delta_1$ above, at the inner boundary located at a particular 
radial distance $r$, we have (recall (\ref{abinner}))
\begin{equation} \label{Deltarone}
\Delta_{(r)}=\Delta_0+\mu^2r^2\frac{2-\mu^2r^2-k^2r^2}{1-\mu^2k^2r^4}\Delta_1\;.
\end{equation}
At the boundary at infinity, given by $r\ra 1/k$, we have, from (\ref{abboundary}) or directly from (\ref{Deltarone})
\begin{equation} \label{DeltaB}
\Delta_B=\lim_{r\ra 1/k} \Delta_{(r)}=\Delta_0+\frac{\mu^2}{k^2}\Delta_1\;.
\end{equation}

\subsection{Separation of variables}
 
To look for eigenfunctions of the laplacian (\ref{LaplacianSquashed}), we separate variables 
in a standard way: 
\begin{equation} \label{Ylmn}
\Ylmn(\theta,\varphi,\psi)=e^{-im\varphi}e^{-in\psi}\eta^{l}_{\;mn}(\theta)
\end{equation}
with $\eta^{l}_{\;mn}(\theta)$ to be determined.
The variables $\varphi$ and $\psi$ separate immediately and we are left with 
\begin{equation}
\nabla_B^2\Ylmn(x)=\frac{1}{a^2}\left[\p_{\theta\theta}+\cot\theta\p_\theta
-\frac{1}{\sin^2\theta}(m^2-2mn\cos\theta+n^2)\right]\Ylmn(x)
-n^2\left[\frac{1}{b^2}-\frac{1}{a^2}\right]\Ylmn(x)\;,
\end{equation}
where $x$ stands for $(\theta,\vf,\psi)$. Renaming $\cos\theta=z$, we obtain 
\begin{equation} \label{SUtwofirst}
\frac{1}{a^2}\left[(1-z^2)\p_{zz}-2z\p_z-\frac{m^2-2mnz+n^2}{1-z^2}\right]\Ylmn(x)-n^2\left[\frac{1}{b^2}
-\frac{1}{a^2}\right]\Ylmn(x)\;.
\end{equation}
The first bracket is a well--known differential equation, arising from representation theory on $\SU(2)$, 
and its eigenfunctions are the Wigner functions (\cite{Wigner}, see also \cite{Varshalovichetal, Vilenkin}), 
sometimes denoted $d^{l}_{\;mn}(z)$.
 We follow \cite{Vilenkin} in denoting them $\Plmn(z)$, a notation that indicates that they
are simply a generalisation of the Legendre functions $\Prm_l(z)$ and associated Legendre functions 
$\Prm^l_m(z)$. 
The eigenvalues of the Wigner functions are simply $-l(l+1)$, depending only on the principal quantum
number $l$. In other words, the first bracket of (\ref{SUtwofirst}) acts on $\Plmn(z)$ as:
\begin{equation} \label{SUtwo}
\left[(1-z^2)\p_{zz}-2z\p_z-\frac{m^2-2mnz+n^2}{1-z^2}\right]\Plmn(z)=-l(l+1)\Plmn(z)\;.
\end{equation}
We can call this equation the ``$\SU(2)$ differential equation'' since it always arises when doing 
representation theory on $\SU(2)$, as for instance when describing the spherical top. 
As discussed in the beginning of this appendix, it is a nice 
(and well--known) result that the Wigner functions are also eigenfunctions of the more general 
equation (\ref{SUtwofirst}) corresponding to the \emph{homogeneously} squashed sphere, or equivalently
to the symmetric top, with very simple eigenvalues\footnote{This is no longer true when we consider the \emph{asymmetric} top, or
equivalently the full mixmaster solution $a^2\sone^2+b^2\stwo^2+c^2\sthree^2$. There the analysis 
becomes much more involved and can only be approached numerically \cite{Hu73}.}. So substituting 
$\eta^{l}_{\;mn}(z)=\Plmn(z)$ in (\ref{Ylmn}), we have that the $\Ylmn(\theta,\varphi,\psi)$ satisfy
\begin{equation}
\nabla_B^2\Ylmn(\theta,\varphi,\psi)=
\left\{-\frac{l(l+1)}{a^2}-\left(\frac{1}{b^2}-\frac{1}{a^2}\right)n^2\right\} \Ylmn(\theta,\varphi,\psi)\;.
\end{equation}
As mentioned earlier, we find it convenient to use the positive laplacian $\Delta_{(r)}$ defined in (\ref{Deltar})
rather that $\nabla^2$.  The operators $\Delta_0$ and $\Delta_1$ we defined above clearly give
\begin{equation}
\Delta_0\Ylmn(\theta,\varphi,\psi)=l(l+1)\Ylmn(\theta,\varphi,\psi)
\end{equation}
and
\begin{equation}
\Delta_1\Ylmn(\theta,\varphi,\psi)=n^2\Ylmn(\theta,\varphi,\psi)
\end{equation}
so that 
\begin{equation}
\Delta_{(r)}\Ylmn(\theta,\varphi,\psi)=\left\{l(l+1)-\frac{\mu^2r^2(2-\mu^2r^2-k^2r^2)}{(1-\mu^2k^2r^4)}n^2\right\}
\Ylmn(\theta,\varphi,\psi)\;.
\end{equation}
Since $\Delta_{(r)}$ is finite as $r\ra 1/k$, if we want to look at the Berger sphere at the boundary at 
infinity (as in 
the eigenfunction expansions of sections \ref{Roundsphere} and \ref{Bergersphere}), 
we can just take $r=1/k$ in $\Delta_B$. We thus obtain\footnote{Note that this is still positive since $\mu\leq k$ and 
for unitary representations $n\leq l$.} :
\begin{equation}
\Delta_B\Ylmn(\theta,\varphi,\psi)=\left\{l(l+1)-\frac{\mu^2}{k^2}\cdot n^2\right\}\Ylmn(\theta,\varphi,\psi)\;.
\end{equation}
If the range of $l$ is taken to be $l=0,\half,1,\ldots\infty$, and that of $m$ and $n$ as $-l,-l+1,\ldots,l-1,l$, the
 $\Ylmn(\theta,\vf,\psi)$ form a complete orthonormal basis on $\Squashedsphere$. They satisfy the following orthogonality relation
\begin{equation}
\int_0^\pi\int_0^{2\pi}\int_0^{4\pi}\mathrm{Y}^{l_1}_{\;m_1n_1}(\theta,\vf,\psi)
\overline{\mathrm{Y}}^{l_1}_{\;m_1n_1}(\theta,\vf,\psi)\diff\psi\diff\vf\sin\theta\diff\theta
=\frac{16\pi^2}{2l+1}\delta_{l_1l_2}\delta_{m_1m_2}\delta_{n_1n_2}\;,
\end{equation}
and also completeness
\begin{equation}
\sum_{lmn}\frac{2l+1}{16\pi^2}\Ylmn(\theta,\vf,\psi)
\overline{\mathrm{Y}}^{l}_{\;mn}(\theta',\vf',\psi')=\delta(\theta-\theta')\delta(\vf-\vf')\delta(\psi-\psi')\;.
\end{equation}
For this reason in the main text we actually use a normalised version:  
\begin{equation}
\Ylmn(\theta,\vf,\psi)=\sqrt{2l+1}e^{-im\phi}e^{-in\psi}\Plmn(\cos\theta)\;.
\end{equation}

\section{Pseudodifferential Operators on Manifolds} \label{Pseudodifferential}

Here we will give just a few basic definitions from the theory of pseudodifferential operators.  
The reader interested in better understanding pseudodifferential operators should consult one of the books on 
the subject (e.g. \cite{Taylor, Kumano-go, Treves}). See also \cite{Gilkey84} for a discussion oriented
towards applications to index theorems.

The definition of a pseudodifferential operator on $\Rset^d$ usually starts from the observation that one
 can think of a differential operator $P$ acting on a function $f(x)$ as defined through its
Fourier transform:
\begin{equation} \label{pseudodiffdefinition}
P f(x)=\frac{1}{(2\pi)^d}\int_{\Rset^d}e^{ix\cdot\xi}p(x,\xi)\hat{f}(\xi)\diff \xi\;.
\end{equation}
Here $\hat{f}$ is the Fourier transform of $f$, and the function $p(x,\xi)$ is called the \emph{symbol} of $P$. 
 Now for differential operators, which for order $m$ can be written
as $P=\sum_{i=0}^{m} a_i(x)(-i\p_x)^i$, the symbol is a polynomial in $\xi$: $p(x,\xi)=\sum_{i=0}^m a_i(x) \xi^i$.
It is natural to generalise by considering symbols that are not polynomials, 
and this leads us to pseudodifferential operators. In that case one has to be more precise about the definition of
the symbol of the operator, and thus its order. We will not need the full definition, 
but only a more restricted one called
a \emph{classical} symbol: We say that a (classical) pseudodifferential operator belongs to $\mathbb{S}^m$ (the set of 
pseudodifferential operators of order $m$) if its symbol $p(x,\xi)$ is smooth in $x$ and $\xi$, and is  
an asymptotic expansion in homogeneous terms, the degree of each term decreasing by an integer. That is, 
\begin{equation} \label{classical}
p(x,\xi)=\sum_{i=0}^{\infty} p_{m-i}(x,\xi)
\end{equation}
where $p_{m-i}$ is a homogeneous function of degree $m-i$. This definition is sufficient for our purposes, since, as 
discussed in \cite{SylvesterUhlmann88}, the Dirichlet--to--Neumann operator is a classical pseudodifferential operator.

Another useful definition is that of the \emph{principal symbol} of a pseudodifferential  operator, which is the 
highest--order part $p_m$ of its symbol. The principal symbol contains a large amount of information about the
operator, for instance it is enough to decide whether the operator is elliptic or not.  

In the theory of pseudodifferential operators, one often encounters \emph{infinitely smoothing} operators, the operators
whose symbols belong to $\mathbb{S}^{-\infty}$ (defined as the intersection of all the $\mathbb{S}^m$). These operators
will completely smoothen out the support of any distribution. They can be arbitrarily added to any asymptotic
series (like (\ref{classical})), so one is often forced to work modulo smoothing operators.  

Probably the major conceptual difference between differential and pseudodifferential operators is that the latter
are nonlocal. However they are a very restricted set of nonlocal operators,  being what is known as \emph{pseudolocal}. 
This means that, although they are defined through the Fourier
transform, which smears out the support of the function they act on, they don't smear out the \emph{singular}
support. In other words, if a function is smooth in some domain, then $P$ acting on it will also be smooth in 
that domain. 

It is a well--known result, due to Seeley, that any complex power of an elliptic operator is pseudodifferential. This
tells us that the operator (\ref{BAdS}) is well defined. As for the expression we found for the full Dirichlet--to--Robin
operator on QTN, we will need the following result (see e.g. \cite{Taylor}, p. 297): A function $p(A_1,A_2)$ of two
self--adjoint operators is pseudodifferential if $A_1$ and $A_2$ (in our case $\sqrt{\Dzero+1/4}$ and $\sqrt{\Done}$) commute, 
$A_1^2+A_2^2$ is elliptic, and the function $p(\grl)$ is a symbol in the appropriate symbol class. The first two conditions
are trivial to show, however we have actually shown the third condition (cf. (\ref{Basymptotic}))  only for a special case. 

So far we have been discussing pseudodifferential operators on $\Rset^d$. However they can be readily
defined on curved manifolds, by first defining them on a coordinate patch (transferring back to 
$\Rset^d$), and then requiring that the definition holds for all the charts in an atlas. We will not
require much knowledge of the technical issues, however it is worth noting that only the \emph{principal
symbol} is diffeomorphism invariant. Clearly this makes it rather tedious to work with 
pseudodifferential operators in a covariant manner, and there has been a lot of effort to try  to overcome this difficulty
(see e.g. \cite{FullingKennedy88} for a discussion).

Fortunately in our case  things are rather simple: As dicussed in appendix \ref{BergerSphere}, the Berger
sphere allows an explicit expansion into known eigenfunctions, and the eigenvalues are simple enough that we can 
directly act on these eigenfunctions with any pseudodifferential operator by naively applying the spectral theorem. 
That means that, assuming we know the eigenfunctions $Y^a$ of some (pseudo)differential operator $A$, we write 
\begin{equation}
A Y^a=\grl_a Y^a\quad \Rightarrow \quad p(A) Y^a=p(\grl_a) Y^a\;.
\end{equation}
Apart from the very important application of pseudodifferential operators in the Atiyah--Singer index theorem (see
e.g. \cite{Gilkey84}), they have long been useful in quantum field theory in curved spacetimes (see e.g. \cite{Fulling89}):
For $P$ an elliptic, positive order operator, the heat kernel $e^{-tP}$ and the zeta function
$Tr(P^{-s})$ are pseudodifferential. As for the square root of the laplacian, it has also appeared in quantum field 
theory in various contexts, mainly related to the construction of non--local actions (some sample references in this field
are  \cite{Lammerzahl93,Junker95,Gorbar97}). \cite{Junker95} actually discusses a factorisation of the (flat) laplacian similar 
to that in section \ref{Definition}. In its lorentzian, flat space edition, the square root of
the wave operator in 2+1 dimensions is known to satisfy Huygens' principle (unlike the wave operator itself)
\cite{BolliniGiambiagi93, Barcietal96} and thus it is natural for it to appear in 2+1--dimensional bosonization
(e.g. \cite{Marino91}) (and thus also invalidating to some extent our comment in a footnote in the introduction that the 
``square root of the laplacian is nonlocal so has nothing to do with the Dirac operator'').

\bibliography{References}
\bibliographystyle{JHEP}

\end{document}